%% file: main.tex
\documentclass[10pt,journal,compsoc]{IEEEtran}

\ifCLASSOPTIONcompsoc
  \usepackage[nocompress]{cite}
\else
  \usepackage{cite}
\fi
\usepackage{graphicx}
\usepackage{balance}
\usepackage{comment}
\usepackage{subfigure}
\usepackage{booktabs}
\usepackage{amsmath}
\usepackage{amssymb}
\usepackage{amsfonts}
\usepackage{bm}
\usepackage{colortbl}
\usepackage{tabularx}
\usepackage{color}
\usepackage{epsfig}
\usepackage{xspace}
\usepackage{hyperref}    % Creates hyperlinks from ref/cite
\hypersetup{pdfstartview=FitH}
\usepackage{graphicx}    % For importing graphics
\usepackage{url}         %
\usepackage{algorithm}
\usepackage{algpseudocode}
\usepackage{placeins}
\usepackage{multirow}
\usepackage{epstopdf}
\usepackage{tabularx}
\usepackage[colorinlistoftodos]{todonotes}
\usepackage{blindtext}

\newcommand{\vtbo}[1]{{\color{black}{#1}}}
\newcommand{\cxbo}[1]{{\color{black}{#1}}}

\DeclareMathOperator*{\argmin}{arg\,min}

\usepackage{lipsum}

\begin{document}

\title{From Real to Complex: Enhancing Radio-based Activity Recognition Using Complex-Valued CSI}
% \title{cRF: Radio-based Device-free Activity Recognition with Radio Frequency Interference}

\author{Bo~Wei,
        Wen~Hu,~\IEEEmembership{Member,~IEEE,}
        Mingrui~Yang,~\IEEEmembership{Member,~IEEE,}
        and~Chun~Tung~Chou,~\IEEEmembership{Member,~IEEE,}% <-this % stops a space

}

\markboth{Journal of \LaTeX\ Class Files,~Vol.~14, No.~8, August~2015}%
{Shell \MakeLowercase{\textit{et al.}}: Bare Demo of IEEEtran.cls for Computer Society Journals}

\IEEEtitleabstractindextext{%
\begin{abstract}

Activity recognition is an important component of many pervasive computing applications. Radio-based activity recognition has the advantage that it does not have the privacy concern and the subjects do not have to carry a device on them. Recently, it has been shown channel state information (CSI) can be used for activity recognition in a device-free setting. With the proliferation of wireless devices, it is important to understand how radio frequency interference (RFI) can impact on pervasive computing applications. In this paper, we investigate the impact of RFI on device-free CSI-based location-oriented activity recognition. We present data to show that RFI can have a significant impact on the CSI vectors. In the absence of RFI, different activities give rise to different CSI vectors that can be differentiated visually. However, in the presence of RFI, the CSI vectors become much noisier and activity recognition also becomes harder. Our extensive experiments show that the performance of state-of-the-art classification methods may degrade significantly with RFI. We then propose a number of counter measures to mitigate the impact of RFI and improve the location-oriented activity recognition performance. We are also the first to use complex-valued CSI to improve the performance in the environment with RFI. 

\end{abstract}

}

\maketitle

\IEEEdisplaynontitleabstractindextext

\IEEEpeerreviewmaketitle

\input{Tex/intro}

\input{Tex/background}

\input{Tex/method}

\input{Tex/experiment}
\input{Tex/related}

\section{Conclusion}\label{sec:conclusion}
In this paper, we investigate the performance of radio based device-free location-oriented activity recognition systems under RFI using complex-valued CSI, and propose a novel fusion algorithm based on SRC that can improve the recognition performance of the systems by up to 10\% when RFI is present. Our prototype robust location-oriented activity recognition systems require only one pair of nodes for a one-bedroom apartment, which enables easy system set-up and maintenance. Finally, we
use an embedded wide band radio device (WASP platform) to emulate and study the recognition performance of popular wireless communication protocols that have different bandwidths under RFI.

\bibliographystyle{abbrv}
\bibliography{sigproc_short}

\begin{IEEEbiography}[{\includegraphics[width=1in,height=1.25in,clip,keepaspectratio]{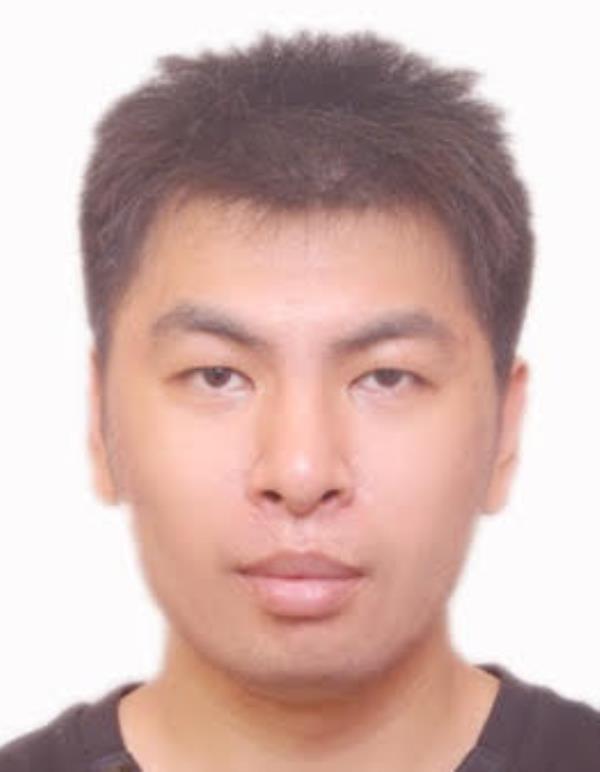}}]{Dr. Bo Wei}
has been a Senior Lecturer in Computer Science at Teesside University since January 2017. Before joining Teesside, he was a postdoctoral research assistant at the University of Oxford. He obtained his PhD degree in Computer Science and Engineering in 2015 from the University of New South Wales, Australia. He obtained his Masters degree in Computer System Architecture in 2011 and his Bachelor degree in Computer Science in 2009, both from Northeastern University, China. He was also a research student in the Commonwealth Scientific and Industrial Research Organisation (CSIRO) Australia from August 2011 to May 2015, and he visited the Swedish Institute of Computer Science (SICS) from April 2013 to November 2013 as a visiting research student. 
\end{IEEEbiography}

\begin{IEEEbiography}[{\includegraphics[width=1in,height=1.25in,clip,keepaspectratio]{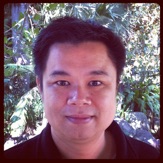}}]{Dr. Wen Hu}
is a senior lecturer at School of Computer Science and
Engineering, the University of New South Wales (UNSW). Much of his
research career has focused on the novel applications, low-power
communications, security and compressive sensing in sensor network
systems and Internet of Things (IoT). Hu published regularly in the
top rated sensor network and mobile computing venues such as ACM/IEEE
IPSN, ACM SenSys, ACM transactions on Sensor Networks (TOSN),  IEEE
Transactions on Mobile Computing (TMC), and Proceedings of the IEEE.

Hu was a principal research scientist and research project leader at
CSIRO Digital Productivity Flagship, and received his Ph.D from the
UNSW. He is a recipient of prestigious CSIRO Office of Chief Executive
(OCE) Julius Career Award (2012 - 2015) and multiple research grants
from Australian Research Council, CSIRO and industries.

Hu is a senior member of ACM and IEEE, and is an associate editor of
ACM TOSN,  as well as serves on technical advisory board (IoT) of ACS
and the organising and program committees of networking conferences
including ACM/IEEE IPSN, ACM SenSys, ACM MobiSys, ACM/IEEE IOTDI, IEEE
ICDCS, IEEE LCN, IEEE ICC, IEEE WCNC, IEEE DCOSS, IEEE GlobeCom, IEEE
PIMRC, and IEEE VTC.
\end{IEEEbiography}

\begin{IEEEbiography}[{\includegraphics[width=1in,height=1.25in,clip,keepaspectratio]{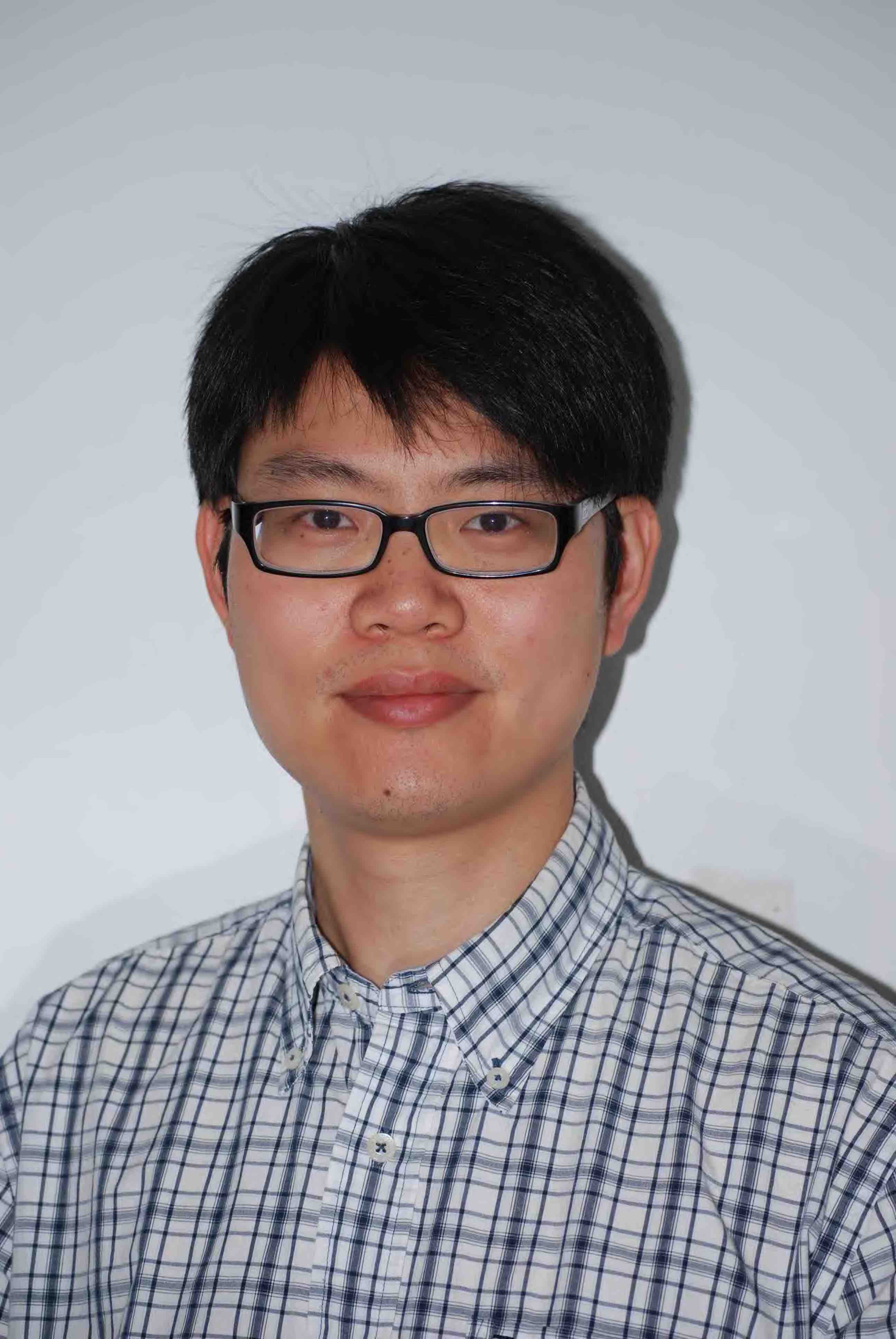}}]{Dr. Mingrui Yang}
is a Senior Research Associate in the department of
Radiology in School of Medicine at Case Western Reserve University. He received his Ph.D. in Mathematics from the University of South Carolina, USA in 2011. After that, he worked as a Postdoctoral research fellow in Digital Productivity Flagship of the Commonwealth Scientific and Industrial Research Organization (CSIRO), Australia. He is interested in interdisciplinary researches in deep learning, compressive sensing, sparse approximation, signal/image processing, greedy algorithms and nonlinear approximation, and their applications in magnetic resonance imaging, sensor networks and hyper-spectral imaging. He has published in high quality journals and conferences in mathematics, computer science, and medicine. He has also served numerous times as a reviewer or TPC member for related journals and conferences. He is a member of ISMRM and IEEE.
\end{IEEEbiography}

\begin{IEEEbiography}[{\includegraphics[width=1in,height=1.25in,clip,keepaspectratio]{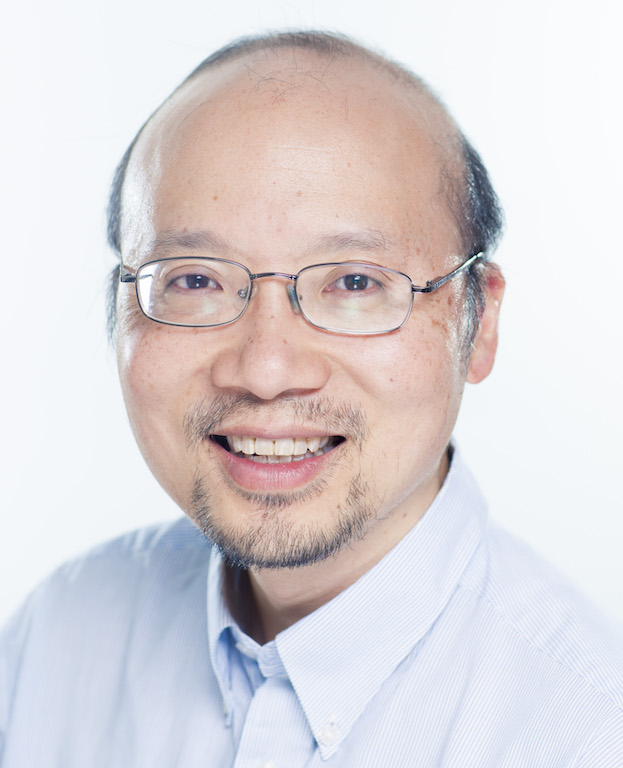}}]{Dr. Chun Tung Chou}
received the BA degree in engineering science from the University of Oxford, UK and the PhD degree in control engineering from the University of Cambridge, UK. He is an associate professor at the School of Computer Science and Engineering, The University of New South Wales, Australia. He is on the editorial board of IEEE Transactions on Molecular, Biological and Multi-Scale Communications; IEEE Wireless Communications Letters and Nano Communication Networks. His current research interests are molecular communications, nano communications and pervasive computing. He is a member of the IEEE.
\end{IEEEbiography}

\end{document}

%% file: Tex/intro.tex
%\vspace{-2mm}

\section{Introduction}\label{intro}
\label{sec:intro}
\let\thefootnote\relax\footnotetext{\cxbo{Part of this research work has been published in the proceedings of the 14th International Conference on Information Processing in Sensor Networks (IPSN '15) \cite{wei2015radio}. 
We propose a new device-free activity recognition method using complex-valued Channel State Information~(CSI) in Section \ref{sec:methods} in this paper. The discussion regarding the evaluation on complex-valued CSI can be found in Section \ref{sec:experiments}}
}
Activity recognition aims to identify what a subject is doing. It is an important component of many pervasive computing applications. For example, the increasing greying population in many countries puts a rising pressure on the health care system. Activity recognition can be used to improve home care for the elderlies. This paper considers the activity recognition problem using radio signals, in the device-free setting, with an emphasis on making  activity recognitions robust to radio frequency interference (\emph{RFI}). 

Activity recognition is a well researched topic. In terms of system components, there are three broad approaches to the activity recognition problem: camera-based \cite{oliver2002layered,harville2004fast,zhao2005real,cohen2003inference,kinect}, sensor-based \cite{kwapisz2011activity,bao2004activity,ravi2005activity,jawboneup,xiaomiwristband,hao2013isleep,yatani2012bodyscope} and 
device-free \cite{wang2014eyes,sigg2014rf,wang2017device}. Cameras are able to provide high resolution data for activity recognition but privacy is a serious concern. Although there is no privacy concern with the sensor-based approach, it imposes the requirement that a subject has to carry sensors on his body. This is inconvenient and activity recognition fails if the subject forgets to carry the device. We have therefore chosen the device-free approach in this paper. 

In the device-free approach to activity recognition, radio devices are placed in the periphery of a monitored area, called the area of interest (AoI). These radio devices send packets to each other regularly and use the received radio signal to obtain information on the radio environment. The key idea is that the radio environment is influenced by the activity taking place in the AoI. The activity recognition problem is to infer the activity from the received radio signal. In general, there are \emph{three requirements} for device-free activity recognition: \emph{informative measurements}, \emph{capability to deal with environment changes}, and \emph{robustness to RFI}. 

Using informative measurements is a pre-requisite for any successful classification problem. Although coarse-grained radio channel measurements, such as radio signal-strength indicators (RSSI), have been successfully used for device-free indoor localisation \cite{youssef2007challenges,WilsonRTI:2010,WeidRTI:2015}, they are no longer informative for activity recognition. Recent work on device-free activity recognition \cite{wang2014eyes,sigg2014rf,wang2017device} has therefore used Channel Frequency Response (CFR) or Channel State Information (CSI) for activity recognition. This is also our observation and our proposed solution therefore uses CSI. 

A challenge for CSI-based device-free activity recognition is that the CSI of the radio channel is sensitive to changes in the environment due to for example new or moved furniture. This is because the CSI-fingerprint of an activity is affected by multi-path effects of the environment. \cxbo{A CSI-based activity recognition method called E-eyes \cite{wang2014eyes} proposed a semi-supervised approach to address the issue of environmental changes. }
When the system detects that the CSI-fingerprint for an activity has changed, E-eyes requires the users to label the new CSI-instances manually. A similar approach was used in \cite{sen2012you} to deal with the impact of environmental changes on CSI-based localisation. 

\begin{figure}[htb]
   \centering
    \includegraphics[scale = .5]{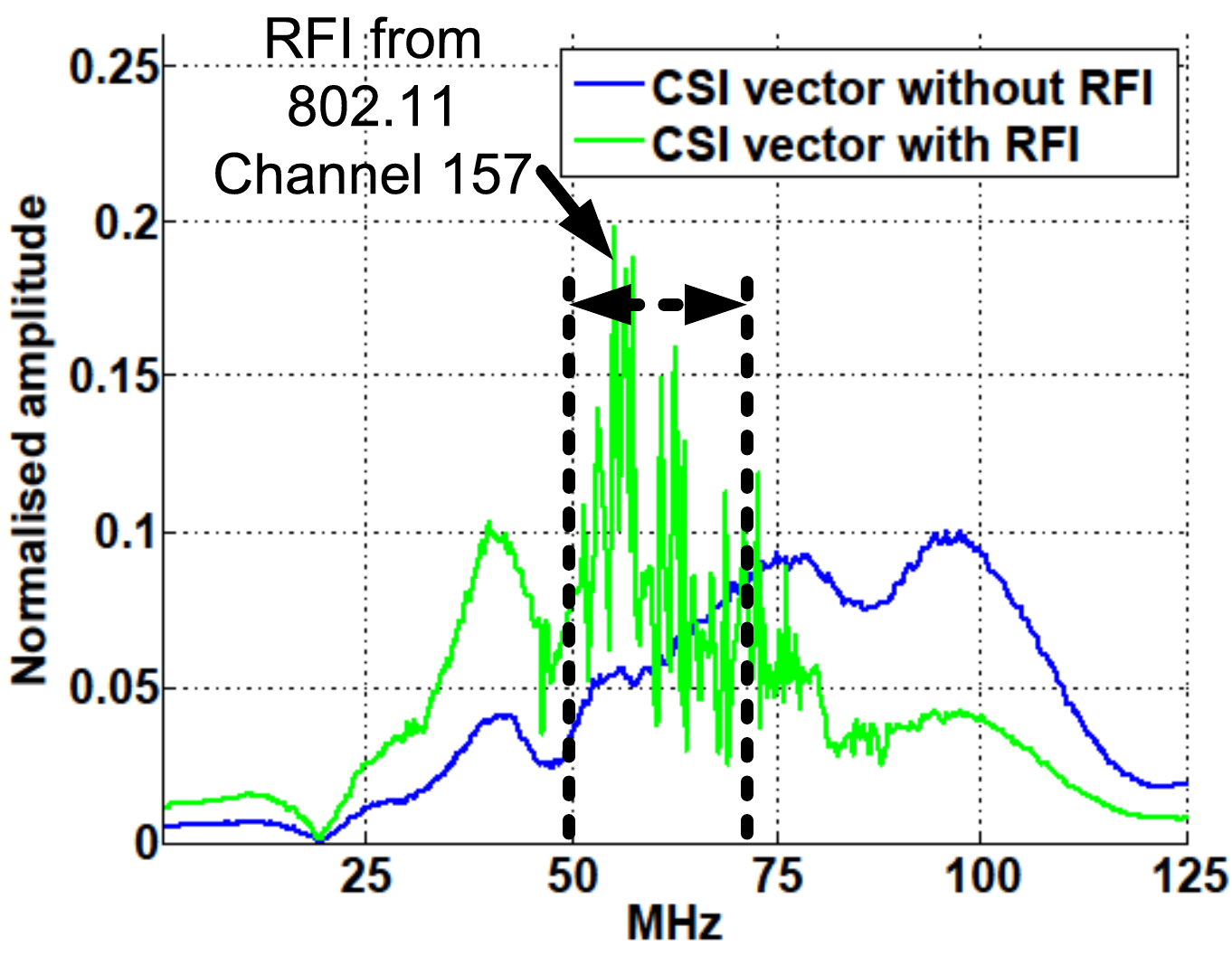}
    \caption{CSI without and with RFI.}\label{fig:sec1:exampleSSNR}
\end{figure} 

Another challenge for CSI-based activity recognition is that the CSI is highly influenced by RFI. The number and types of radio devices have proliferated in the last decade. It is hardly possible to find a frequency band that is clean or without RFI. Our observation, which is depicted in Fig.~\ref{fig:sec1:exampleSSNR}, shows that CSI is {\bf highly impacted} by RFI. The figure shows a CSI without RFI as well as a CSI with RFI in a particular radio channel. The impact of RFI on CSI is conspicuous. This implies that CSI-based activity recognition must be robust to RFI. This is the key topic of this paper. 
% \cxbo{Our previous research \cite{wei2015radio} proposed the Sparse Representation Classification (SRC) \cite{Wright:09} based method on the amplitudes of CSI vectors (real-valued CSI) to mitigate the influence of RFI. In this paper, we further improve the recognition performance by using complex-valued CSI.}

\begin{figure}[thb]
   \centering
    \includegraphics[scale = .7]{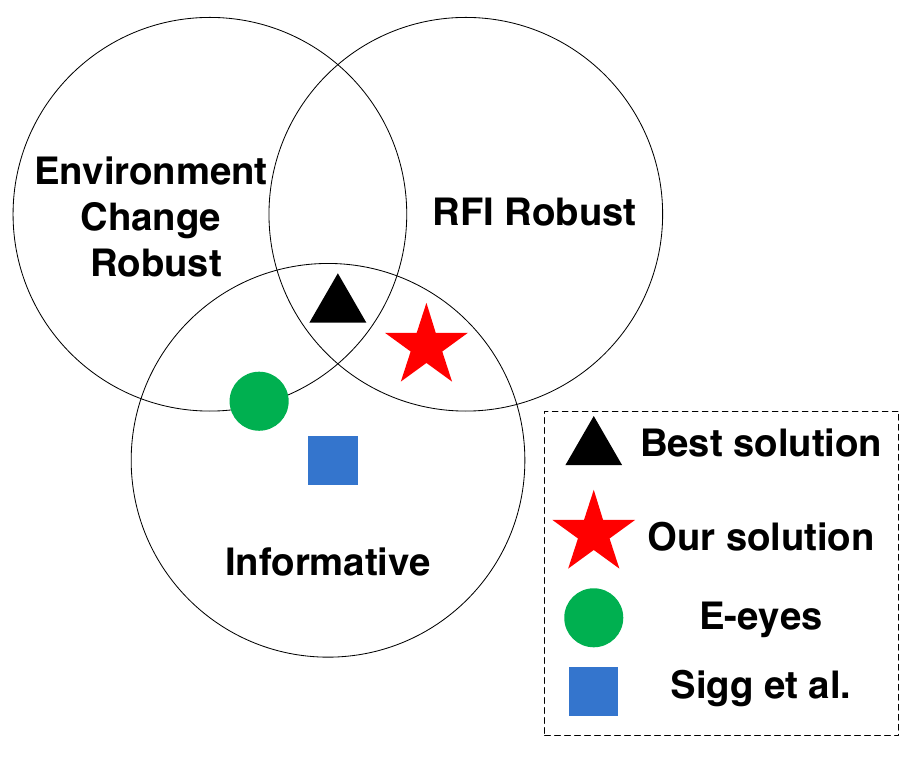}
    \caption{Differentiating our work from other recent work on CSI-based activity recognition.}\label{fig:research_scope}
\end{figure}

As a summary of the above discussion, we have drawn a Venn diagram with the three requirements of informative measurements, robustness to environmental changes, and robustness to RFI in Fig.~\ref{fig:research_scope}. This diagram is used to differentiate our work (indicated by a star) from two other CSI-based activity recognition solutions: E-eyes \cite{wang2014eyes} and Sigg et al.~\cite{sigg2014rf}. It shows which solution is able to deal with which requirements. 
\cxbo{
% Two important remarks are appropriate here.
Neither of \cite{wang2014eyes,sigg2014rf} addresses the impact of RFI on CSI. 
% Second, our proposed solution can use the semi-supervised approach of \cite{wang2014eyes} to deal with the impact of environmental changes. This means our proposed solution can deal with all the three requirements.  
}

% what we do
\cxbo{In order to enable the robustness of CSI-based location-oriented activity classification, our previous research \cite{wei2015radio} exploits the SRC approach based on $\ell_1$-optimisation.} SRC has been shown to be robust to noise. It has also been shown to significantly improve the classification performance in face recognition \cite{Wright:09,Shen2014Face} and acoustic classification \cite{wei2013real}, outperforming other classification approaches such as support vector machine (SVM) and $k$-nearest neighbours ($k$NN). \cxbo{Our previous research \cite{wei2015radio} also proposed SRC based method on the amplitudes of CSI vectors (real-valued CSI) to mitigate the influence of RFI and increases the performance of real-valued CSI based activity recognition.
In this paper, we propose a new classification method to further enhance the robustness of activity recognition using \textbf{complex-valued} CSI. }
Location-oriented activity recognition is able to indicate the location information as well as types of activities.

% Our previous research \cite{wei2015radio} proposed the Sparse Representation Classification (SRC) \cite{Wright:09} based method on the amplitudes of CSI vectors (real-valued CSI) to mitigate the influence of RFI. In this paper, we further improve the recognition performance by using complex-valued CSI.

To summarise, the contributions and novelties of this paper are: 
\begin{itemize}
	\item We demonstrate by using measurements that CSI is highly impacted by RFI. We show that, while the CSI vectors for different activities in a RFI-free environment are clearly distinguishable even by naked eyes, this is no longer the case for an environment with RFI. 
	
  \item \cxbo{To address the above challenge, we propose a novel complex-valued CSI based classification algorithm for CSI-based location-oriented activity recognition. The algorithm fuses the results from a number of $\ell_1$-optimisation according to their signal-to-noise ratios (SNRs). We show that this method can boost recognition accuracy and outperforms $k$NN and other SRC-based methods by up to 10\%. }
  
  \item \cxbo{ We are also the first to use complex-valued CSI for a SRC-based classification algorithm. Complex valued CSI contains phase information additional to amplitude information. We show this can improve the performance compared with using only real-valued CSI. }

\item We study the impact of channel bandwidth on the accuracy of activity recognition. We use 4 different bandwidths: 5, 10, 15 and 20 MHz, which cover low bandwidth devices (e.g. ZigBee) and high bandwidth devices (e.g. WiFi). We show that our proposed classification algorithm produces good classification accuracy for activities with 20~MHz of bandwidth. 

\end{itemize}

The rest of this paper is organised as follows. Section \ref{sec:background} presents background materials on CSI and SRC. We then study the impact of RFI on CSI and propose our complex-valued CSI based activity recognition method in Section  \ref{sec:methods}. Next, we present evaluation results in Section \ref{sec:experiments} using experimental data collected from an apartment. Section \ref{sec:related} presents related work. Finally, Section \ref{sec:conclusion} concludes the paper. 

%shows the background on radio channel state information and sparse representation classification. We then present the noise robust activity recognition method in Section  shows the evaluation results of experiments in actual indoor environments. We present the Related work in and conclude the paper in Section

%\begin{figure}[htb]
%   \centering
%    \includegraphics[scale = .5]{figure/research_question_example_c.pdf}
%    \caption{CSI based device free activity recognition}\label{fig:sec1:examplefrequencyresponse}
%\end{figure}

%% file: Tex/background.tex
%\vspace{-2mm}
\section{Background} 
\label{sec:background}

\subsection{Wireless Platform and Channel State Information}
\label{subsec:radiocsi}
We use a platform called Wireless Ad hoc System for Positioning (WASP) \cite{sathyan2011wasp} for our experiments. The WASP nodes are originally designed for high resolution localisation and use a much wider bandwidth than many off-the-shelf wireless devices. WASP can operate in both 2.4~GHz and 5.8~GHz industrial, scientific and medical (ISM) bands, using a bandwidth of, respectively, 83~MHz and 125~MHz. WASP uses Orthogonal Frequency-Division Multiplexing (OFDM)  at the physical layer and time division multiple access (TDMA) at the media access control (MAC) layer. In fact, the physical layer of WASP is \cxbo{implemented} by using commercial off-the-shelf IEEE 802.11 radio chips. 

OFDM is a multi-carrier modulation technique. At the 5.8~GHz band, the WASP nodes use 320 sub-carriers. Each of these sub-carriers has a different centre frequency. This means the received sub-carriers at two frequencies can experience different amount of phase shifts, giving different frequency response. For a sub-carrier with centre frequency $f_i$, the complex channel response $C(f_i)$ is related to the transmitted symbol $T(f_i)$ and received symbol $R(f_i)$ by $R(f_i) = C(f_i)T(f_i)$. 
% The complex number $C(f_i)$ captures both relative change in signal amplitude and phase shift. 
\cxbo{The complex number $C(f_i)$ captures, respectively, the sub-carrier gain and phase shift.}
Let $f_1, f_2, ..., f_{320}$ denote the centre frequencies of the 320 sub-carriers that WASP nodes use. Then the CSI is a complex vector $[C(f_1), C(f_2), ..., C(f_{320})]$.  More details about CSI can be found in \cite{Yang2013From}.
\cxbo{Our previous research \cite{wei2015radio} only uses the amplitude~(gain) of CSI of each sub-carrier, while in this paper complex-valued CSI is applied, which contains both the \emph{amplitude} and \emph{phase} shift, to improve the classification accuracy. }
%Since we will only be using the \emph{amplitude} of CSI for classification and in order to keep the language simple, we assume that CSI is the real vector $[ |C(f_1)|, |C(f_2)|, ..., |C(f_{320})|]$. More details about CSI can be found in \cite{Yang2013From}.

\subsection{Sparse Representation Classification}
\label{subsec:sparseapproximation}
Sparse Representation Classification (SRC) is first proposed in \cite{Wright:09} for face recognition. It has subsequently been applied to other areas, such as acoustic classification~\cite{wei2013real} and visual tracking~\cite{mei2011robust}. A key feature of SRC is its use of $\ell_1$ minimisation to make the classification robust to noise. 

We give a brief description of SRC here, the details can be found in \cite{Wright:09}. We assume that the classification problem has $s$ classes. Class $i$ is characterised by the sub-dictionary $D_i = [d_{i1},...,d_{in_{i}}]$ where $d_{ij}$ ($j = 1, ..., n_i$) are the $n_i$ feature vectors derived from training data. For classification, the $s$ sub-dictionaries are concatenated to form a dictionary $D = [D_1, D_2, ..., D_s]$. Ideally, a test sample $y$ should sit in a subspace spanned by the dictionary, i.e. there exists a \emph{coefficient vector} $x$ such that $y = Dx$. However, due to noise, such an $x$ cannot be found or is perturbed. Instead, the SRC method solves for the coefficient vector $x$ using the following $\ell_1$ optimisation problem: 
\begin{equation}
\hat{x} = \argmin_x \| x \|_1 \quad \text{ subject to } \|y - Dx\|_2 < \epsilon,
\label{eqn:l1minimization}
\end{equation}
where $\epsilon$ is the noise level. Note that instead of requiring that $y = Dx$, the constraint requires only that the vectors $y$ and $Dx$ are sufficiently close to each other. The estimated coefficient vector $\hat{x}$ is used for the classification algorithm. Note that the length of $\hat{x}$ is $\sum_{i = 1}^s {n_i}$. Let $\hat{x}_{i}$ denote the $n_i$-dimension sub-vector in $\hat{x}$ that corresponds to the sub-dictionary $D_i$. We calculate the residual $r_i = \parallel y - D\hat{x}_{i} \parallel_2$ for Class $i$. The class that gives the minimum residual is returned as the classification result. 
\cxbo{It is important to point out that this SRC formulation can be used with real or complex-valued $y$ and $D$, and $\ell_1$ optimisation is also capable of handling complex-valued vectors \cite{van2011sparse,van2008probing}. }

SRC has a main superiority: featureless. This provides us an opportunity to build a training set from the CSI measurements rather than exacting feature from them. Moreover, SRC is known to be robust to noise. 
% \cxbo{ $\ell_1$ optimisation is also capable to handle complex-valued vectors\cite{van2011sparse,van2008probing}. }
As our work is to study the performance of activity recognition when RFI is present, we investigate how to explore SRC to boost SNR for improving recognition performance. \cxbo{As far as we know, we are the first to use a SRC classification method on complex numbers.}

%% file: Tex/method.tex
%\vspace{-2mm}
\begin{figure}[htb]
   \centering
    \subfigure[Examples of multiple paths caused by different activities]{
            \includegraphics[scale = .55]{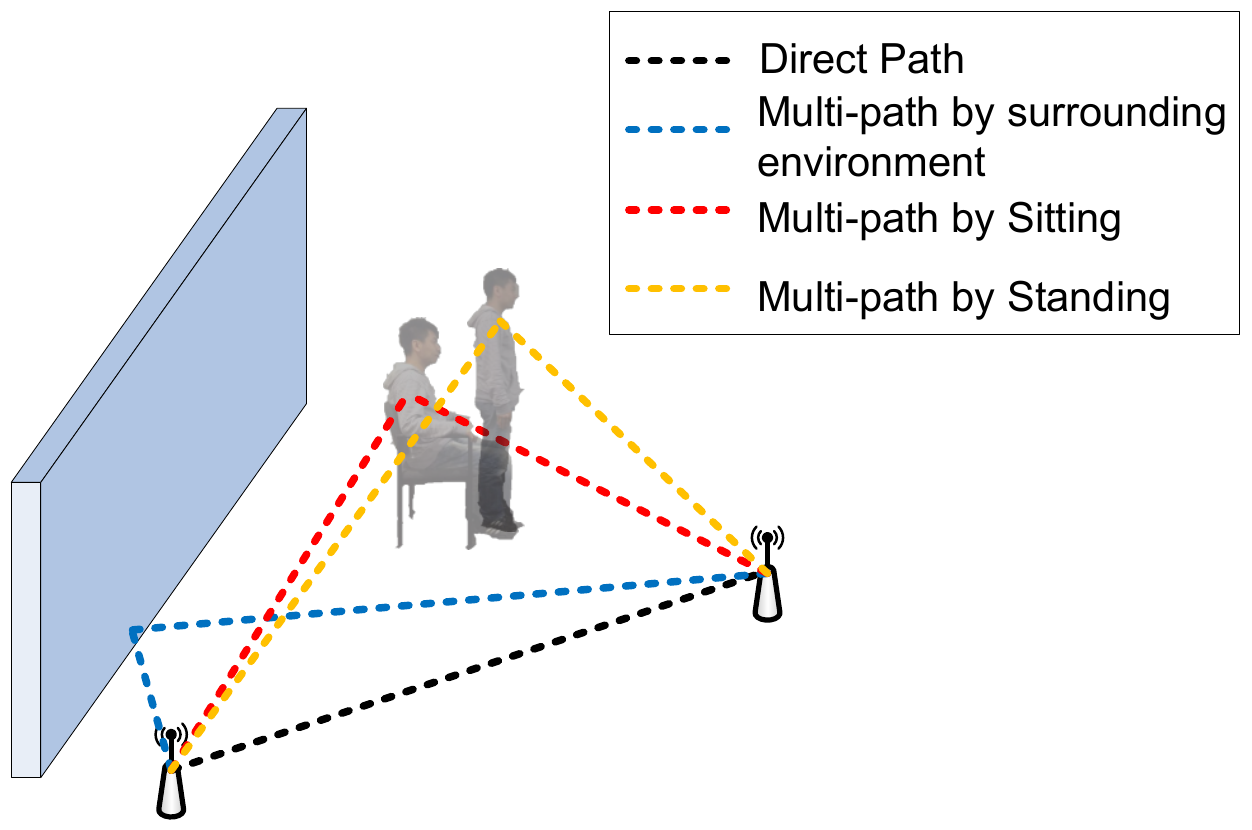}
            \label{fig:multipathexample}
        }
    \subfigure[Examples of CSI vectors cause by different activities]{
               \includegraphics[scale = .4]{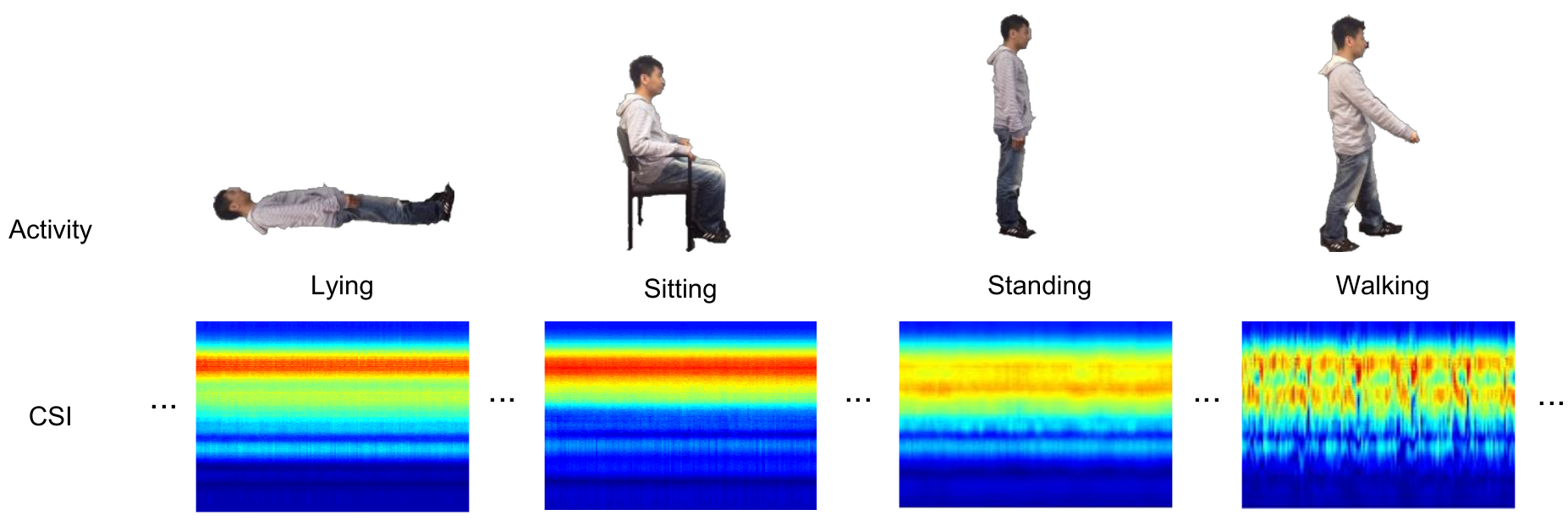}
               \label{fig:exampleCFR}
           }
    \caption{Examples of multiple paths and CSI vectors (Best view in colour)}
                   \label{fig:intuition}
\end{figure}
\section{Activity Recognition using Complex-valued CSI}\label{sec:methods}
This section presents our method to recognise a set of location-oriented activities using CSI in the device-free setting. We first demonstrate that complex-valued CSI is influenced by activities taking place in a room and can be used to identify location-oriented activities. We demonstrate the challenge of CSI based activity recognition when RFI is present. Finally, we present our SRC based classification method which takes RFI into consideration. 

\subsection{CSI contains location-oriented activity information}\label{subsec:feasibility}
We first present some intuition on why device-free CSI-based location-oriented activity recognition is possible. Fig.~\ref{fig:multipathexample} depicts an indoor environment with two wireless nodes and the multi-paths that the radio propagation may take. It shows that different multi-path effects can be obtained if a person is sitting or standing. Its results in different CSI vectors at the receiver and can be used to identify the activity (normalised CSI amplitude vector shown in Fig.~\ref{fig:exampleCFR}). Furthermore, different locations of a monitored person also differentiate multi-paths, \emph{which makes location-oriented activity recognition feasible}.  

In order to demonstrate the feasibility of CSI-based activity recognition, we set up two WASP nodes in an apartment with one living room and one bedroom. The nodes are 5 metres apart with 3 walls in the line-of-sight path between the nodes. The subject is positioned in an AoI between the two nodes but is not in the direct path between the 2 nodes. The subject carry out 4 different activities: sitting, lying, standing, and walking. A WASP node is used as the transmitter and the other as a receiver. The transmitter sends to the receiver at 10 packets per seconds, and it needs a 2.5 milliseconds slot to send a packet. This means only \emph{2.5\% of running time} is occupied for sending data for activity recognition, which does not occupy bands too much to affect the radio communication of other wireless devices. For each packet received, the receiver uses the WASP interface to obtain the CSI vector and SNR for that packet. It is also important to point out that the data in this experiment are collected in a \emph{clean} environment without any RFI. 

%\cxbo{TODO: phase information? How to tell the story}

\begin{figure*}[htb]
{
\centering
       \subfigure[CSI vectors in clean environment]{
        \includegraphics[scale = .3]{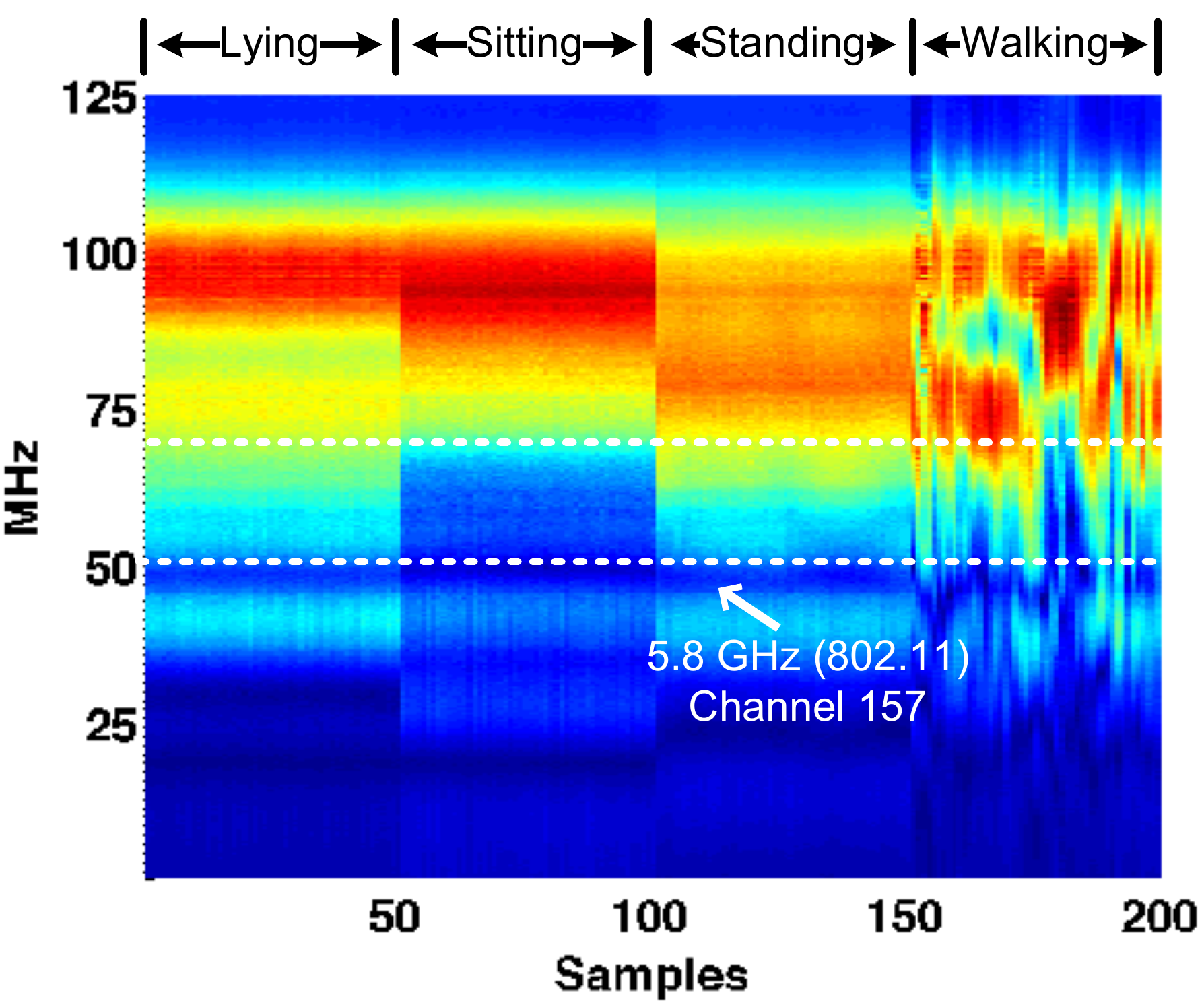}
        \label{fig:examplefrequencyresponse}
    }
    \subfigure[CSI vectors in Channel 157 without RFI]{
            \includegraphics[scale = .3]{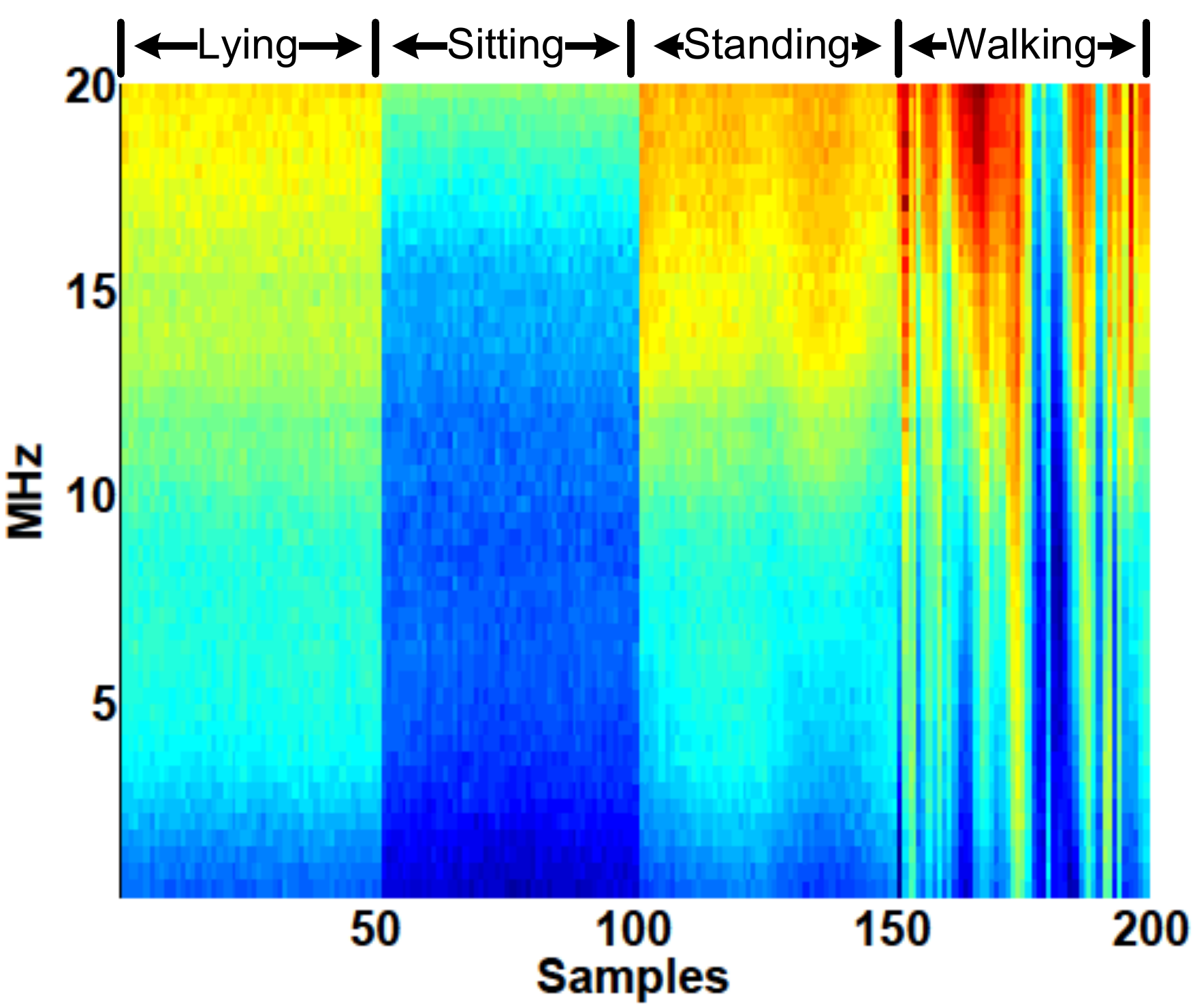}    
            \label{fig:examplefrequencyresponsecleanpart}
            }
    \subfigure[SNR without RFI]{
        \includegraphics[scale = .3]{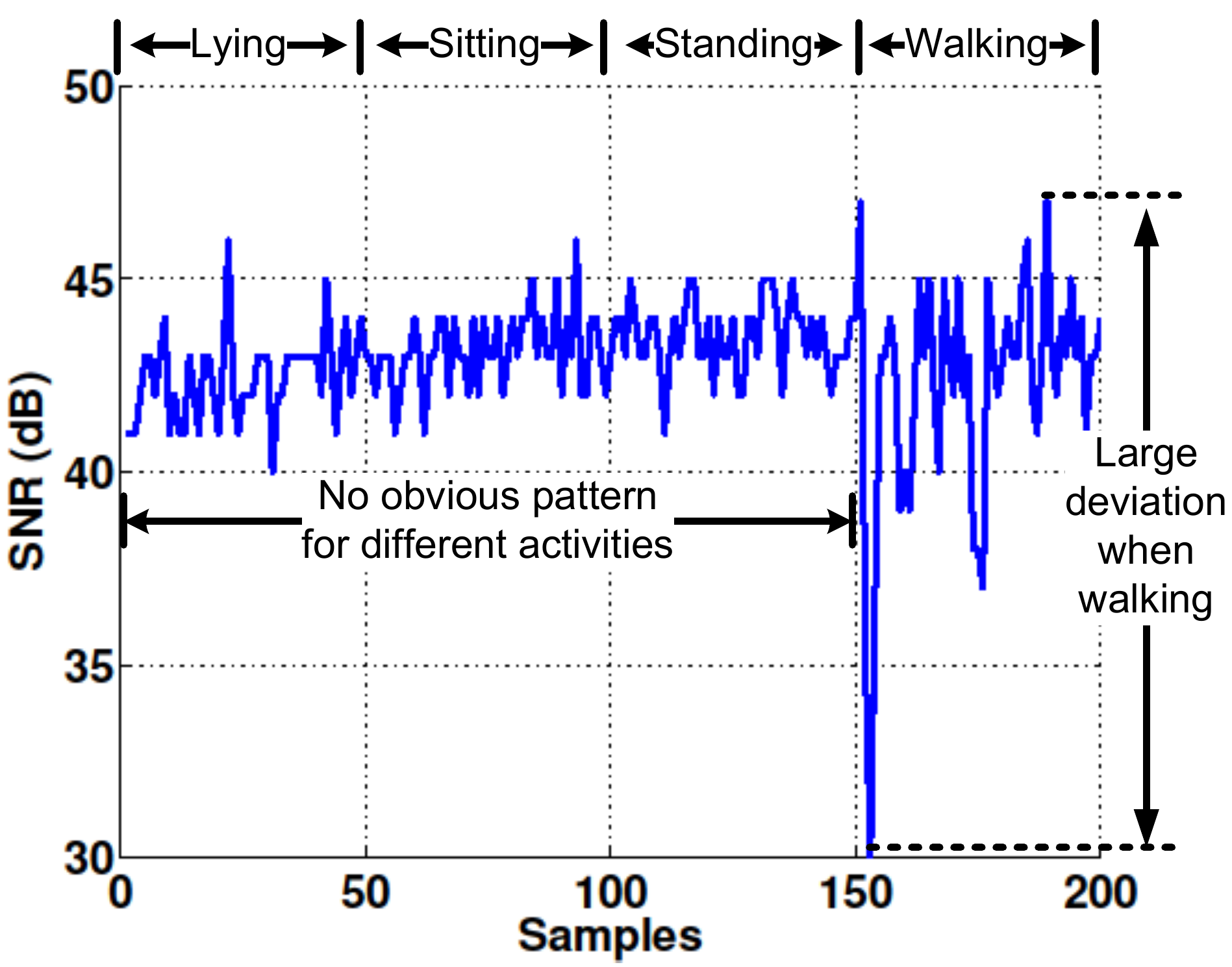}    
        \label{fig:exampleRSS}
        }
     \subfigure[CSI vectors in RFI environment]{
             \includegraphics[scale = .3]{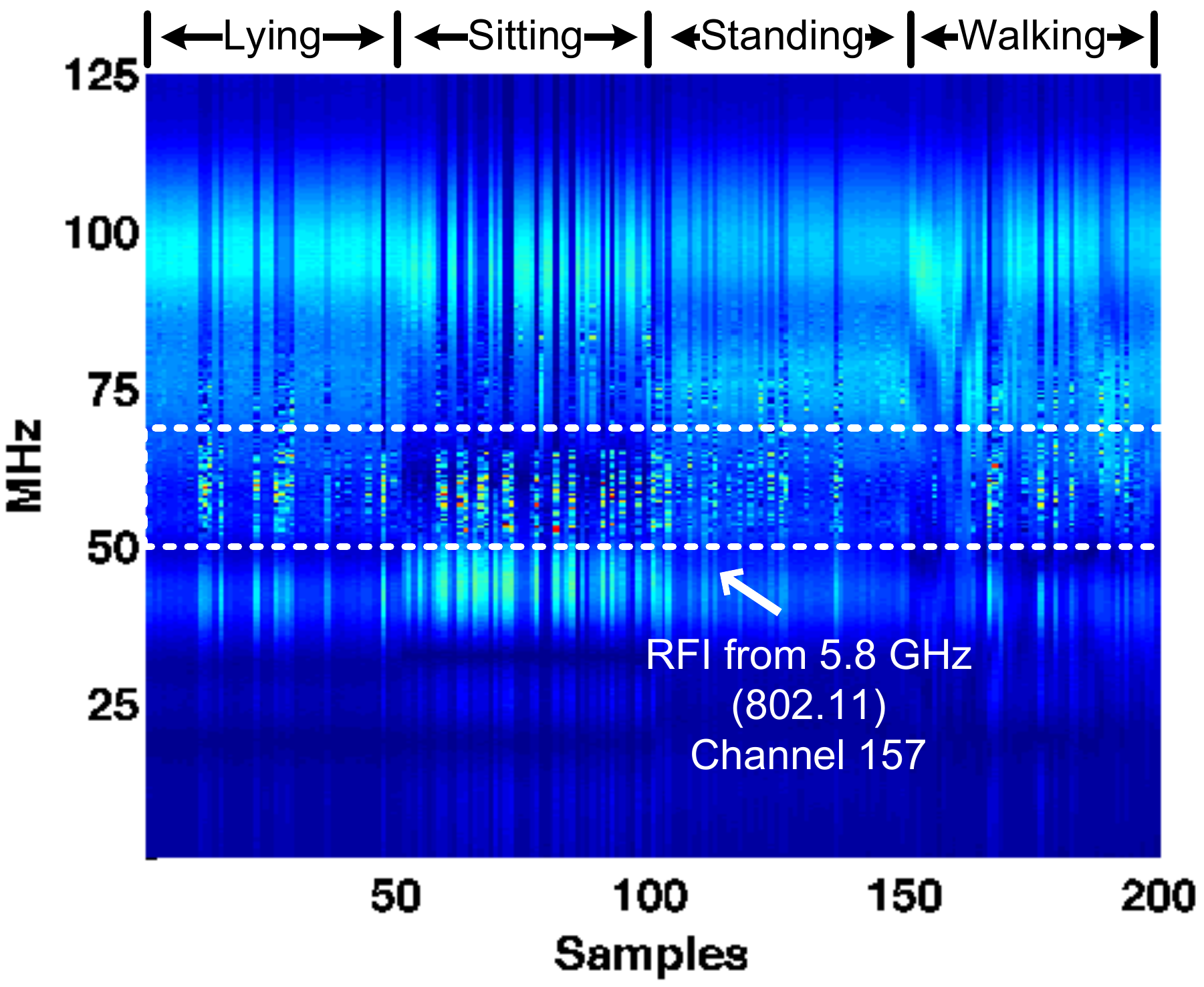}
             \label{fig:examplefrequencyresponsenoisewhole}
         }
     	\subfigure[CSI vectors in Channel 157 with RFI]{
             \includegraphics[scale = .3]{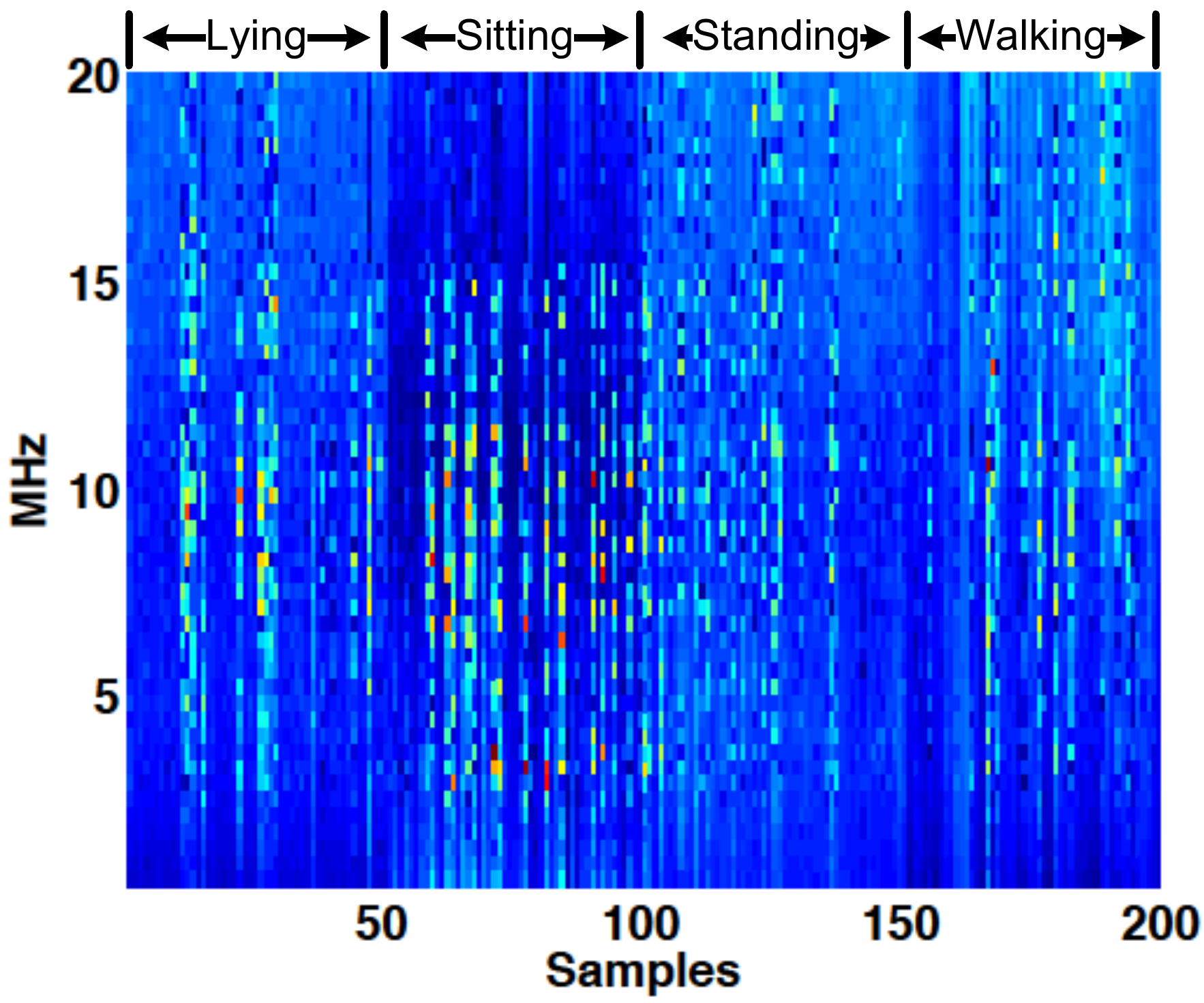}
             \label{fig:examplefrequencyresponsenoisepart}
         }    
         \subfigure[SNR with RFI]{
             \includegraphics[scale = .3]{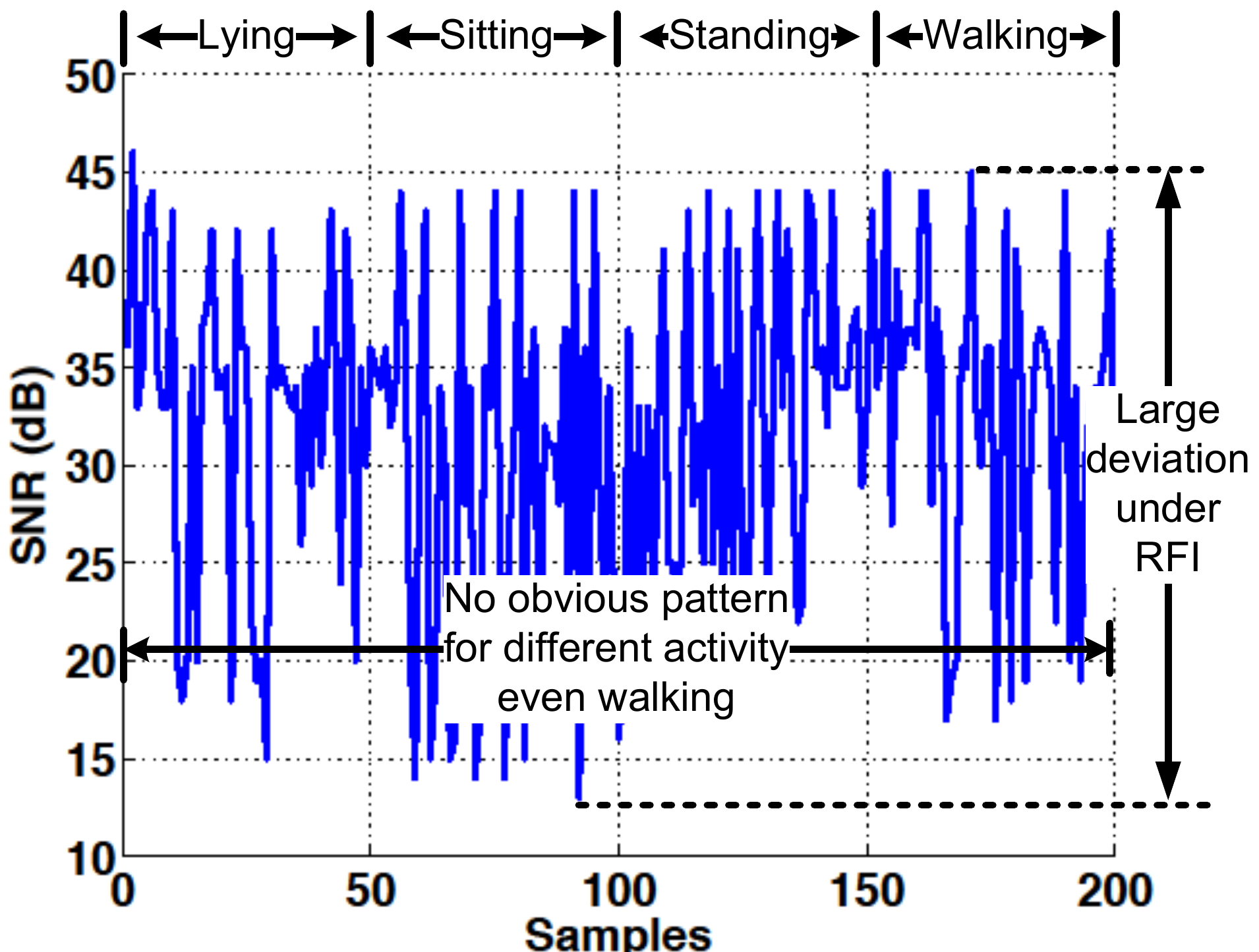}    
             \label{fig:exampleRSSnoise}
             }    
         \caption{CSI and SNR performance in different activities (Best view in colour)}
               \label{fig:dynamicRSSexample}               
}
\end{figure*}

Fig.~\ref{fig:examplefrequencyresponse} shows the normalised CSI \cxbo{amplitude} vectors under the four different activities. The horizontal axis shows the sample number where a sample corresponds to a packet. There are 320 values in the vertical axis which corresponds to the 320 sub-carriers. The magnitude is shown as a heat plot. We have put four blocks of data side-by-side in the figure, which corresponds to the four activities of lying, sitting, standing and walking. It can readily be seen that the four activities have highly distinguishable CSI. 
% \cxbo{Besides, we can also observe distinguishable patterns from CSI phase vectors.} 
This confirms that CSI contains information on activity. Another observation is that the CSI fluctuates a lot when the subject is walking. This is due to different multi-path effects created by the person walking. Fig.~\ref{fig:examplefrequencyresponse} shows the CSI when a 125~MHz bandwidth is used. We now show that the same observations also apply when we use a 20~MHz channel. The box in Fig.~\ref{fig:examplefrequencyresponse} is Channel 157 in the 802.11 standards with 20~MHz bandwidth. We have enlarged the CSI in the box and plotted it in Fig.~\ref{fig:examplefrequencyresponsecleanpart}. It can be seen that the CSI for the four activities are very distinguishable and walking creates more fluctuations in CSI. 

Fig.~\ref{fig:exampleRSS} shows the SNR of the corresponding samples (packets). It shows that the SNR has a slightly larger fluctuation when the subject is walking. However, there does not appear to be any noticeable differences in the SNR data series among lying, sitting, and walking. These observations suggest that it may be possible to use SNR to distinguish between walking from the other three activities where the subject is stationary. However, it does not seem to be possible to use SNR to distinguish between the three stationary activities. 

\begin{figure}[htb]
   \centering
    \includegraphics[scale = .40]{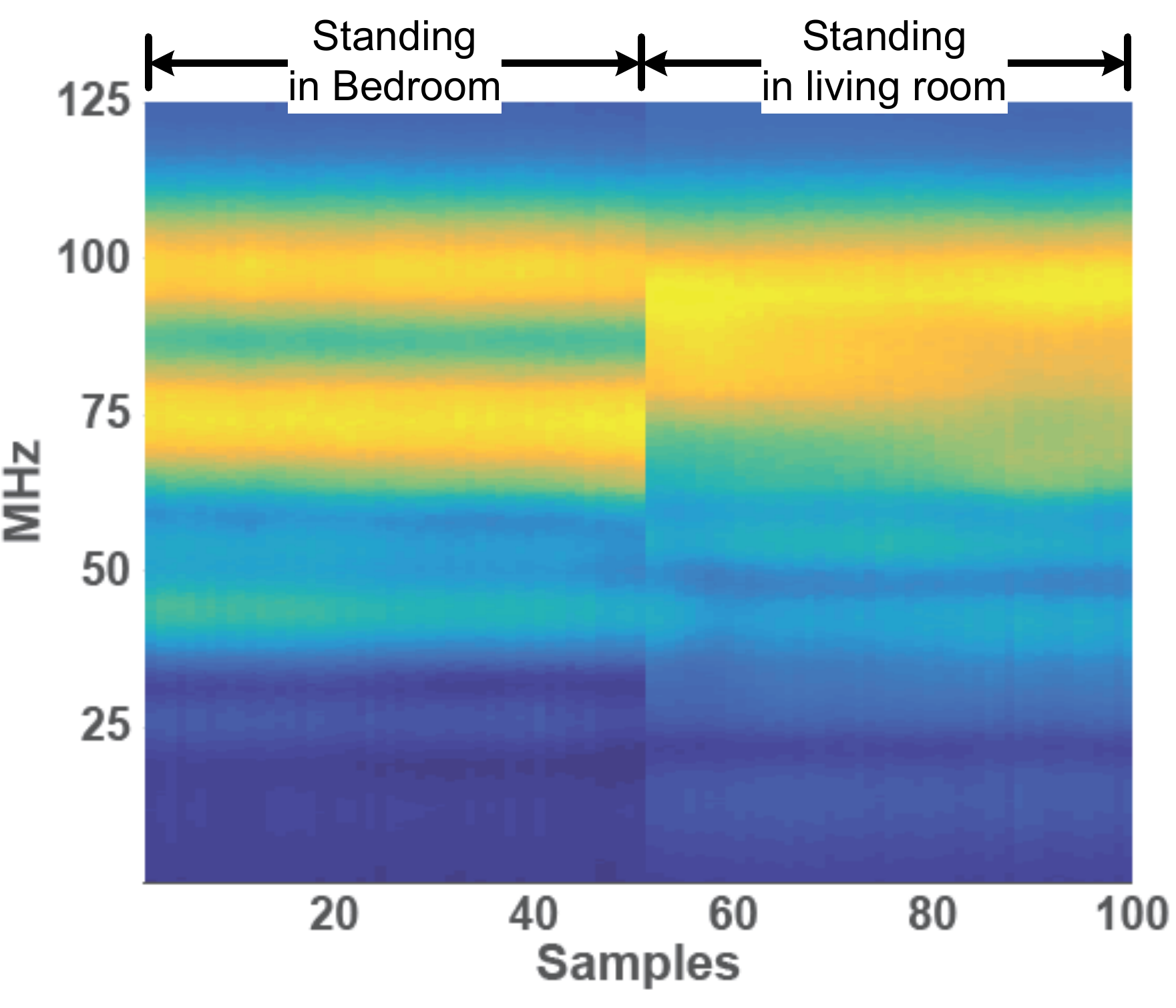}
    \caption{CSI of standing in different locations.}\label{fig:sec3:differnt_location}
\end{figure}

Since CSI is sensitive to the multi-path effect, same activity in different locations can have different CSI. Training in each interesting location must be performed for location-orientated activity recognition, and this is a limitation of CSI-based finger-printing activity recognition. Fig.~\ref{fig:sec3:differnt_location} demonstrates the different CSI \cxbo{amplitudes} as a result of the same activity ``standing'' in two locations. This fact requires additional training for same activity in different locations, but it helps location-orientated activity recognition systems locate activities. \cite{Xu:2012,Xu:2013,wang2014eyes,sigg2014rf,melgarejo2014leveraging} also apply a similar strategy, i.e. conducting training in various locations, for improving the performance of radio-based pattern recognition.

%\begin{figure*}[htb]
%{
%\centering
%       \subfigure[CSI vectors in RFI environment]{
%        \includegraphics[scale = .3]{figure/Example_frequency_response_noise/Example_frequency_response_noise_whole_c.pdf}
%        \label{fig:examplefrequencyresponsenoisewhole}
%    }
%	\subfigure[CSI vectors of 802.11 5.8 GHz Channel 157 under RFI]{
%        \includegraphics[scale = .3]{figure/Example_frequency_response_noise/Example_frequency_response_noise_part_c.pdf}
%        \label{fig:examplefrequencyresponsenoisepart}
%    }    
%    \subfigure[SNR of under RFI]{
%        \includegraphics[scale = .3]{figure/Example_frequency_response_noise/Example_RSS_noise_whole_c.pdf}    
%        \label{fig:exampleRSSnoise}
%        }
%         
%         \caption{Frequency response and SNR performance in different activities}
%               \label{fig:dynamicRSSexamplenoise}               
%}
%\end{figure*}

\subsection{Challenges of CSI-based activity recognition} 
The results above  are obtained when the two WASP nodes are in a clean environment without RFI. We conduct another experiment using the same set up but we add a pair of IEEE 802.11a devices that communicates in Channel 157 (a 20 MHz band). Our goal is to understand the impact of RFI on CSI and SNR.  

Fig.~\ref{fig:examplefrequencyresponsenoisewhole} shows the CSI for the four activities when RFI is present in Channel 157. It shows that the CSI in Channel 157 (enclosed by the white rectangle) is fairly noisy but the four activities still have distinguishable CSI outside of Channel 157. This suggests that if wide-band devices are used to obtain the CSI for activity recognition, then we can use the part of CSI with little RFI to identify the activity. 
%However, wide-band devices such as WASP are not common off-the-shelf devices. 
%Furthermore, with the increasing number of the usage of wireless radio device in the future, it will be difficult or even impossible to have any totally clean channel. 
\vtbo{\emph{However, with ubiquitous use of wireless technologies such as WiFi, Bluetooth, IEEE 802.15.4, etc., it becomes more and more difficult or even impossible to find RFI-free bandwidth, particularly in ISM bands, for radio-based activity recognition systems.}}
\emph{We therefore consider the possibility of using CSI in an interfered channel to perform activity recognition.} 

In order to examine the effect of RFI, we plot the CSI of Channel 157 in Fig.~\ref{fig:examplefrequencyresponsenoisepart}. It shows that the CSI vectors of different activities are no longer highly distinguishable. 
We now examine the impact of RFI on SNR. Fig.~\ref{fig:exampleRSSnoise} shows the SNR of the four activities when RFI is present in Channel 157. We see that the SNR of all four activities are highly fluctuating. We suggested earlier that it would be possible to tell walking from the static activities using SNR when RFI is absent, however, this does not appear to be feasible once RFI is present. 

\begin{figure}[]
{
\centering
       \subfigure[a CSI subcarrier without RFI]{
        \includegraphics[scale = .17]{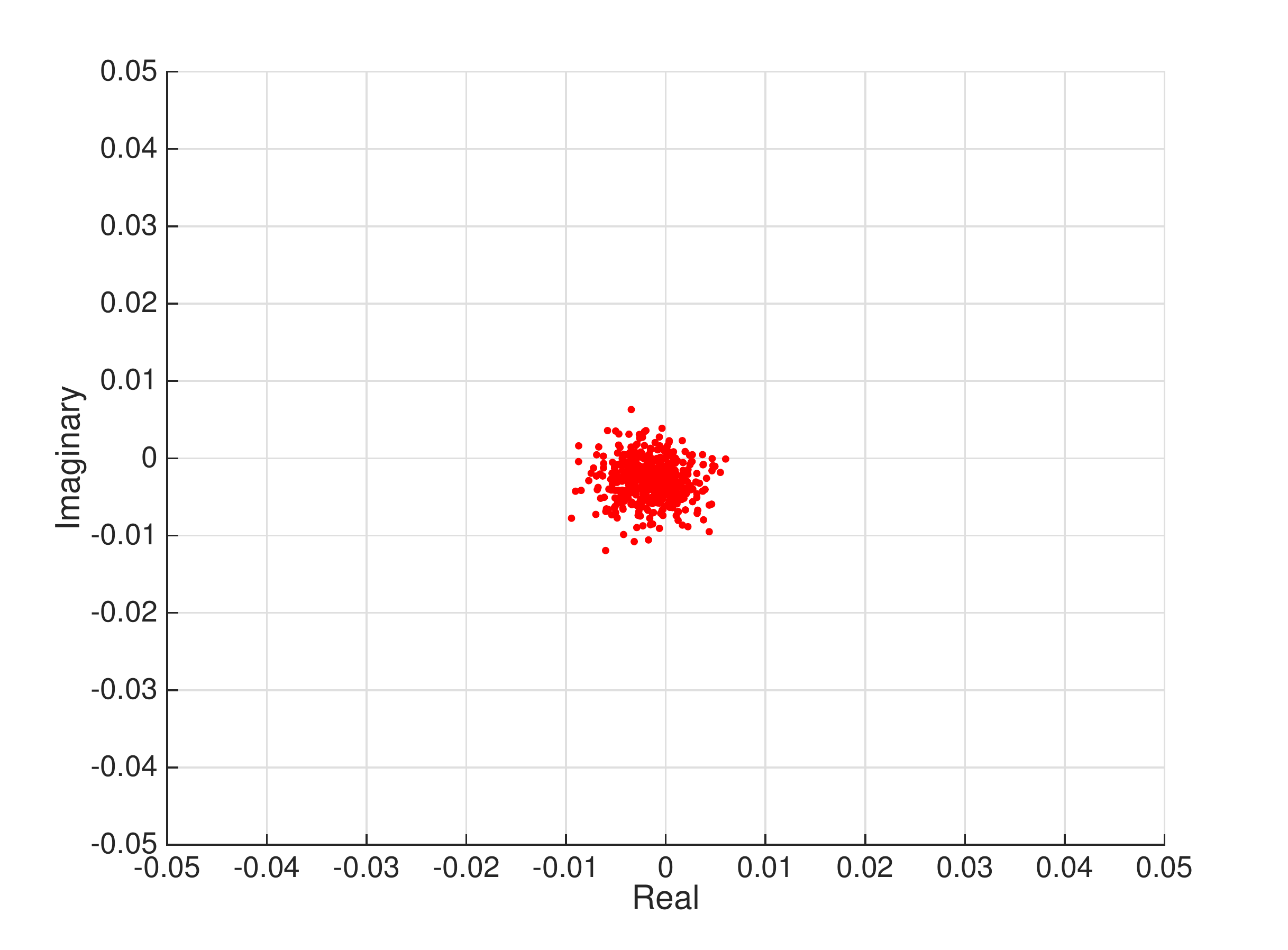}
        \label{fig:complexcluster_clean}
    }
    \subfigure[a CSI subcarrier with RFI]{
            \includegraphics[scale = .17]{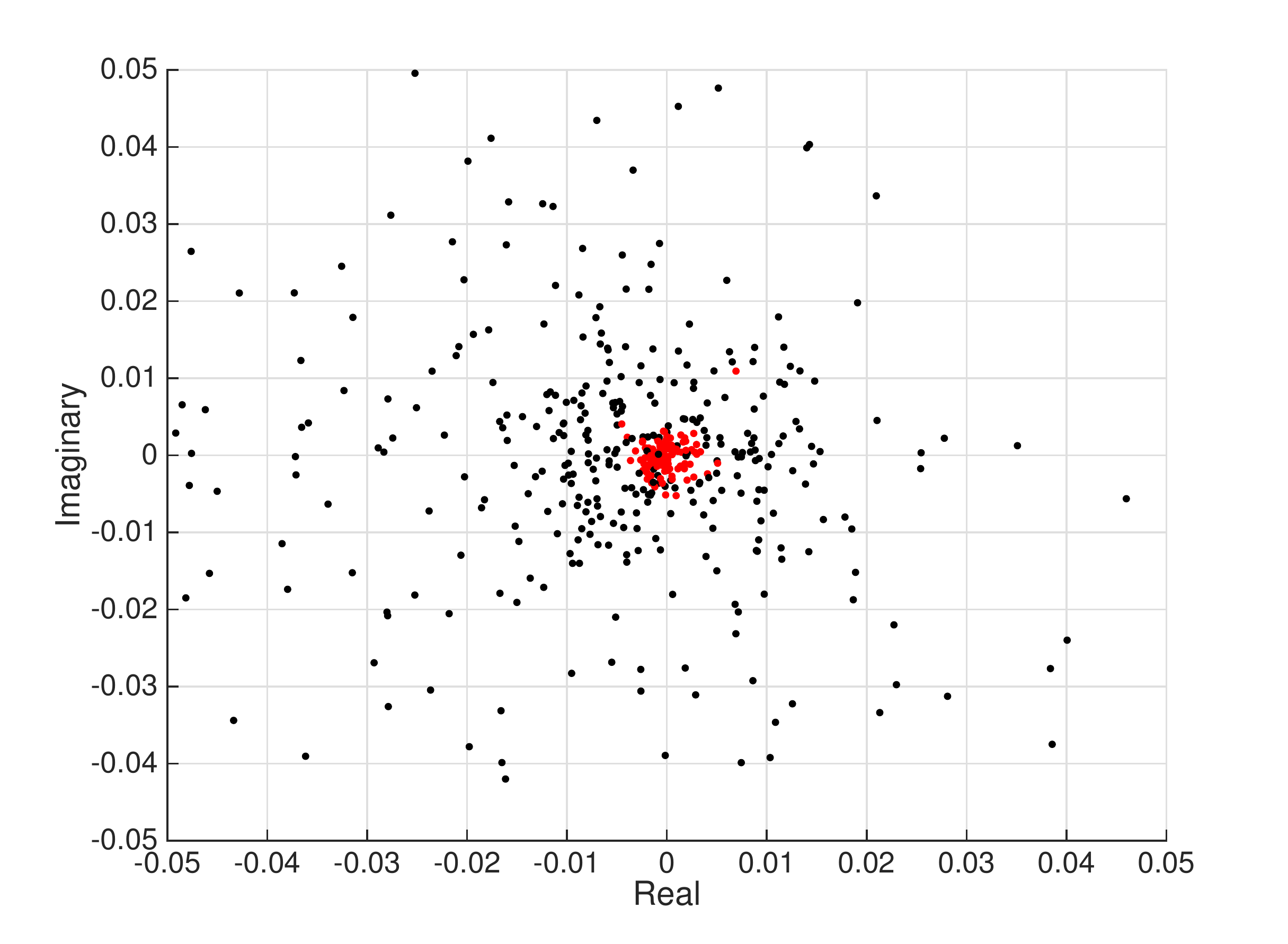}    
            \label{fig:complexcluster_noisy}
            } 
         \caption{CSI performance in complex plane (Best view in colour)}
               \label{fig:complexcluster}               
}
\end{figure}

\cxbo{
To further show the challenge with RFI, Fig.~\ref{fig:complexcluster} shows clusters of complex-valued CSI from the same sub-carrier with or without RFI in complex plane. 
The red spots represent the CSI samples without RFI affected. 
To better show the scatter of CSI, we shift the centre to (0,0). In clean environment without RFI, all the samples have high SNR and concentrate in a clear cluster. However, in the environment with RFI, the samples scatter in a much larger area, and less pattern can be explored compared with that in clear environment.

However, if we look closer at the CSI vectors of each activity, we can see that a number of CSI vectors among one activity are almost the same. This recurrence of CSI vectors suggests that we may use a block of CSI vectors for classification instead of individual CSI vectors (see Fig.~\ref{fig:examplefrequencyresponsenoisewhole} and \ref{fig:examplefrequencyresponsenoisepart} ). However, this classification is going to be challenging because the CSI vectors appear to be fairly noisy. We will propose a few different classification methods in Section \ref{subsec:CoefficientFusion} to address this challenge. 
}

To sum up, it is a challenge to use CSI to perform activity recognition when RFI is present.

\subsection{Location-Oriented Activity Recognition with RFI}
\label{subsec:ar_with_rfi}

We now describe our proposed complex-valued CSI-based activity recognition in the presence of RFI. The goal of activity recognition is to identify four daily activities: sitting, standing, lying and walking in different rooms, as well as whether the AoI is empty.

\begin{figure}[]
	\centering
	\includegraphics[scale = .6]{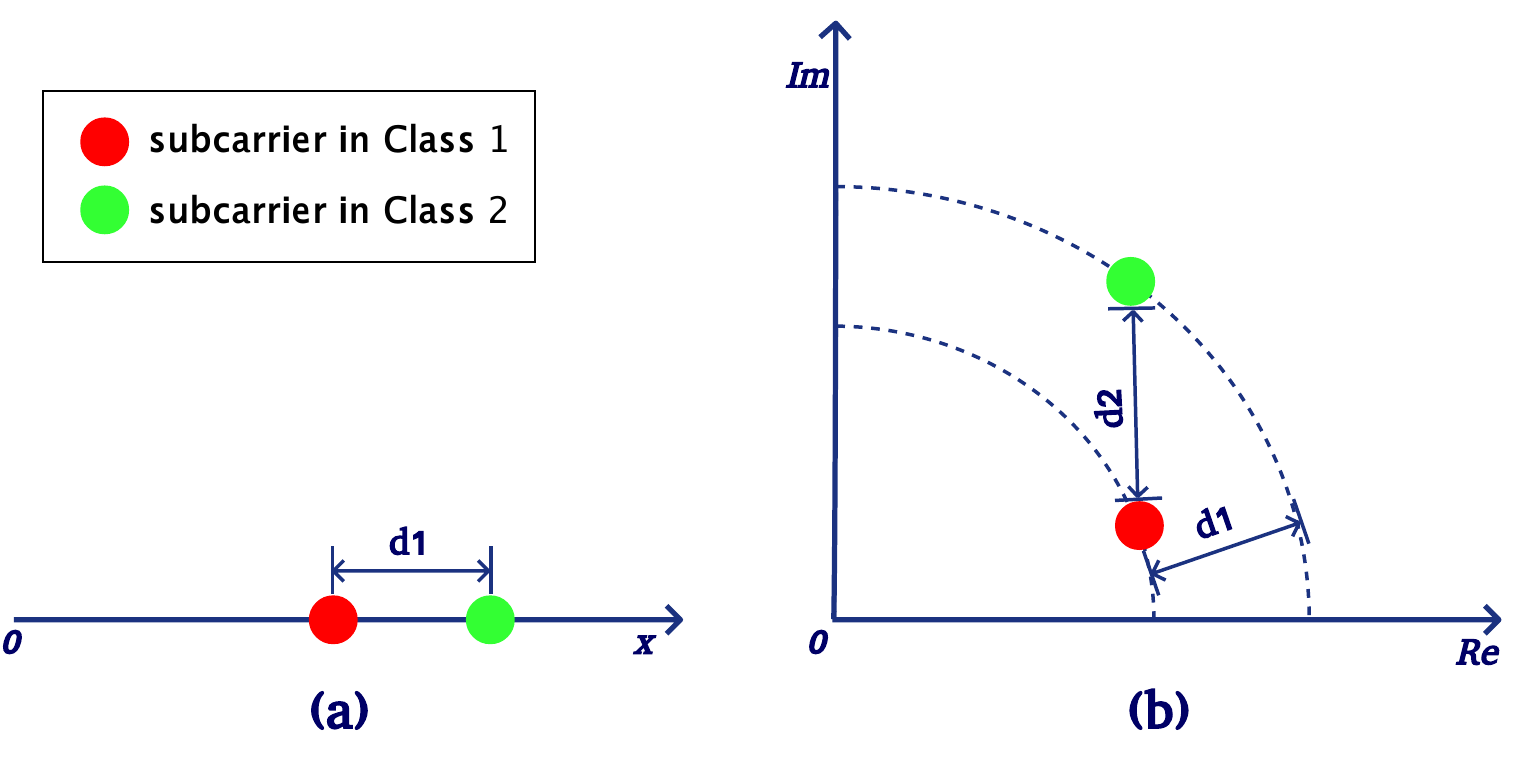}
	\caption{Distance difference between two classes using real value and complex value based CSI. (a) distance using amplitude value CSI (b) distance using complex value CSI }\label{fig:com_example}
\end{figure}

\subsubsection{Data collection}

Our method fingerprints the activities using CSI complex vectors. The first procedure is to record the CSI measurements and use a CSI data sanitisation method for building a training set, and use the training set to fingerprint the test data by using our proposed machine learning algorithm.

% For each CSI vector in a training set or a test set, we first apply smoother to remove the high frequency noise and decrease the noise level. However, the smoother cannot take effect for all CSI vectors when experiencing high SNR. Therefore, we still need further mechanism to improve the recognition performance.

% Besides, 
RFI causes the unexpected change to CSI vectors. There is no existing model for the performance of CSI vector under RFI. Therefore, we need to consider the RFI environment when training the dictionary. In other words, the CSI vectors of one specific activity under different physical or radio environments vary significantly. We have to update training set for a new RFI environment. The method in \cite{wang2014eyes} can be applied for updating the dictionary when RFI is present.

\cxbo{
\subsubsection{CSI  Data Sanitisation}\label{subsussec:csisan}
We use a CSI data sanitisation method to obtain both CSI amplitude and phase information. A CSI vector with $n$ sub-carriers, i.e. 
$C = [C(f_1), C(f_2), ..., C(f_{n})]$.
% $C = [C_1, C_2, ..., C_{n}]$
The CSI amplitude vector is the absolute values of each elements.
% , i.e. $C_{am} = [|C_1|, |C_2|, ..., |C_{n}|]$. 
	%The tricky part is to obtain CSI phase information. Sen et al. proposed the challenge that 
Since the receiver and transmitter do not attempt to synchronise time in OFDM and there is an unknown random phase shift in each CSI vector, a data sanitisation method to calculate the phase information for OFDM is required. We use the data sanitisation method proposed in \cite{sen2012you} for calculating this unknown phase shift. The details of this algorithm can be found in \cite{sen2012you}. 
% is shown in Algorithm~\ref{alg:DataSan}.
Having CSI phase information, we are ready to calculate CSI vectors. The CSI for $i$-th sub-carrier is a complex number $(|\hat{C}_i|\cos(\hat{C}_i), {|\hat{C}_i|}\sin(\hat{C}_i))$, where $\hat{C}_i$ is the CSI complex value of $i$-th sub-carrier after sanitisation and $|\hat{C}_i| = |C_i|$. 
We apply this algorithm to all CSI vectors to remove the unknown phase shift. From now onwards, unless otherwise stated, all CSI vectors are assumed to have been sanitised. 

The complex values of CSI will contains both amplitude and phase information. This is extremely important for improving performance in the environments with RFI where limited useful data can be explored. 
% Fig.~\ref{fig:com_example} shows an example of the distance of one sub-carrier between two classes. The distance is d1 only using amplitude (shown in Fig.~\ref{fig:com_example}(a)). When using complex value CSI, the amplitude difference stays d1, but the distance in complex plane increases to d2 (as shown in Fig.~\ref{fig:com_example}(b)). In other words, the complex value CSI can enlarge the distance between two classes. The increased separation will help classier distinguish two classes and improve the classification performance. 
Fig.~\ref{fig:com_example} gives an intuitive explanation of this improvement. Consider the CSI of the same sub-carrier for two different classes. Fig.~\ref{fig:com_example}(b) depicts the situation when complex CSI is used, the distance d2 between the two classes is the distance between two complex numbers on the plane. However, if only the CSI amplitudes are used, then we get the situation in Fig.~\ref{fig:com_example}(a) where the distance between the two classes is d1. It can be shown that d2$\geq$d1, which means complex CSI enhances the separation between different classes and therefore giving better classification performance.

}

\subsubsection{Classification algorithms} 
\label{subsec:CoefficientFusion}
We have seen that RFI causes the CSI vector to be very noisy. In order to deal with RFI, we introduce a window size $ws$ where $ws$ consecutive complex-valued CSI vectors are used for classification. One possible method is to stack $ws$ complex-valued CSI vectors into a long feature vector and use it for classification. However, this will be computationally intensive because the feature vector has a very high dimension. Instead, we will use the one complex-valued CSI vector at a time and investigate different fusion methods. 

Let $y_1, y_2, ..., y_{ws}$ denote the complex-valued CSI vectors in the time window, and $D$ be the dictionary. We first solve the following $\ell_1$-optimisation problem for $i = 1,.., ws$: 
\begin{equation}
\hat{x}_i = \argmin_x \| x \|_1 \quad \text{ subject to } \|y_i - Dx\|_2 < \epsilon,
\label{eqn:l1minimizationwin}
\end{equation}

We now present three different fusion methods which use $\hat{x}_i$ ($i = 1,.., ws$) in different ways. 

The first method is to use decision fusion and will be referred to as \emph{$\ell_1$-voting}. For this method, the algorithm uses each $\hat{x}_i$ to arrive at a decision class using the standard SRC algorithm described in Section \ref{subsec:sparseapproximation}. This method then uses majority voting to arrive at a decision. 

The second method is to fuse the $\hat{x}_i$ vectors by computing their mean: $\hat{x}_{\rm sumup} = \frac{1}{ws} \sum_{i = 1}^{ws} \hat{x}_i$. The mean vector $\hat{x}_{\rm sumup}$ is then use to compute the residuals for each class as in the standard SRC algorithm described in Section \ref{subsec:sparseapproximation}. This method returns the class that minimises the residual. Note that this fusion method was proposed by Misra et al.~in \cite{misra2014energy} where they showed that such method could improve the GPS recovery accuracy. We will call this method \emph{$\ell_1$-sumup}.

The method \emph{$\ell_1$-sumup} applies equal weights to all $\hat{x}_i$ by computing a simple average of them. However, it is possible that some CSI vectors in the window are less affected by noise. This can also be seen from Fig.~\ref{fig:dynamicRSSexample} where the SNR fluctuates. We therefore propose to use SNR of a sample to derive a weighting for that sample. Let $S_i$ denote the SNR of the $i$-th sample in the window. We compute the weighted mean of $\hat{x}_i$ using: 
\begin{equation}
\hat{x}_{\rm weighting}  =  \sum {w_{i}} \hat{x}_i ,
\label{eqn:weighting}
\end{equation}
\begin{equation}
w_i =  \frac{A_i}{ \sum_{j = 1}^{ws} {A_j} }, \quad   A_i = 10^{( \frac{S_i}{20} )},
\label{eqn:weightingelement}
\end{equation}

The mean vector $\hat{x}_{\rm weighting}$ is then use to compute the residuals for each class as in the standard SRC algorithm described in Section \ref{subsec:sparseapproximation}. This method returns the class that minimises the residual. We call this method as \emph{$\ell_1$-weighting}.

%% file: Tex/experiment.tex
%\vspace{-2.5mm}
\section{Evaluation} \label{sec:experiments}
\begin{figure}[]
   \centering
    \includegraphics[scale = .25]{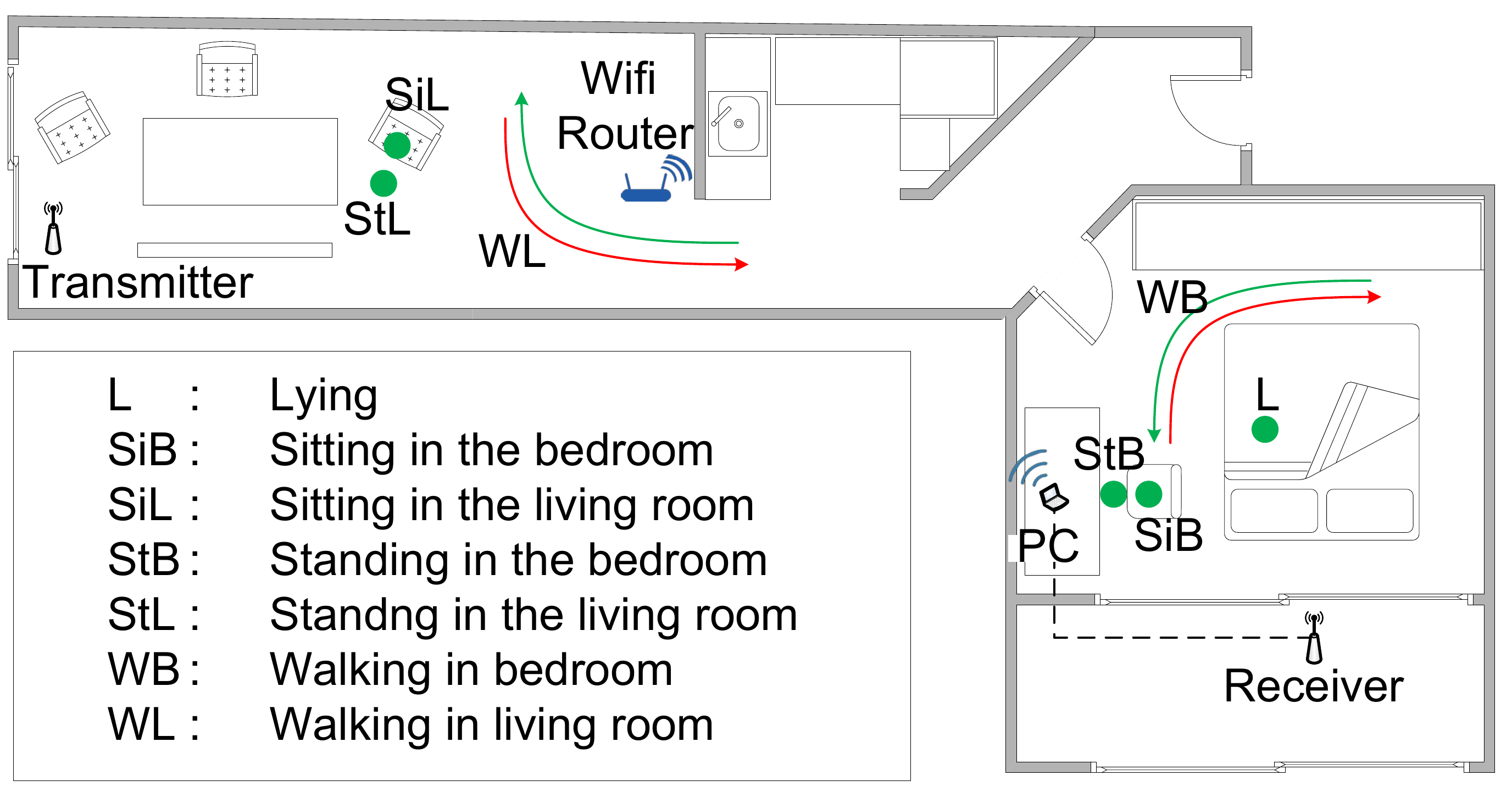}
    \caption{Floor plan of the experiment environment }\label{fig:floorplan}
\end{figure}

\begin{figure*}[]
{
\centering
       \subfigure[``\emph{whole bandwidth without RFI}"]{
        \includegraphics[scale = .30]{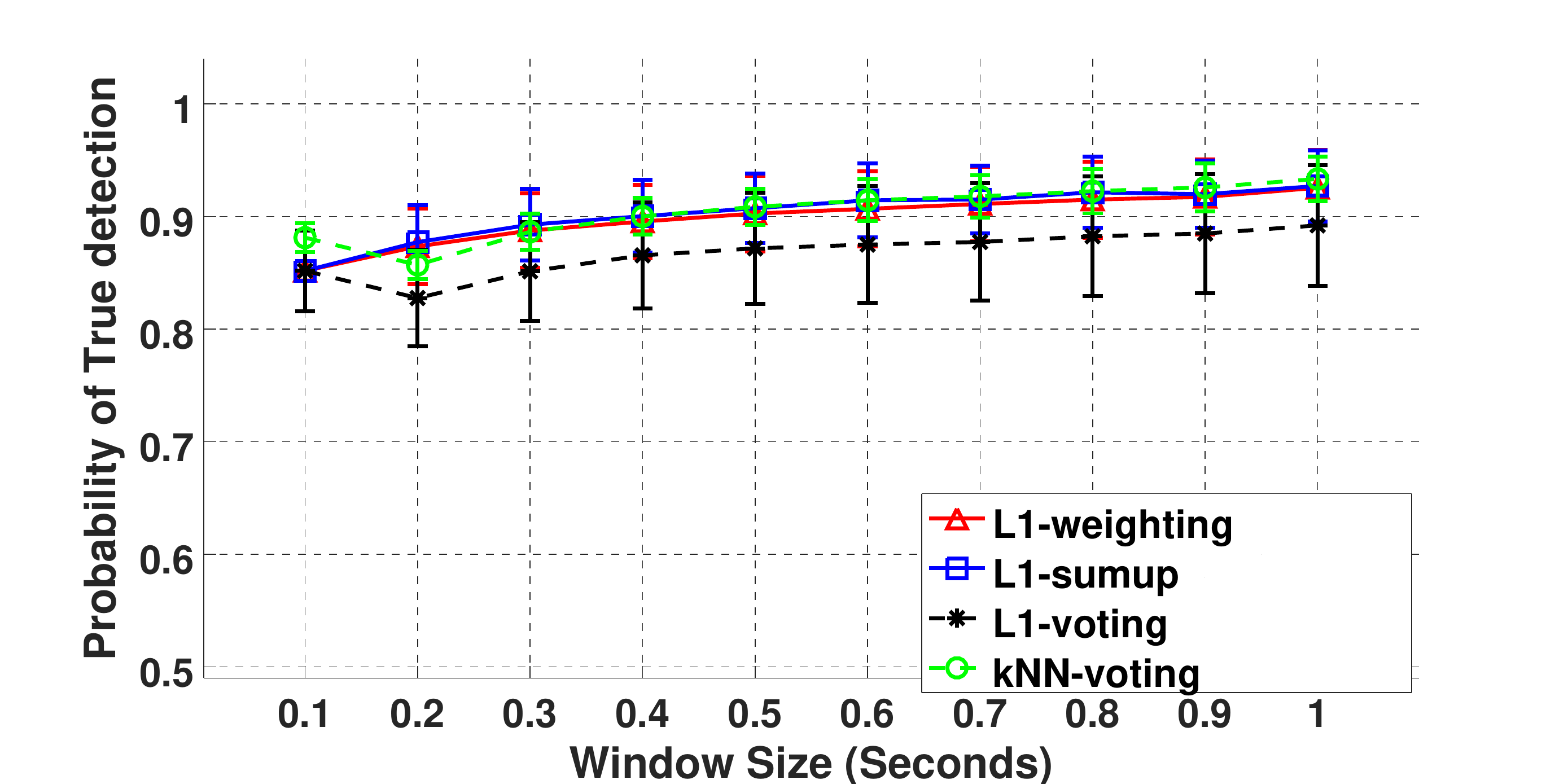}
        \label{fig:win_whole_clean}
    }
    \subfigure[``\emph{whole bandwidth with RFI}"]{
            \includegraphics[scale = .30]{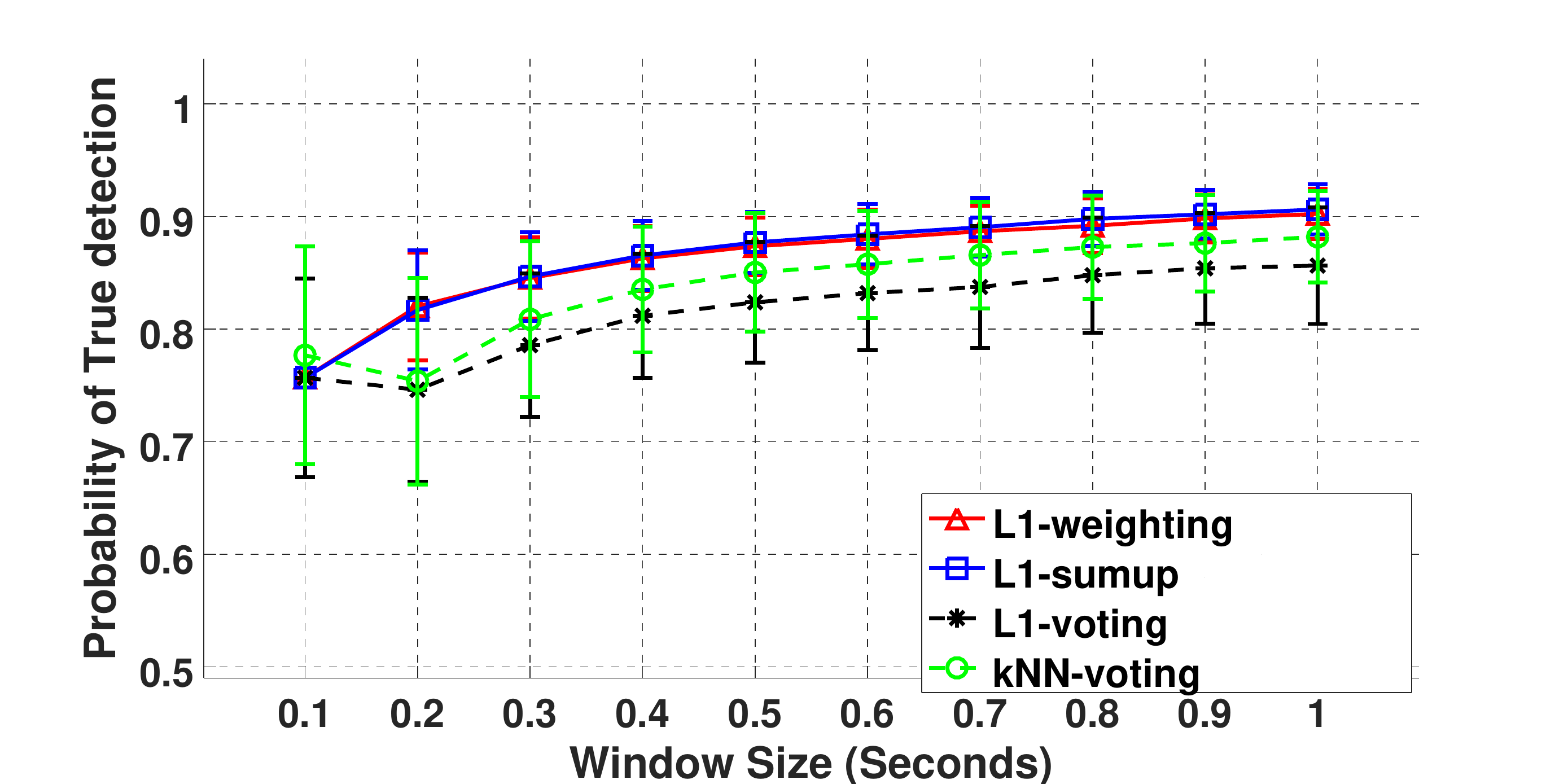}    
            \label{fig:win_whole_noise}
            }
	 \subfigure[``\emph{Channel 157 without RFI}"]{
	         \includegraphics[scale = .30]{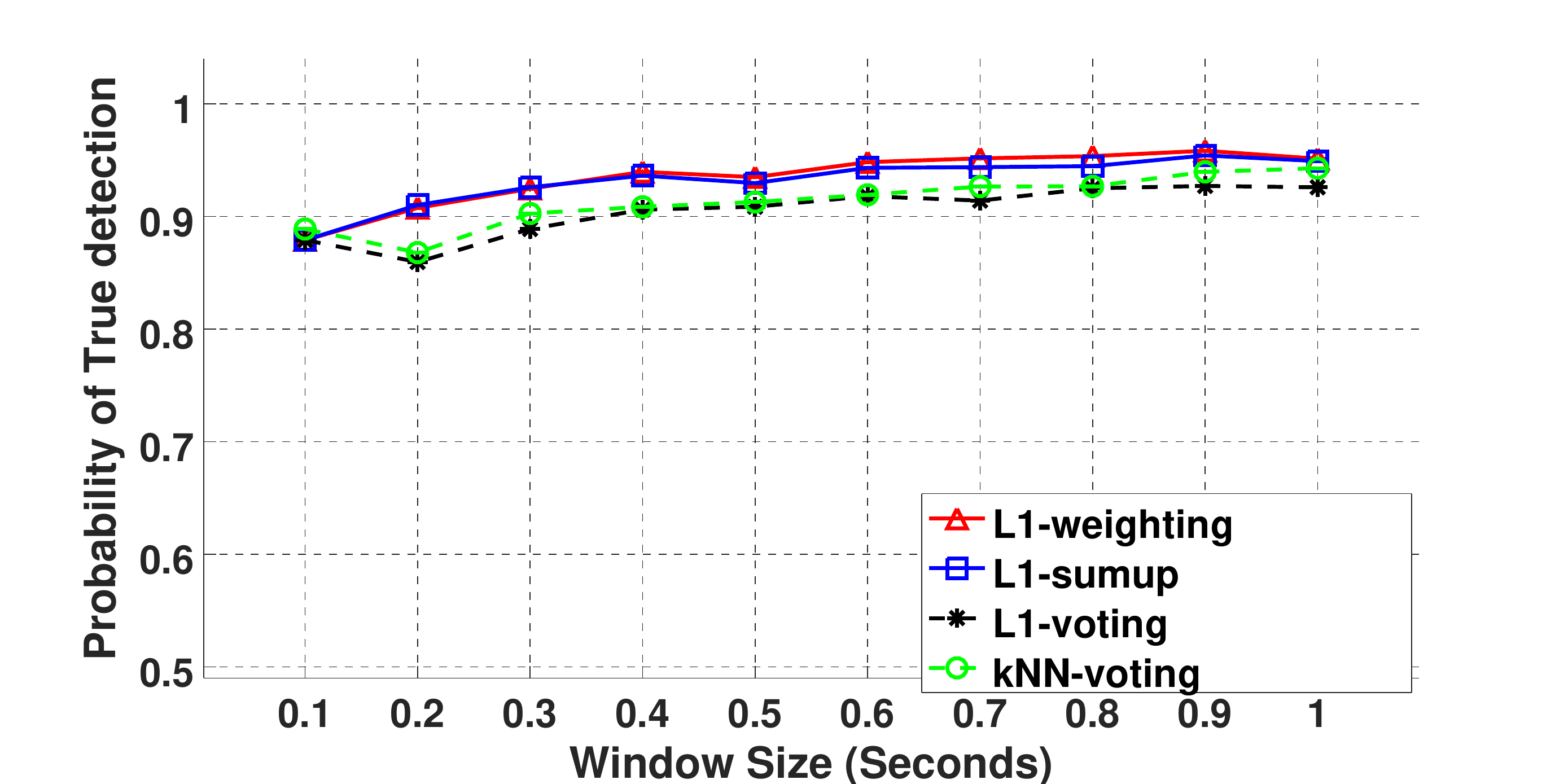}
	         \label{fig:win_part_clean}
	     }   	
    \subfigure[``\emph{Channel 157 with RFI}"]{
        \includegraphics[scale = .30]{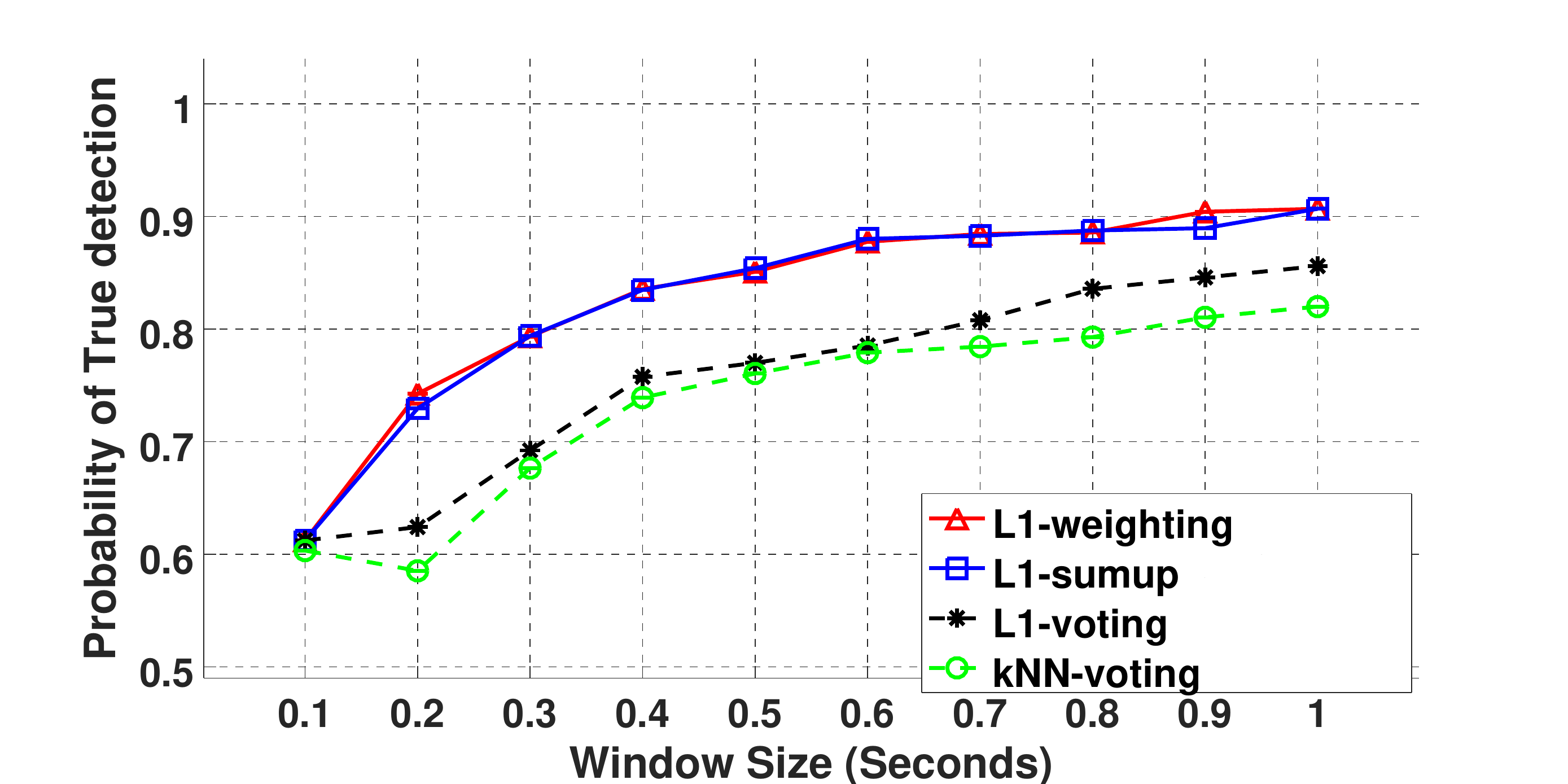}    
        \label{fig:win_part_noise}
        }       
         
         \caption{The Performance vs Window Size}
               \label{fig:win}               
}
\end{figure*}

\subsection{WASP nodes}
We use a pair of WASP nodes in our experimental evaluation. We provided some basic information on WASP nodes in Section \ref{subsec:radiocsi}. We provide further background information and explain some design choices here.  

We choose WASP because it is a software-defined radio and there is an API to obtain CSI. WASP can operate in both 2.4~GHz and 5.8~GHz. We choose to perform our evaluation in 5.8~GHz because this band is less used compared to the 2.4~GHz band. It is therefore easier to find places where RFI is absent across the entire 125~MHz bandwidth that WASP operates in. This allows us to do two things. First, we can experiment in a clean radio frequency (RF) environment and use CSI from the clean environment to establish benchmark for the classification algorithms. Second, this allows us to control the amount of RFI present in our experiments and we can be sure that any RFI present in the environment is added by us. We can therefore study the impact of RFI on activity recognition. Another reason for choosing the 5.8~GHz band is that WASP nodes have a bandwidth of 125~MHz in this band. This allows us to emulate protocols with wider bandwidth, e.g. 80~MHz bandwidth for 802.11ac. 

WASP is a low-power wireless platform. The energy cost is 2~W when WASP is receiving beacons, and 2.5~W when transmitting. A pair of WASP nodes only consumes 4.5~W \cite{sathyan2011wasp}. %The application scenario is indoor activity recognition for elder monitoring, and 
WASP nodes can be powered by cable, which means the deployment of a pair of WASP nodes for activity recognition only cost no more than 4~kWh per month.

	\begin{table*}[htb]
	\caption[Table caption text]{Bandwidth for different wireless protocols}\label{protocols}
	 \centering
	    \begin{tabular}{|l|l|l|}
	    \hline
	    Wireless Protocol     & Bandwidth per Channel	& Number of Subcarrier in OFDM	\\ \hline
	    2.4 GHz (ZigBee)      & 5 MHz                 	& 13					\\ \hline
	    2.4 GHz (802.11b/g/n) & 20 MHz                	& 52					\\ \hline
	    3.6 GHz (802.11y)     & 5/ 10/ 20 MHz         	& 13/ 26/ 52			\\ \hline
	    4.9 GHz (802.11y)     & 20 MHz                	& 52					\\ \hline
	    5 GHz (802.11a)       & 20 MHz                	& 52					\\ \hline
	    5 GHz (802.11n)       & 20/ 40 MHz            	& 52/ 104				\\ \hline
	    5 GHz (802.11ac)      & 20/ 40/ 80 MHz   	& 52/ 104/ 208		\\ \hline
	    \end{tabular}

	\end{table*}

\subsection{Experiment Setup}
The experiment is conducted in an apartment whose floor plan is shown in Fig.~\ref{fig:floorplan}. 
%The dimensions of the apartment is indicated in the floor plan. 
The AoI includes one living room (top half of the floor plan) and one bedroom (the room on the right). Two WASP nodes are deployed at the edge of the AoI. One node works as the transmitter and sends a beacons once every 0.1 second. This node is near the left-hand end of the apartment and is marked as transmitter in the floor plan. The other WASP node acts as a receiver and this is where the CSI data is collected. This node is located in the balcony just outside the bedroom. This node is marked as the receiver in the floor plan. The receiver is connected to a computer (labelled as PC in the floor plan) in the bedroom and this computer is the sink for the CSI data. The distance between the transmitter and receiver is about 5 metres. 

%(??? Did you mean the computer acted as the sink ?)  

%For this pair of WASP nodes, one is performed as transmitter ,and the role of the other one is receiver which is connected to a computer whose role is base station. 
%The sampling rate of the pair of WASP platforms is 10~Hz.

We consider 8 different location-oriented activity classes: (1)~E: empty environment (2)~L: lying on the bed in the bedroom (3)~SiB: sitting in the bedroom (4)~SiL: sitting in the living room (5)~StB: standing in the bedroom (6)~StL: standing in the living room (7)~WB: walking in the bedroom (8)~WL: walking in the living room. The location of the activities are marked in the floor plan in Fig.~\ref{fig:floorplan}. 
%Except different activities, the monitored person in different locations can also cause the CSI deviation, so we differentiate the same activities from different locations for evaluation.   
\vtbo{We perform standing and sitting in the same location, because we also want to focus on activity classification without considering different locations to evaluate the efficiency of our method. We also differentiate sitting and standing in different locations for evaluating the location-oriented activity recognition.} For each activity class, CSI data is collected for 1 minute, so that no physical environmental changes take place during this time. For a given data set, we have 600 CSI samples for each in-place activity, resulting in $600 \times 8 =4,800$ CSI samples in total.

% We collect the CSI data for 1 minute for each setting to guarantee no physical environmental changes in the surrounding environment. 

We use a computer (PC) and a WiFi router to create RFI in the environment. Their locations are marked in the floor plan in Fig.~\ref{fig:floorplan}. They use 802.11a protocol, which operates in 5.8 GHz, to communicate in Channel 157 (a 20~MHz channel). The computer communicates with the router using the echo request \verb|ping| command as fast as possible with the default packet size 7 kilobytes; the router responses the request with echo reply packet containing the exact data of request packet. The average transmission rate will arrive at more than 30 Mbit/s. In order to simulate the RFI that the activity recognition system may actually experience, we place the WiFi router in the middle of the apartment, which is a natural location that people will use in order to provide WiFi coverage to their apartment. The distance between the WiFi router and the receiver is about 4 metres,  but in one experiment, the router is moved to different locations to create different amount of RFI at the receiver. The distance between the PC and the receiver is about 1.5 metres.

% using amplitude in IPSN submission
\begin{figure*}
{
\centering
\subfigure[$ws = 5$ and $B = 20~\text{MHz}$ in ``\emph{Channel 157 without RFI}'']{
	         \includegraphics[scale = .26]{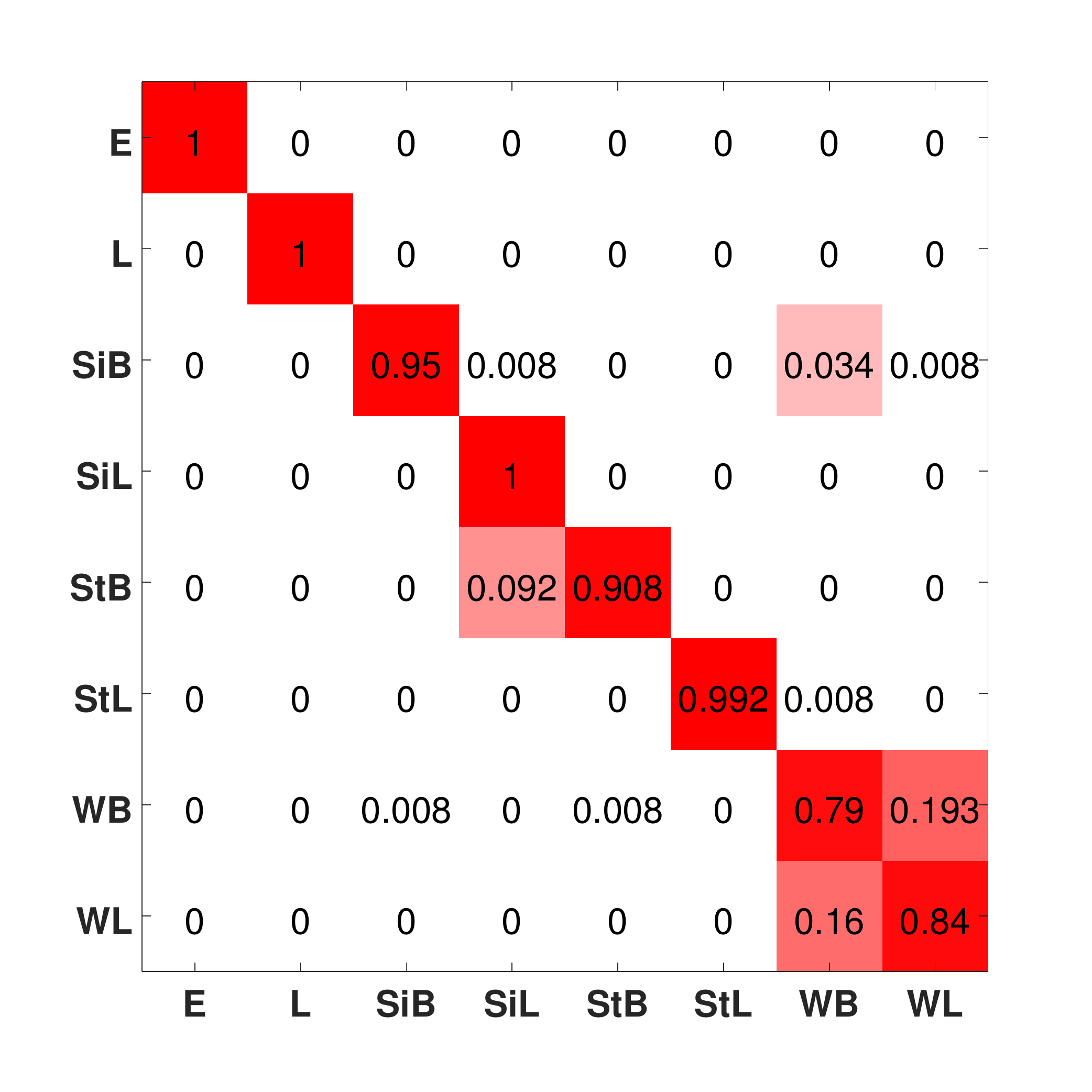}
	         \label{fig:cm_win10_clean}
	     } 
       \subfigure[$ws = 1$ and $B = 20~\text{MHz}$ in ``\emph{Channel 157 with RFI}'']{
        \includegraphics[scale = .26]{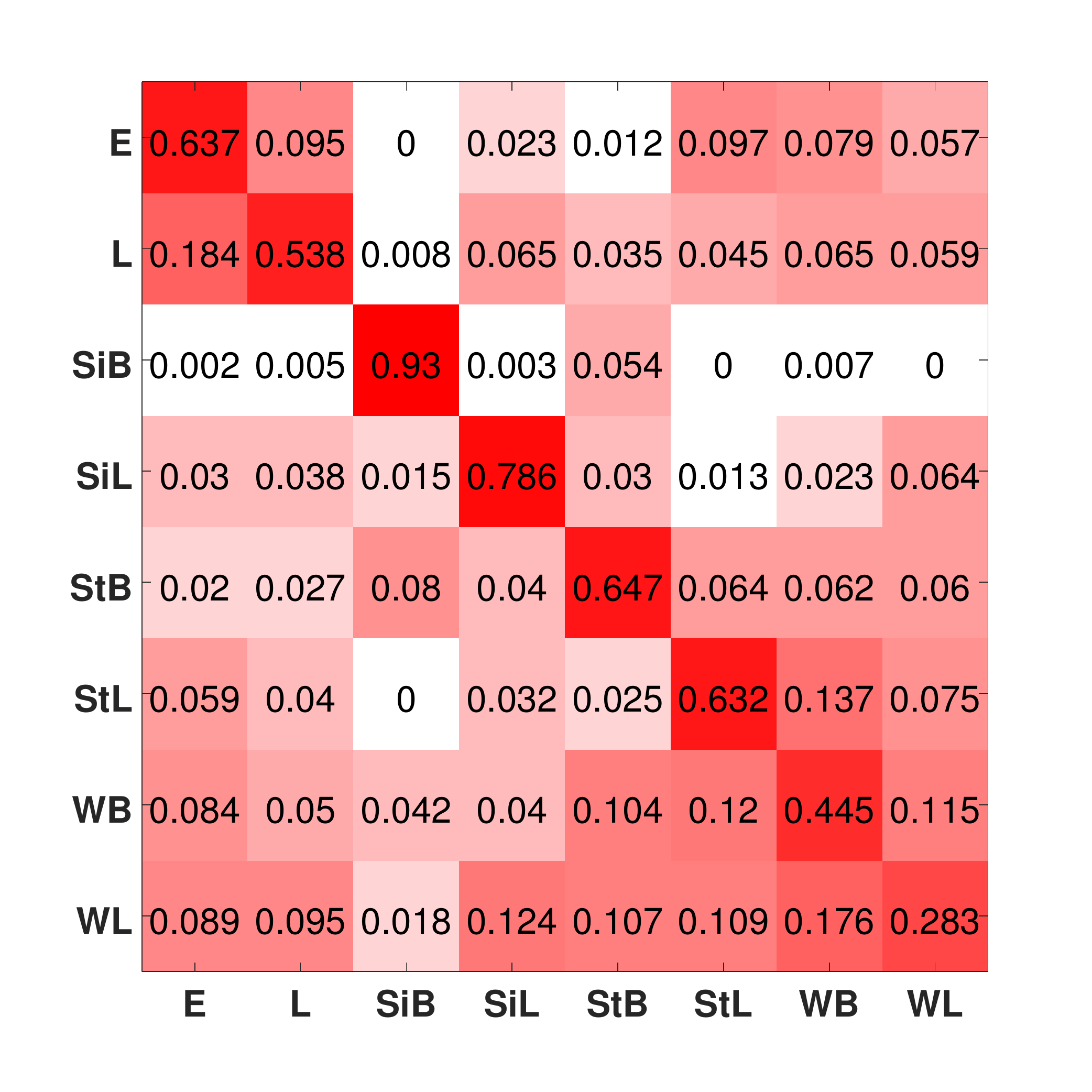}
        \label{fig:cm_win1_noise}
    }
    \subfigure[$ws = 5$ and $B = 20~\text{MHz}$ in ``\emph{Channel 157 with RFI}'']{
            \includegraphics[scale = .26]{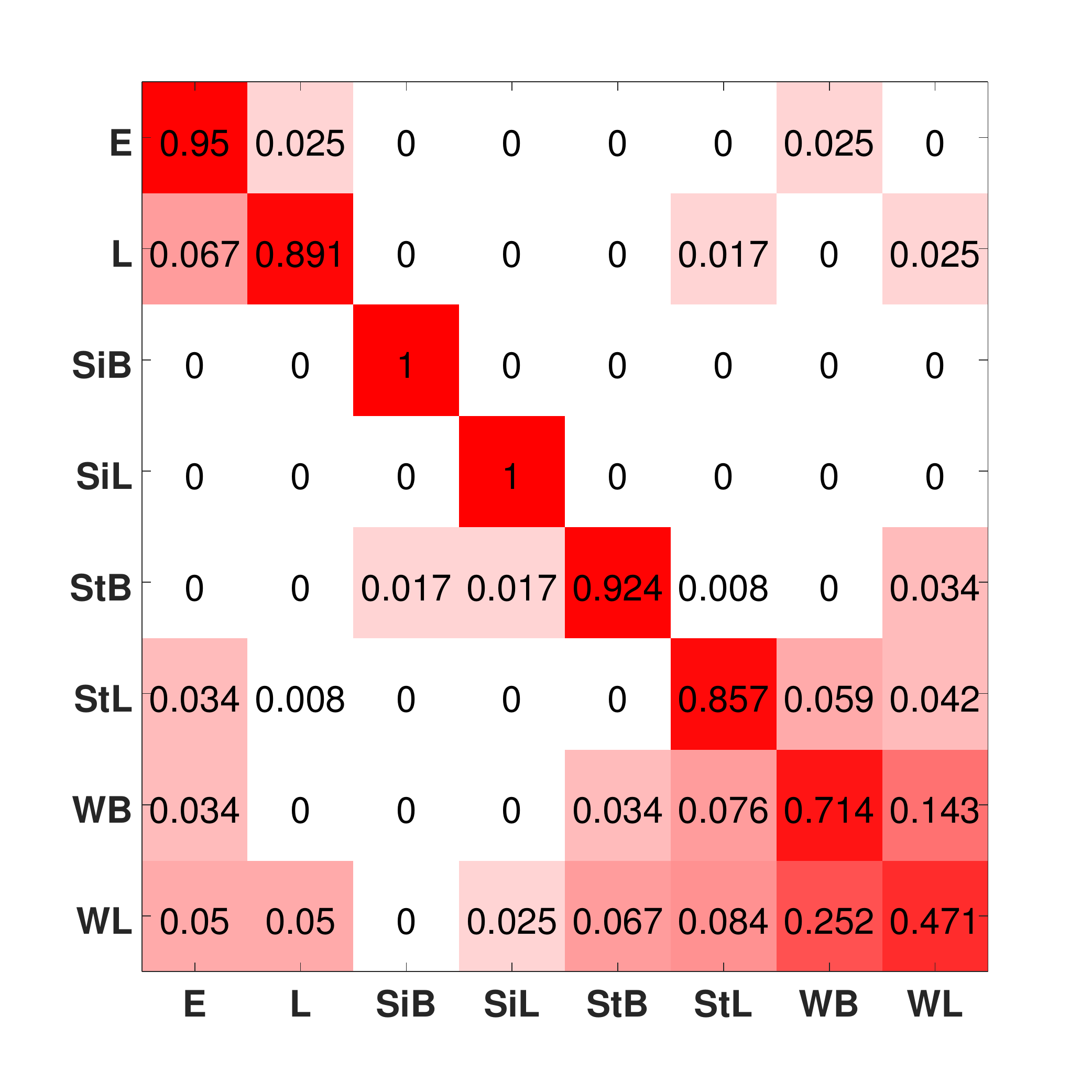}    
            \label{fig:cm_win10_noise}
            }
            
    \subfigure[$ws = 5$ and $B = 5~\text{MHz}$ in ``\emph{Channel 157 with RFI}'']{
            \includegraphics[scale = .26]{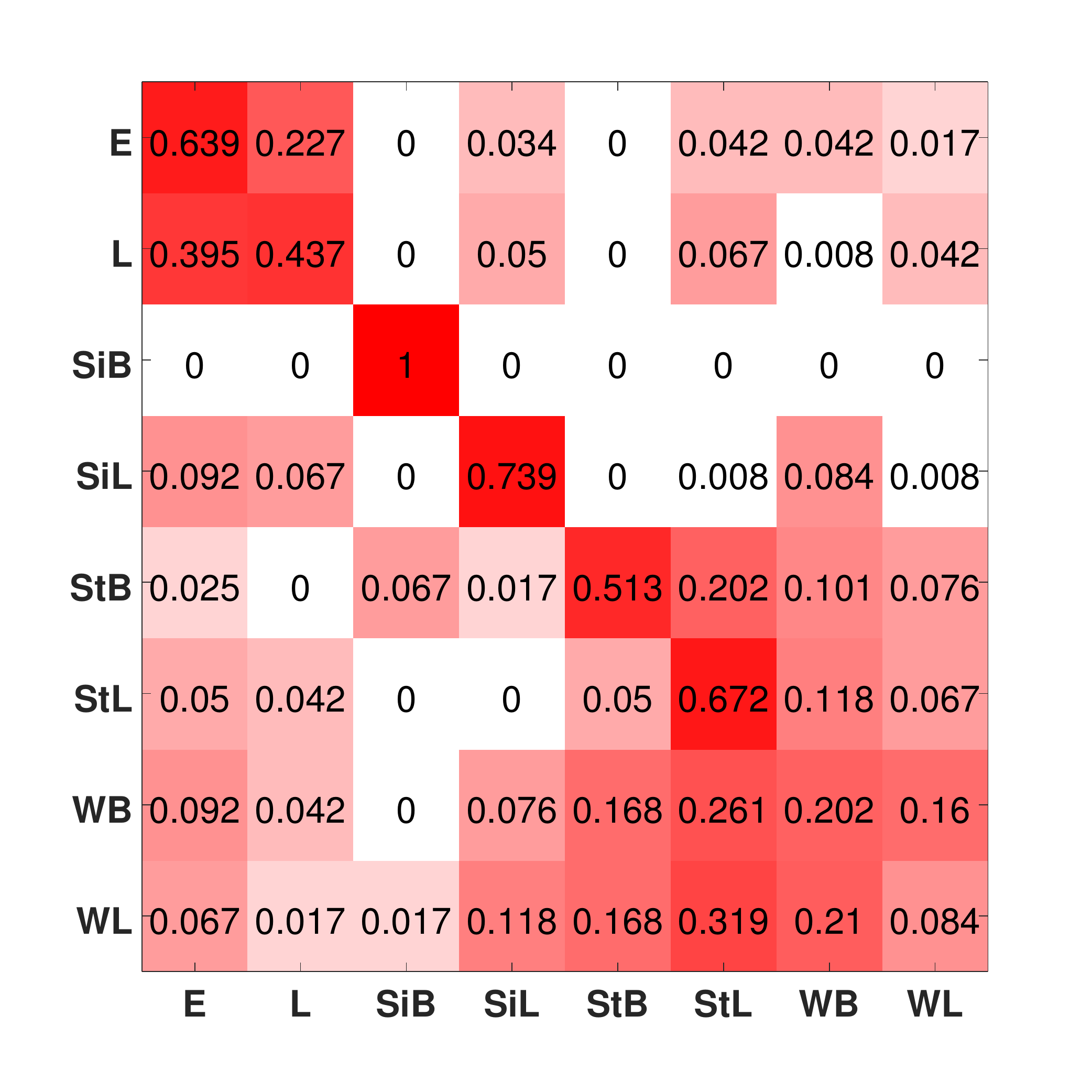}
            \label{fig:subcarrier_part_noise_cm_13}
        }
        \subfigure[$ws = 5$ and $B = 10~\text{MHz}$ in ``\emph{Channel 157 with RFI}'']{
                \includegraphics[scale = .26]{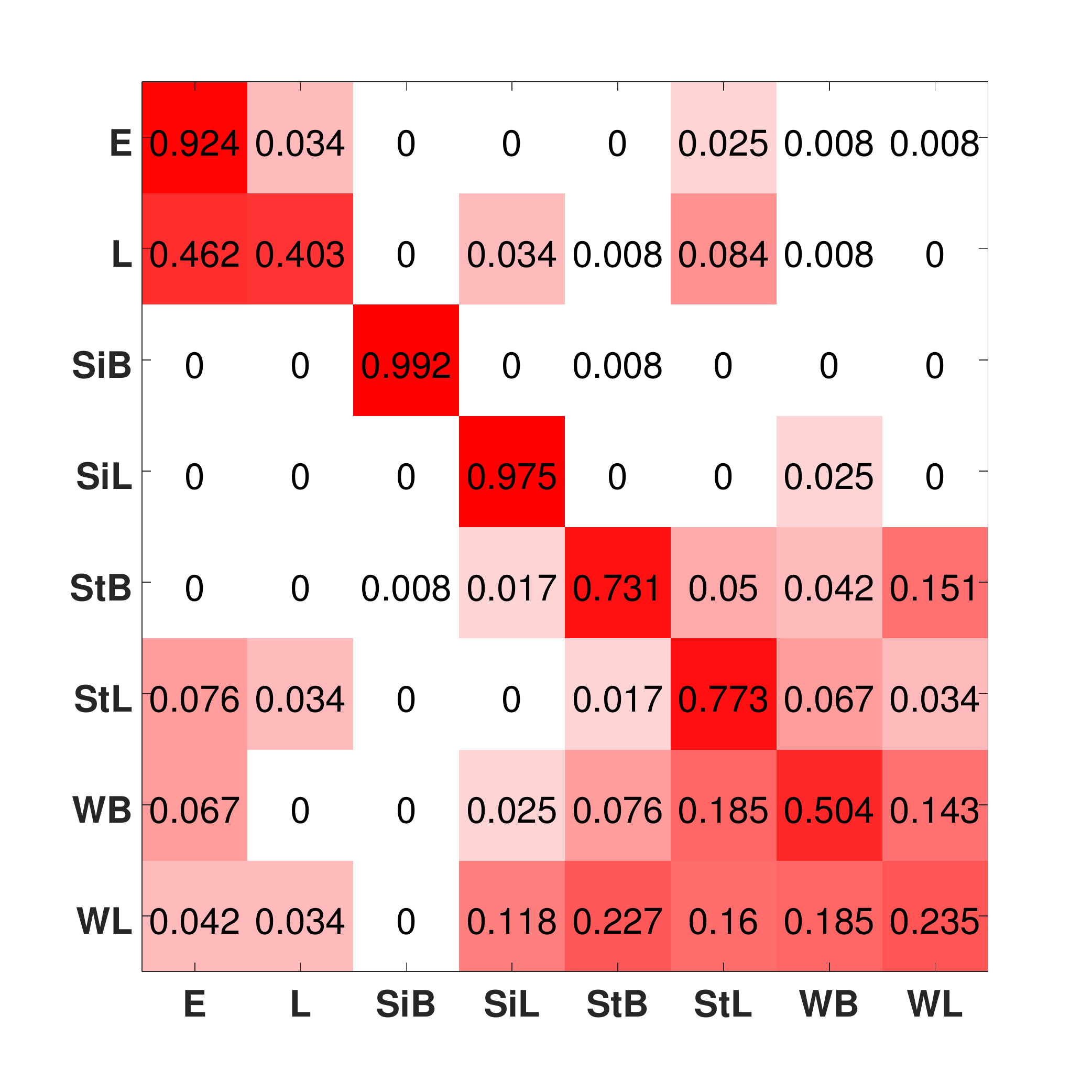}    
                \label{fig:subcarrier_part_noise_cm_26}
                }
    	 \subfigure[$ws = 5$ and $B = 15~\text{MHz}$ in ``\emph{Channel 157 with RFI}'']{
    	         \includegraphics[scale = .26]{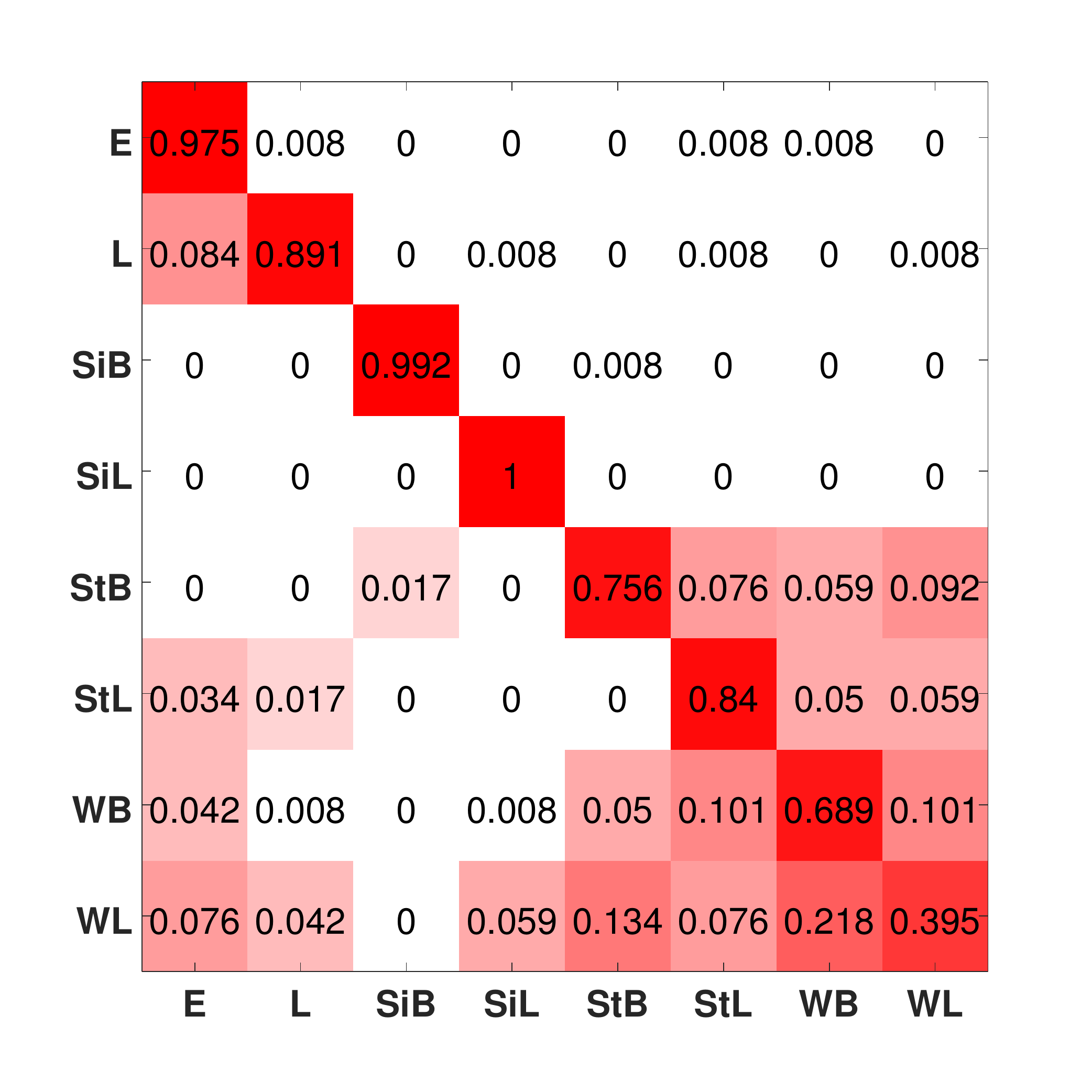}
    	         \label{fig:subcarrier_part_noise_cm_39}
    	     }
             \subfigure[Real-valued CSI, $ws = 5$ and $B = 20~\text{MHz}$ in ``\emph{Channel 157 without RFI}'']{
    	         \includegraphics[scale = .26]{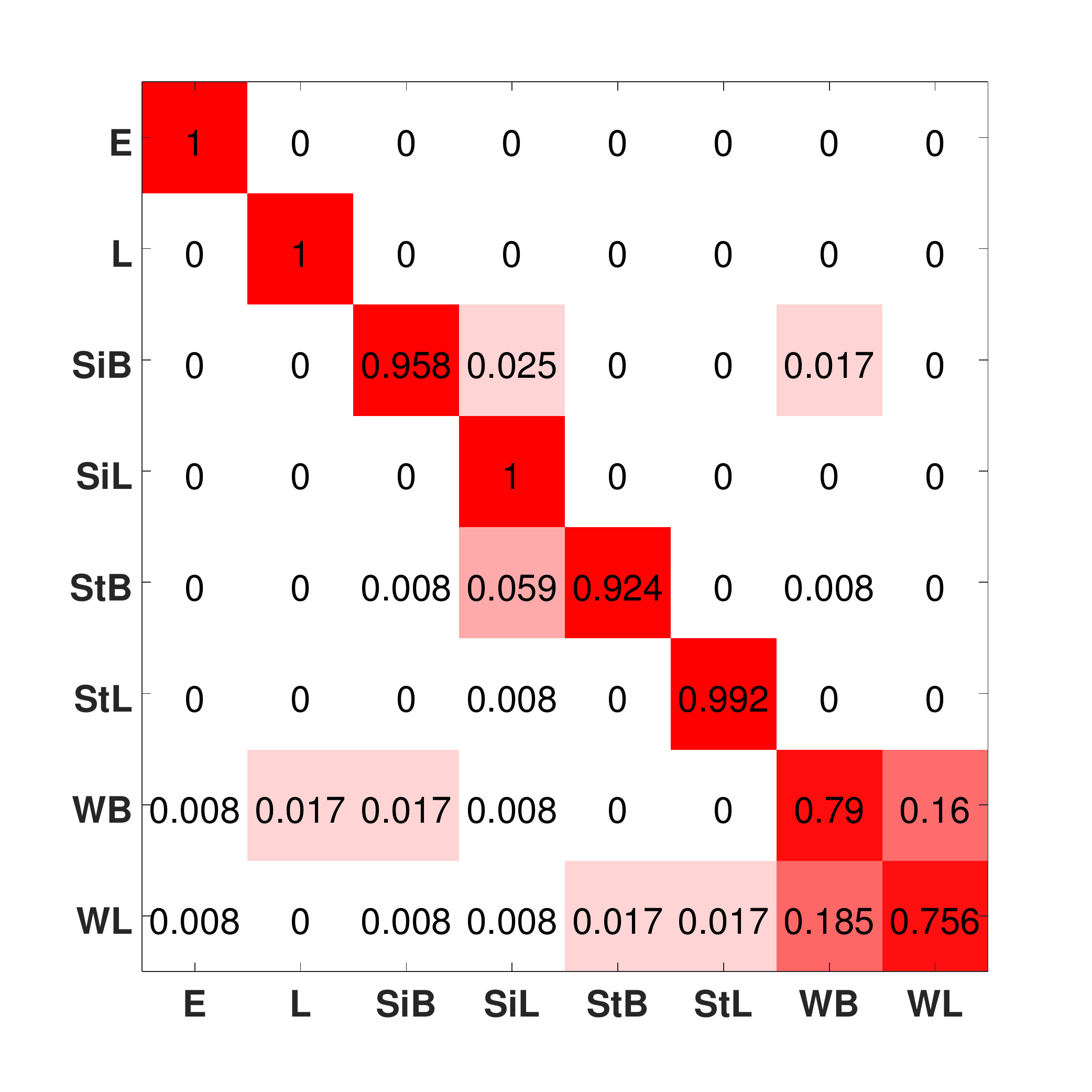}
    	         \label{fig:cm_win10_clean_real}
    	     }
             \subfigure[Real-valued CSI, $ws = 5$ and $B = 20~\text{MHz}$ in ``\emph{Channel 157 with RFI}'']{
    	         \includegraphics[scale = .26]{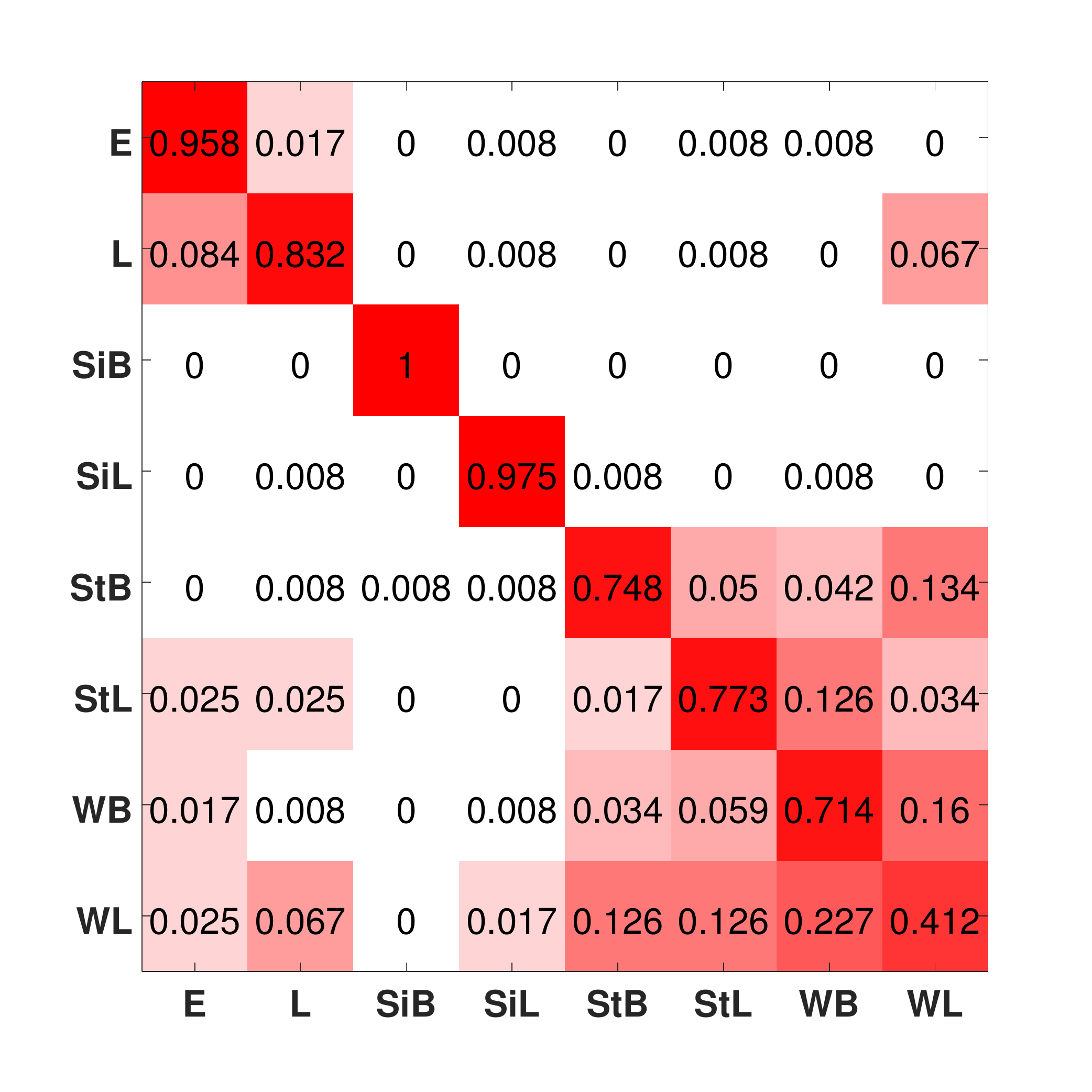}
    	         \label{fig:cm_win10_noise_real}
    	     }
       \caption{Confusion Matrix vs different settings}
               \label{fig:cm_win}               
}
\end{figure*} 

\subsection{Evaluation methodology and metrics}
We apply 10-fold cross validation to each data set to evaluate our proposed method. The results from the 10 folds are averaged to obtain the final result. We use both the probability of true detection and confusion matrix to present our results. 

%When evaluating the performance of our method for static activity recognition, we only the samples of static activities. When evaluating the walking detection, we add the samples of walking activities.

%In the evaluation of the performance of the static activities, the accuracy is the probability of true detection (i.e. the number of true detection over the summation of the total number of the true detection and false detection) for our metric to evaluate our method. Moreover, we use confusion matrix to show the performance of activity recognition.
%As the walking detection is binary classification, we use True Positive Rate (TPR) and False Positive Rate (FPR) to show the efficiency of our method.

We have two \emph{primary} data sets. One data set is collected under clean environment while the other is collected when there is RFI in Channel 157. We use these data sets to investigate the effect of bandwidth on location-oriented activity recognition. In particular, we investigate what happens if we use a bandwidth of 5~MHz, 10~MHz, 15~MHz, 20~MHz, 40~MHz, 80~MHz and 125~MHz. Let us assume that we use a bandwidth window size $B$ MHz where $B$ is one of 5, 10, 15, 20, 40, 80 or 125. Recalling that WASP nodes have a bandwidth of 125 MHz. We first select the first $B$ MHz of the 125~MHz-band and use the sub-carriers in the $B$ MHz to perform classification. We then shift the bandwidth window by 5~MHz. If a \emph{complete} $B$ MHz can be found in the data, we perform another calculations. We iterate until the whole 125~MHz is covered. We will refer to the results obtained by sliding bandwidth window over the 125~MHz band as ``\emph{whole bandwidth without RFI}" and ``\emph{whole bandwidth with RFI}". 

Instead of using the whole 125~MHz in the primary data sets. We also created two \emph{secondary} data sets, from the with and without RFI cases, which include over those sub-carriers in Channel 157. These secondary data sets span a bandwidth of 20~MHz. Note that, when interference sources exist, all sub-carriers in the secondary data sets are with RFI while only some of the sub-carriers in the primary data sets are with RFI. By using the secondary data sets, we investigate what happens when we use a bandwidth window of 5~MHz, 10~MHz, 15~MHz, 20~MHz. The methodology of shifting the bandwidth window is the same as that for primary data sets. We will refer to the results obtained from the secondary data sets as ``\emph{Channel 157 without RFI}" and ``\emph{Channel 157 with RFI}". 

Our classification algorithm uses a window size of $ws$ consecutive CSI samples for classification, as discussed in Section \ref{subsec:sparseapproximation}. We will also vary this window size in our investigation. 

We consider the following 4 classification algorithms: $k$NN with majority voting ($k$NN-$voting$), $\ell_1$-$voting$ , $\ell_1$-$sumup$ and $\ell_1$-$weighting$. In order to demonstrate the improvement in using a window size $ws$, we sometimes also show the result of using one CSI sample or a window size of 1; we will use ``$k$NN-$win1$" and ``$\ell_1$-$win1$" to denote the algorithms that use a $ws = 1$. 
\cxbo{The default SRC algorithms in this section is complex-valued unless we state otherwise. We also use complex-valued $k$NN for comparison.}

\subsection{Effect of Window Size}\label{subsec:win_size}

% \begin{figure*}[]
% {
% \centering
%        \subfigure[``\emph{whole bandwidth without RFI}"]{
%         \includegraphics[scale = .2]{figure/smoother1_walking/subcarrier_tier2/cx_win5/subcarrier_tier2_52_band_1_320_clean/l1vsknn_cx.eps}
%         \label{fig:subcarrier_whole_clean}
%     }
%     \subfigure[``\emph{Channel 157 without RFI}"]{
%     	         \includegraphics[scale = .2]{figure/smoother1_walking/subcarrier_tier2/cx_win5/subcarrier_tier2_52_band_128_52_clean/l1vsknn_cx.eps}
%     	         \label{fig:subcarrier_part_clean}
%     	     }
%     	     
%     \subfigure[``\emph{whole bandwidth with RFI}"]{
%             \includegraphics[scale = .2]{figure/smoother1_walking/subcarrier_tier2/cx_win5/subcarrier_tier2_52_band_1_320_noise/l1vsknn_cx.eps}    
%             \label{fig:subcarrier_whole_noise}
%             }  	
%     \subfigure[``\emph{Channel 157 with RFI}"]{
%         \includegraphics[scale = .2]{figure/smoother1_walking/subcarrier_tier2/cx_win5/subcarrier_tier2_52_band_128_52_noise/l1vsknn_cx.eps}    
%         \label{fig:subcarrier_part_noise}
%         }       
%          
%          \caption{The performance vs bandwidth window size}
%                \label{fig:subcarrier}               
% }
% \end{figure*} 
 \begin{figure*}[]
 {
 \centering
        \subfigure[``\emph{whole bandwidth without RFI}"]{
         \includegraphics[scale = .3]{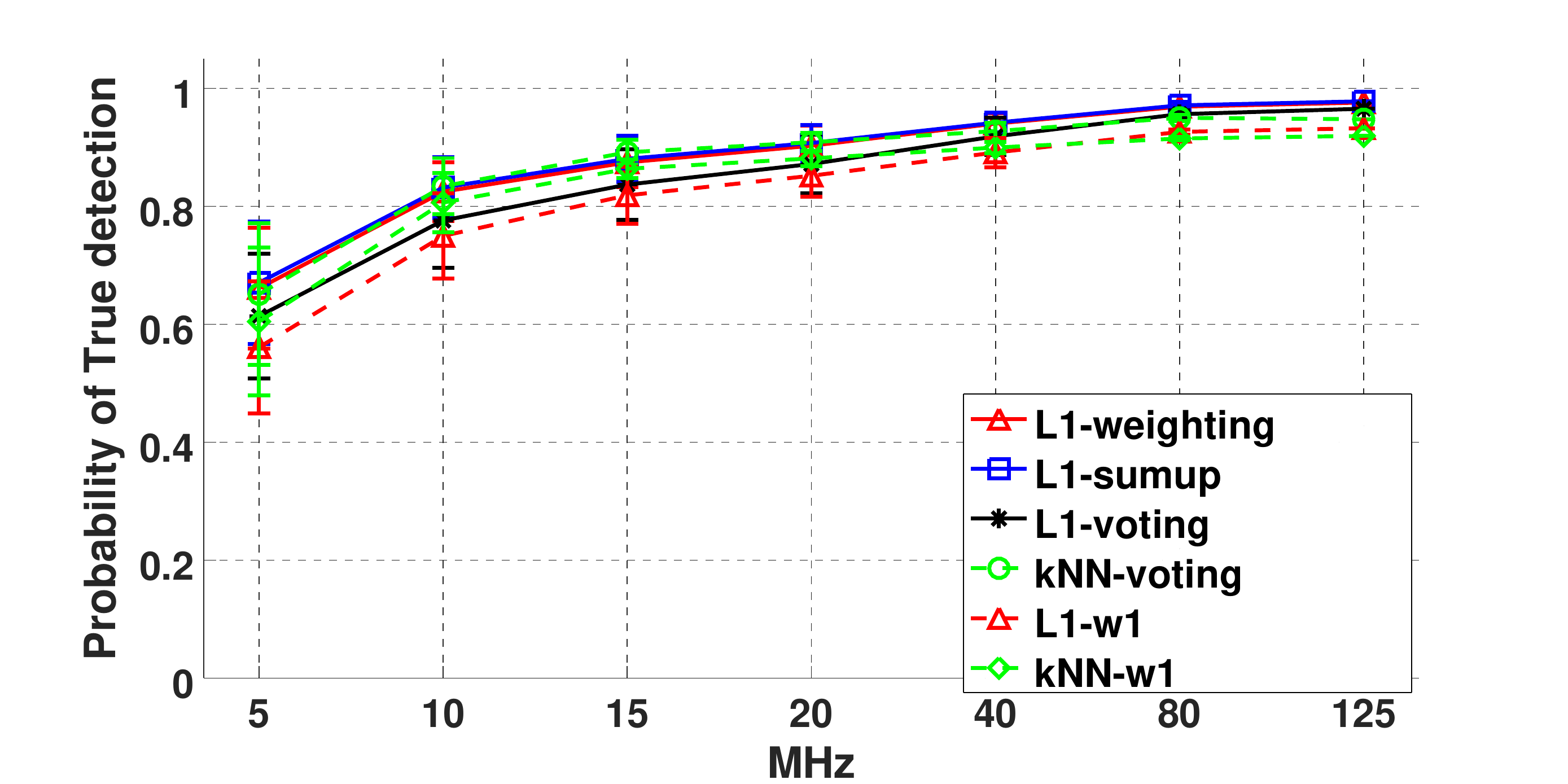}
         \label{fig:subcarrier_whole_clean}
     }
     \subfigure[``\emph{Channel 157 without RFI}"]{
     	         \includegraphics[scale = .3]{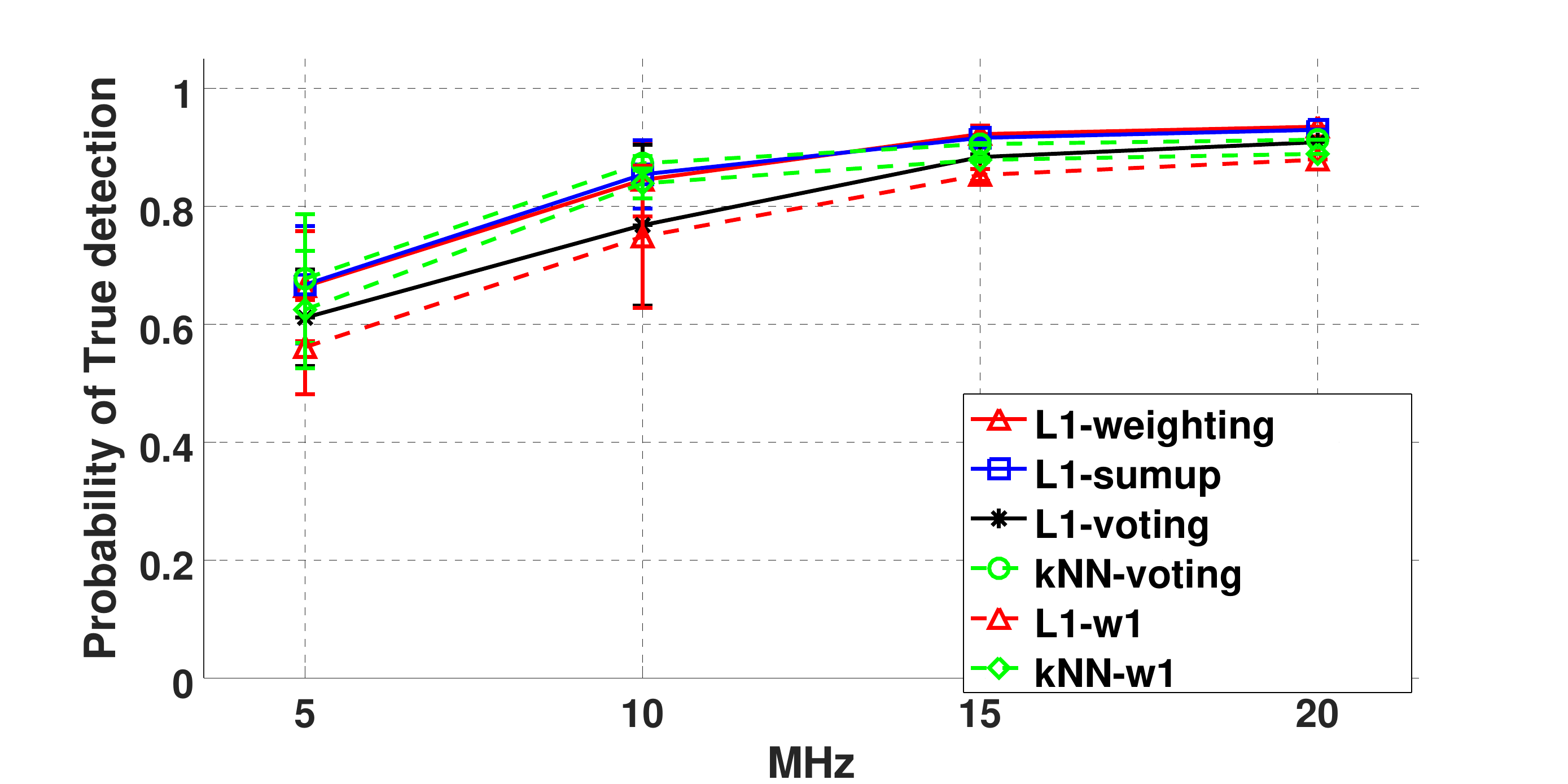}
     	         \label{fig:subcarrier_part_clean}
     	     }
     	     
     \subfigure[``\emph{whole bandwidth with RFI}"]{
             \includegraphics[scale = .3]{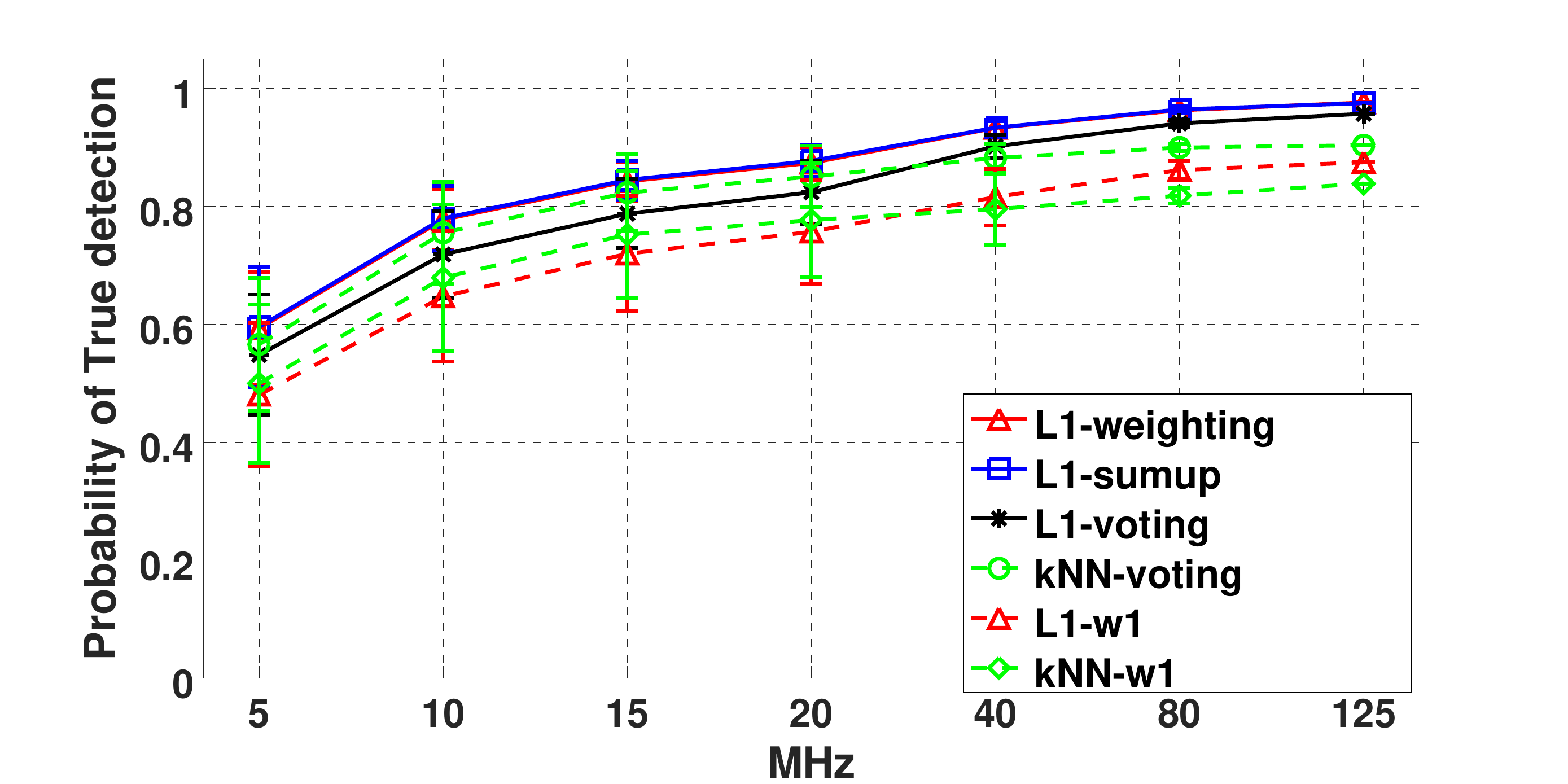}    
             \label{fig:subcarrier_whole_noise}
             }  	
     \subfigure[``\emph{Channel 157 with RFI}"]{
         \includegraphics[scale = .3]{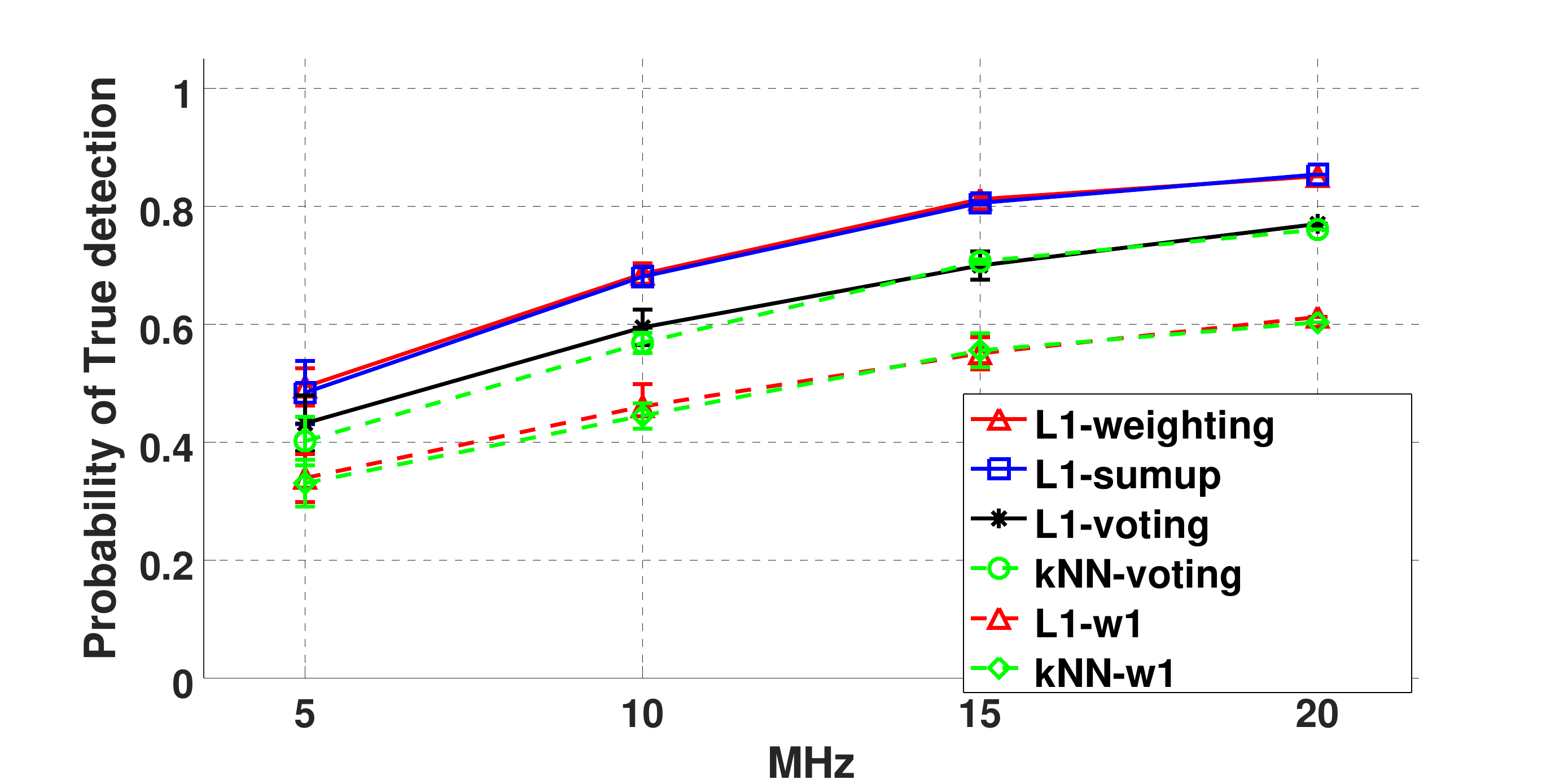}    
         \label{fig:subcarrier_part_noise}
         }       
          
          \caption{The performance vs bandwidth window size}
                \label{fig:subcarrier}               
 }
 \end{figure*} 
 
In this section, we study the impact of window size $ws$ on  activity recognition performance. We use $ws$ from 1 to 10, which correspond to a time of 0.1~s and 1~s, because the transmitter sends beacons at a frequency of 10~Hz. We will show that window size can improve accuracy but this is at the expense of decreasing the temporal resolution of activity recognition.
%\cxbo{, and also compare the performance of only using CSI amplitude vectors with that of using CSI vectors.}
We will use both primary data sets (``\emph{whole bandwidth without RFI}" and ``\emph{whole bandwidth with RFI}") and both secondary data sets (``\emph{Channel 157 without RFI}" and ``\emph{Channel 157 with RFI}") in this study. We assume a bandwidth window size $B = 20$, which is the bandwidth of one 5.8~GHz 802.11 channel.
\cxbo{Fig.~\ref{fig:win} shows performance in different window size using different data fusion methods. 
%Fig.~\ref{fig:win_2} shows performance by using CSI vectors and CSI amplitude vectors , and '-am' suffix means the responding methods only using CSI amplitude vectors. 
%In Fig.~\ref{fig:win_2}, we compare the performance by using the algorithm $\ell_1$-$weighting$.
}
 
First, we discuss the performance of using ``\emph{whole bandwidth without RFI}". 
Fig.~\ref{fig:win_whole_clean} shows the probability of true detection of the 4 different algorithms. When there is no RFI and with a 20~MHz bandwidth window size, a window size $ws$ of 1 can already achieve an accuracy of approximately 90\% for all four classification algorithms. The accuracy gradually increases to 95\% when the window size is increased to 10. The results are similar if we use ``\emph{Channel 157 without RFI}", as shown in Fig.~\ref{fig:win_part_clean}. The algorithms $\ell_1$-$weighting$ and $\ell_1$-$sumup$ show similar performance but are slightly better than $k$NN$-voting$ \cxbo{and $\ell_1$-$voting$}. This shows that, without RFI, very good classification accuracy can be obtained. 

As we discussed earlier, the challenge is to perform classification when there is RFI. Fig.~\ref{fig:win_whole_noise} shows the probability of true detection for the data set ``\emph{whole bandwidth with RFI}". It shows that the performance increases with larger window size. Among the four algorithms used, $\ell_1$-based algorithms outperform $k$NN and the best performing algorithms are $\ell_1$-$weighting$ and $\ell_1$-$sumup$. If we compare the classification accuracy between without and with RFI in Fig. \ref{fig:win_whole_clean} and \ref{fig:win_whole_noise}, we see a significant drop in accuracy when RFI is present especially when the window size is small. For example, for $ws = 1$, accuracy decreases from 85\% to 76\% because of RFI. 
%\cxbo{In the scenario, the difference between the performance of using CSI vectors and CSI amplitude vectors is small.}  
We now turn to the data set ``\emph{Channel 157 with RFI}" \cxbo{whose results shown in Fig.~\ref{fig:win_part_noise}}. Again,the accuracy increases when the window size is increased, and both $\ell_1$-$weighting$ and $\ell_1$-$sumup$ perform the best. The most telling observation is that, for $ws = 1$, the classification accuracy is merely 63\% but if $ws = 5$ is used, an accuracy of almost 85\% can be obtained. This shows that window size can have a significant effect on performance when RFI is present.  

We now present the confusion matrices for location-oriented activity recognition using our proposed $\ell_1$-$weighting$. Fig.~\ref{fig:cm_win10_clean} shows the confusion matrix for ``\emph{Channel 157 without RFI}'' with a window size of \cxbo{5 (0.5 second)}. It shows that perfect accuracy is achieved with 5 activities. The accuracy for static activities is extremely high. The accuracy for the two walking activities are also very good. We now present the confusion matrix for ``\emph{Channel 157 with RFI}" with a window size of \cxbo{0.1 second and 0.5 second}, in respectively, Fig.~\ref{fig:cm_win1_noise} and \ref{fig:cm_win10_noise}. It can be seen that a big window size has significantly improved classification accuracy of many activities. In the following sections, we will use a window size \cxbo{$ws = 5$ (0.5 second) by default.}

\subsection{Effect of the Bandwidth Window Size}

In this section, we discuss the influence of the bandwidth window size on the location-oriented activity recognition performance. This study takes advantage of the available wide bandwidth from WASP nodes to simulate different kinds of protocols shown in Table~\ref{protocols}.

The result of using different bandwidth window size $B$ for data set ``\emph{whole bandwidth without RFI}" is shown in Fig.~\ref{fig:subcarrier_whole_clean}. As expected, increasing $B$ gives a better classification accuracy. In particular, $\ell_1$-$weighting$ achieves an accuracy of \cxbo{70\%, 90\% and 95\% }when using a bandwidth window $B$ of 5~MHz, 20~MHz and 125~MHz respectively.  Similar trend is also observed for the data set ``\emph{Channel 157 without RFI}" (shown in Fig.~\ref{fig:subcarrier_part_clean}). The algorithms $\ell_1$-$weighting$, $\ell_1$-$sumup$, $\ell_1$-$voting$  and $k$NN-$voting$  show similar performance. 

Fig.~\ref{fig:subcarrier_whole_noise} shows the probability of true detection for the data set ``\emph{whole bandwidth with RFI}''. It shows the performance increases when the bigger bandwidth window size increases. Comparing four algorithms using window size $ws$ \cxbo{5}, $\ell_1$-$weighting$ and $\ell_1$-$sumup$ perform the best, \cxbo{which outperform $\ell_1$-$voting$ and $k$NN-$voting$}. When looking at the data set ``\emph{Channel 157 with RFI}'' (shown in Fig.~\ref{fig:subcarrier_part_noise}), the accuracy decreases significantly, especially when the bandwidth window sizes $B$ are 5~MHz and 10~MHz whose accuracy decreases from \cxbo{72\% and 88\% to 50\% and 70\%} using $\ell_1$-$weighting$ compared with the data set ``\emph{Channel 157 without RFI}". The performance of $\ell_1$-$weighting$ and $\ell_1$-$sumup$ is close, but they show their superiority over $\ell_1$-$voting$   and $k$NN-$voting$, for example, the accuracy of $\ell_1$-$weighting$ and $\ell_1$-$sumup$ is 5\% better than $\ell_1$-$voting$   and $k$NN-$voting$ when the bandwidth window size $B$ is 20~MHz. This shows $\ell_1$-$weighting$ and $\ell_1$-$sumup$ algorithms increase the recognition performance when there is RFI. 

Fig.~\ref{fig:subcarrier_part_noise_cm_13}, Fig.~\ref{fig:subcarrier_part_noise_cm_26}, Fig.~\ref{fig:subcarrier_part_noise_cm_39} and  Fig.~\ref{fig:cm_win10_noise} illustrate the confusion matrix using $\ell_1$-$weighting$ with the bandwidth window size $B$ 5~MHz, 10~MHz, 15~MHz and 20~MHz respectively in the data set ``\emph{Channel 157 with RFI}''. It shows that the using larger bandwidth window size $B$ helps increase the accuracy and robustness to RFI. When the bandwidth window size $B$ is 20~MHz, \cxbo{2 static activities have perfect accuracy, 2 static activities have accuracy more than 90\%, the other 2 static activities have accuracy more than 85\%, and ``walking in bedroom'' has the accuracy more than 70\%. }

% TODO 20180105 6:10pm

\begin{figure}[htb]
	\centering
	\includegraphics[scale = 0.5]{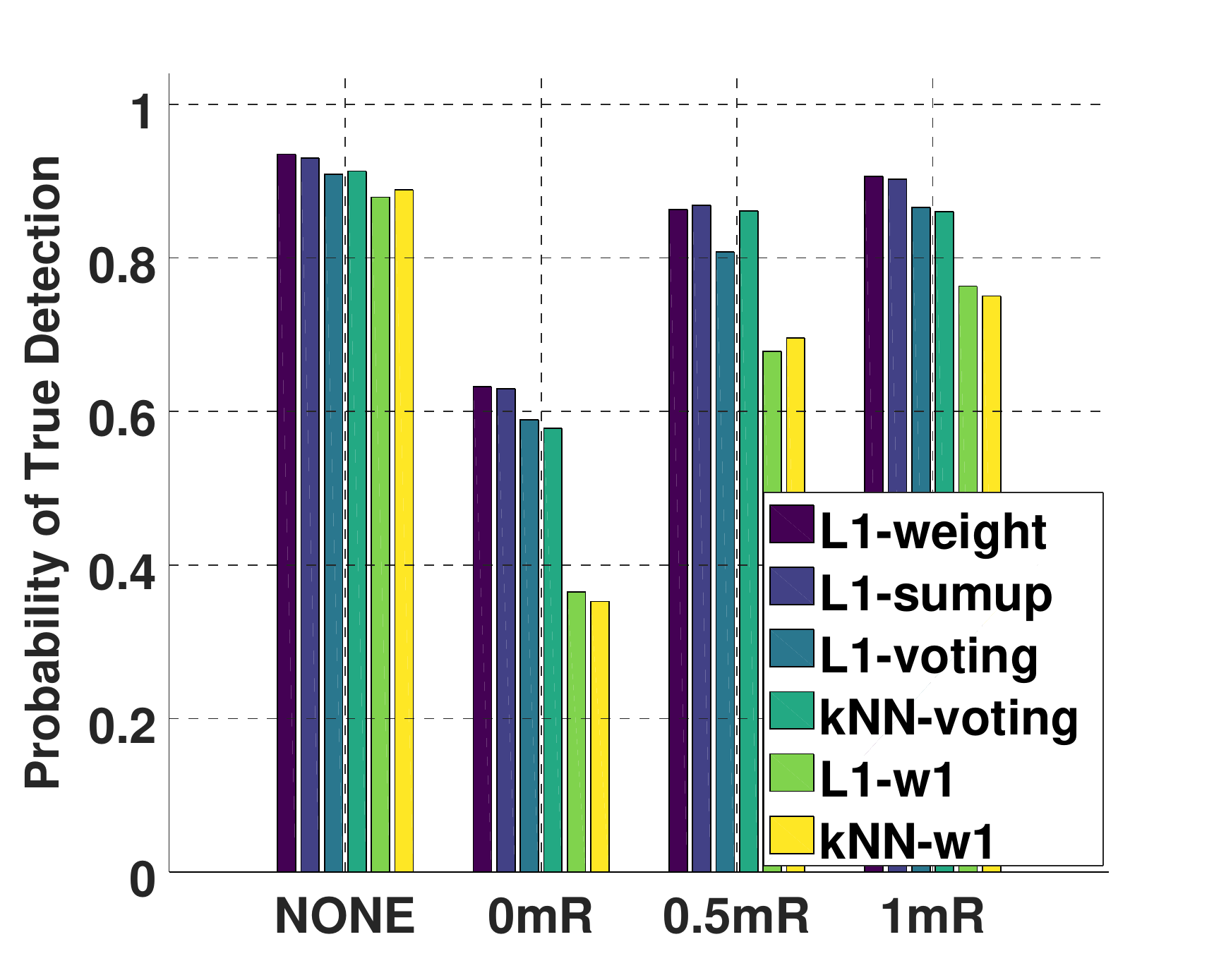}
	\caption{The performance influenced by the distances between the router and the receiver }\label{fig:dis}
\end{figure}

\subsection{Effect of Different Distances Between the Router and Receiver}\label{subsec:dis}
In order to evaluate the impact of the amount of interference on classification. We vary the distance between one interferer (the WiFi router in the floor plan in Fig.~\ref{fig:floorplan}) and the WASP receiver. We consider 4 cases: no interference (NONE), interferer just next to the the receiver (0mR), 0.5~metre away from the receiver (0.5mR), and 1~metre away form the receiver (1mR). The classification uses only Channel 157. The computer also communicates with the router using the echo request \verb|ping| command as fast as possible. 

Fig.~\ref{fig:dis} shows the influence of receiver-interferer distance on the probability of true detection. It shows that for 0mR, the accuracy can drop by more than 10\%. However, by using a window size \cxbo{$ws = 5$}, $\ell_1$-$weighting$ has an accuracy about 60\%. Note that there is not much difference between the 0.5mR and 1mR cases. Also, there is only a slight drop in performance for the $\ell_1$ algorithms between NONE and 0.5mR cases. This shows that if an interferer is not present within 0.5m of the receiver, then the activity recognition accuracy is still high. 

\begin{figure}[htb]
   \centering
    \includegraphics[scale = 0.5]{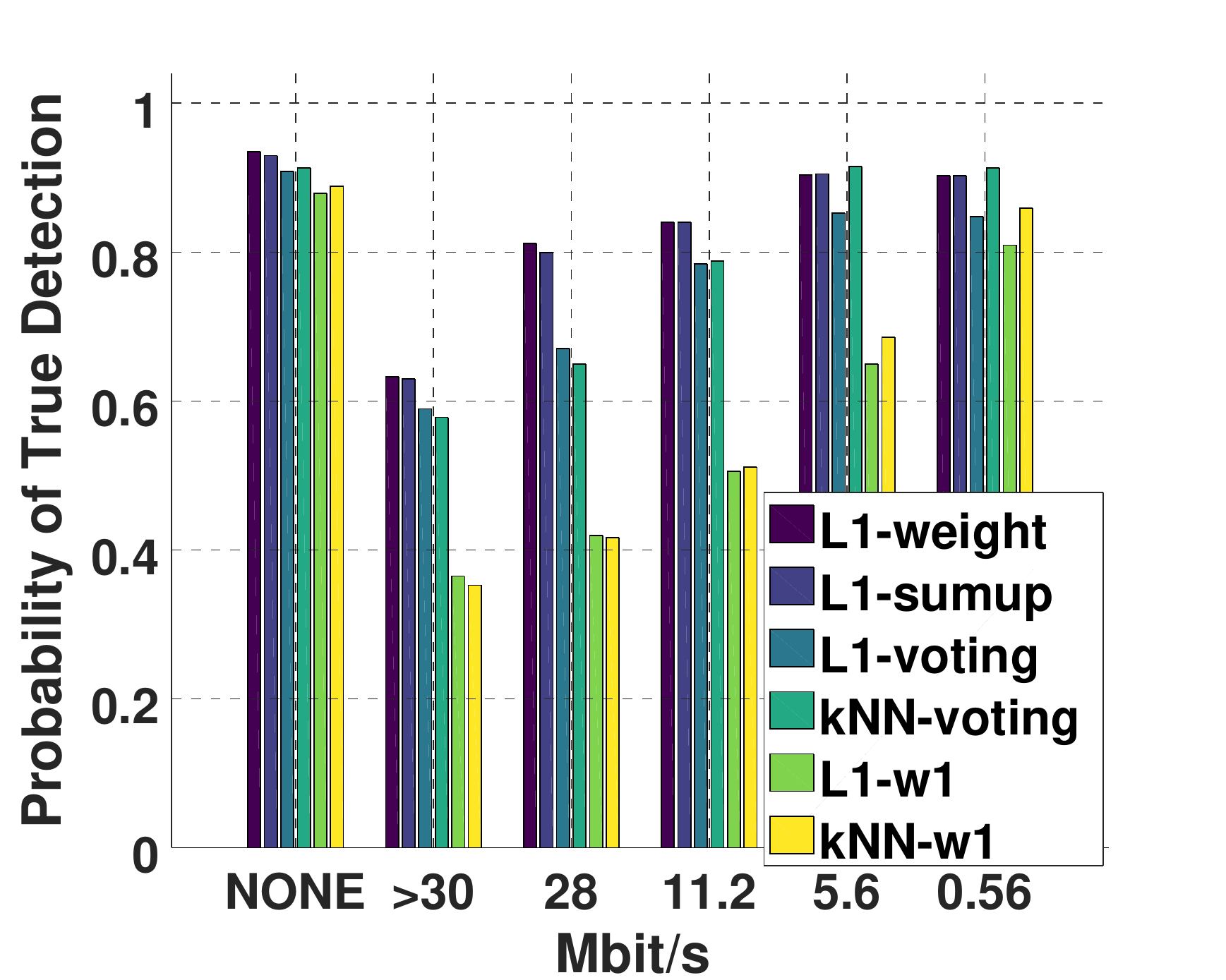}
    \caption{The performance under different transmission rates between the router and the receiver }\label{fig:traffic}
\end{figure}

\subsection{Effect of Different traffic Between Interference Source and Receiver}\label{subsec:traffic}

In this section, we study the effect of traffic sending rates on the classification performance. We keep one interference source (the WiFi router in the floor plan in Fig.~\ref{fig:floorplan}) next to the receiver and adjust the \verb|ping| rates of the interferer.  
The \verb|ping| rate is set to ``as fast as possible'' (average transmission rate more than 30 Mbit/s ), 500 packets per second (transmission rate 28 Mbit/s), 200 packets per second(transmission rate 11.2 Mbit/s), 100 packets per second (transmission rate 5.6 Mbit/s) and 10 packets per second (transmission rate 0.56 Mbit/s). These settings roughly correspond to the bit rates of watching online videos with frame rates 1080p, 480p and 360p, which give rise to bit rates of 8~Mbit/s, 5~Mbit/s and 1~Mbit/s respectively.

\cxbo{Fig.~\ref{fig:traffic} shows the classification performance under different transmission rates. Our proposed $\ell_1$-$weighting$ method reaches 80\% accuracy when transmission rate is 28 Mbit/s, which means it is robust to the transmission rate 28 Mbit/s. In contrast, the accuracy of both $\ell_1$-$voting$   and $k$NN-$voting$ algorithm is no more than 70\%. \emph{This means our proposed $\ell_1$-$weighting$ is more robust to RFI than the other methods}. Moreover, $\ell_1$-$weighting$ achieves an accuracy of 85\% when the transmission rate is 11.2~Mbit/s. The accuracy stays almost the same for lower transmission rates. } 

\begin{table}
\caption{Features for SNR based walking detection}\label{features}
    \begin{tabular}{|l|l|}
        \hline
        Feature                  & Equation \\ \hline
        Standard Deviation       & $\sigma = \frac{1}{n-1}  \sqrt{ E[( S(k)- \mu )^2]  }$         \\ \hline
        Peak                     & $\rho = max ( S ) - min ( S )$          \\ \hline
        Head Size                & $\eta = max ( S ) - median ( S )$        \\  
        \hline
        3rd Order Central moment & $\gamma = E[(S(k)- \mu)^3 ] $        \\ \hline
    \end{tabular}
    
\end{table}

\subsection{SNR based Walking Detection Discussion} \label{subsec:WalkingDetection}

In Section \ref{sec:methods}, we discuss the possibility of using SNR to differentiate walking from non-walking, i.e. a detection or binary classification problem. We see from Fig.~\ref{fig:exampleRSS} that, when RFI is absent, this is probably feasible because walking gives rise to highly fluctuating \cxbo{RSS} while non-walking does not. However, in the presence of RFI, the distinction between walking and non-walking is not so conspicuous, as seen in Fig.~\ref{fig:exampleRSSnoise}. In this section, we want to investigate what classification performance we can get if we use SNR for walking detection. 
This study is also motivated by the fact that many device-free localisation methods \cite{ZhaoNoise2011,KaltiokallioBP12,ZhaoKRTI:2013} use SNR as a feature to find the location of the person in AoI. 

The detection problem is to detect whether the person is walking. This covers the classes of WL and WB instead of walking detection with location information. A commonly used feature for device-free localisation is the variance of the SNR, which is used in for example \cite{ZhaoNoise2011}. However, in order to minimise the possibility that walking detection fails due to poor choice of features, we have chosen to use 4 different features listed in Table \ref{features}: Standard Deviation, Peak, Head Size, and 3rd Order Central moment where $S$ is SNR vectors and $\mu$ is the mean of SNR values in a vector. For training, we use a logistic regression model. 

The alternative to using SNR for walking detection is to use CSI with $\ell_1$ classifier. We will compare these two methods. The comparison uses 10-fold cross validation and measurements from Channel 157. We consider two data sets, ``\emph{Channel 157 without RFI}" and ``\emph{Channel 157 with RFI}". Also, for both data sets, we use bandwidth window sizes $B$ of 5~MHz, 10~MHz and 20~MHz bands. 

Since walking detection is binary classification, we use the following metrics:

\textbf{True Positive Rate (TPR)}: $TPR = TP/(TP + FN)$

\textbf{False Positive Rate (FPR)}: $FPR = FP/(FP + TN)$

\textbf{F1 score}: $F1~score = 2TP/(2TP + FP + FN)$\\
where $TP$, $TN$, $FP$ and $FN$ are the number of true positives, true negatives, false positives, and false negatives respectively.   
% is the number of true positives, $TN$ is the number of  $FN$ is the number of false negatives, $FP$ is the number of false positives. 

Table \ref{walkingdetectionRFIno} shows the comparison between detection using SNR and using CSI. Each column illustrates one bandwidth window size $B$ in each data set and we highlight the better statistics.
A number of observations can be made: (1) In the absence of RFI, detection using SNR has a higher TPR compared to using CSI for a bandwidth window of \cxbo{5~MHz and 10~MHz; however, for a bandwidth window of 20~MHz}, detection using CSI has a higher TPR. (2) In the absence of RFI, it is viable to use either SNR or CSI based detector for walking detection. (3) RFI causes the performance of both detectors to decrease. However, the SNR-based detector has a sharper drop in TPR. (4) Overall, the CSI-based detector is more robust in the presence of RFI.

\begin{table*}
 \centering
\caption[Table caption text]{Performance of walking detection \vtbo{(Better statistics in each column is highlighted)}} \label{walkingdetectionRFIno}
\begin{tabular}{|l|l|l|l|l|l|l|}
	\hline
	& \multicolumn{3}{c|}{Without RFI}                    & \multicolumn{3}{c|}{With RFI}                       \\ \hline
	Bandwidth      & 5 MHz           & 10 MHz          & 20 MHz          & 5 MHz           & 10 MHz          & 20 MHz          \\ \hline
	TPR (SNR)      & \textbf{0.7059} & \textbf{0.8740} & 0.9076          & 0.0924          & 0.1975          & 0.4370          \\
	FPR (SNR)      & 0.0420          & \textbf{0.0187} & \textbf{0.0070} & \textbf{0.0256} & \textbf{0.0490} & 0.0506          \\
	F1 score (SNR) & \textbf{0.7648} & \textbf{0.9052} & 0.9407          & 0.1446          & 0.2885          & 0.5428          \\ \hline
	TPR (CSI)      & 0.4706          & 0.8313          & \textbf{0.9913} & \textbf{0.3277} & \textbf{0.5336} & \textbf{0.8393} \\
	FPR (CSI)      & \textbf{0.0056} & 0.112           & 0.0084          & 0.0938          & 0.0574          & \textbf{0.0309} \\
	F1 score (CSI) & 0.6328          & 0.8314          & \textbf{0.9833} & \textbf{0.4073} & \textbf{0.6256} & \textbf{0.8393} \\ \hline
\end{tabular}
\end{table*}

\cxbo{

\subsection{Effect on Complex-valued CSI}\label{subsec:evulcom}
\subsubsection{ Effect of Complex-Valued CSI Based Classification Discussion}\label{subsec:expcxclean}

%\begin{table*}
%	\centering
%	\caption{True Detection Rates using Complex CSI vs Real CSI: "whole bandwidth without RFI" }\label{tab:cxrealcomcleanwhole}
%	\begin{tabular}{|l|l|l|l|l|l|l|}
%		\hline
%		& L1\_weight & L1\_sumup & L1\_voting & L1\_w1 & kNN\_voting & kNN\_w1 \\ \hline
%		Complex & 0.902      & 0.907     & 0.872      & 0.852  & 0.909       & 0.852   \\ \hline
%		Real    & 0.887      & 0.893     & 0.853      & 0.834  & 0.894       & 0.866   \\ \hline
%		
%	\end{tabular}
%	
%\end{table*}
%
%\begin{table*}
%	\centering
%	\caption{True Detection Rates using Complex CSI vs Real CSI: "whole bandwidth with RFI" }\label{tab:cxrealcomnoisewhole}
%	\begin{tabular}{|l|l|l|l|l|l|l|}
%		\hline
%		& L1\_weight & L1\_sumup & L1\_voting & L1\_w1 & kNN\_voting & kNN\_w1 \\ \hline
%		Complex & 0.873      & 0.877     & 0.824      & 0.757  & 0.834       & 0.777   \\ \hline
%		Real    & 0.848      & 0.852     & 0.798      & 0.730  & 0.855       & 0.730   \\ \hline
%	\end{tabular}
%	
%\end{table*}	
\begin{table*}
	\centering
	\caption{True Detection Rates using complex-valued CSI vs Real CSI: "Channel 157 without RFI" }\label{tab:cxrealcomclean}
	\begin{tabular}{|l|l|l|l|l|l|l|}
		\hline
		& L1\_weight & L1\_sumup & L1\_voting & L1\_w1 & kNN\_voting & kNN\_w1 \\ \hline
		Complex & 0.935	 	& 0.930     & 0.909      & 0.879  & 0.913       & 0.889   \\ \hline
		Real    & 0.928      & 0.929    & 0.888      & 0.856  & 0.912       & 0.885   \\ \hline
	\end{tabular}
	
\end{table*}

\begin{table*}
	\centering
	\caption{True Detection Rates using complex-valued CSI vs Real CSI: "Channel 157 with RFI" }\label{tab:cxrealcomnoise}
	\begin{tabular}{|l|l|l|l|l|l|l|}
		\hline
		& L1\_weight & L1\_sumup & L1\_voting & L1\_w1 & kNN\_voting & kNN\_w1 \\ \hline
		Complex & 0.851      & 0.854     & 0.770      & 0.612  & 0.760       & 0.603   \\ \hline
		Real    & 0.801      & 0.807     & 0.752      & 0.558  & 0.748       & 0.575   \\ \hline
	\end{tabular}
	
\end{table*}

In this section, we explore the advantage of complex-valued CSI based classification. 
As illustrated in Fig.~\ref{fig:com_example}, complex-valued CSI can enlarge the separation between  classes. To further support this, we use the data from the datasets ``\emph{Channel 157 without RFI}'' and ``\emph{Channel 157 with RFI}'' and calculate the distances \emph{between} and \emph{within} classes. 
We use the metric, which was introduced to calculate the distance between two classes and tune SVM hyperparameters \cite{sun2010analysis}. The distance is defined as 
\begin{multline}
D(C_1,C_2) =  \frac{2}{n_{1}n_{2}} \sum_{i=1}^{n_1} \sum_{j=1}^{n_2} 
{d( \frac{x_{1,i}}{||x_{1,i}||}, \frac{x_{2,i}}{||x_{2,i}||} )}  \\
-  \frac{1}{n_{1}^2}  \sum_{i=1}^{n_1} \sum_{j=1}^{n_1} {d(\frac{x_{1,i}}{||x_{1,i}||},\frac{x_{1,j}}{||x_{1,j}||})} \\
-  \frac{1}{n_{2}^2}  \sum_{i=1}^{n_2} \sum_{j=1}^{n_2} {d(\frac{x_{2,i}}{||x_{2,i}||},\frac{x_{2,j}}{||x_{2,j}||})} \label{equ:dis}
\end{multline}
, where $C_1$ and $C_2$ are samples for two classes, $n_1$ and $n_2$ are the numbers of samples in each class. $x_{k,i}$ is the $i$-th sample in Class $k$. The function $d$ calculates the Euclidean distance between two vectors.  
This metric considers distances between two classes as well as the distance within each class. We average distances between each pair of classes to calculate \emph{class distance}. The longer class distance will help improve the classification performance as a result.

\begin{figure}[htb]
	\centering
	\includegraphics[scale = 0.4]{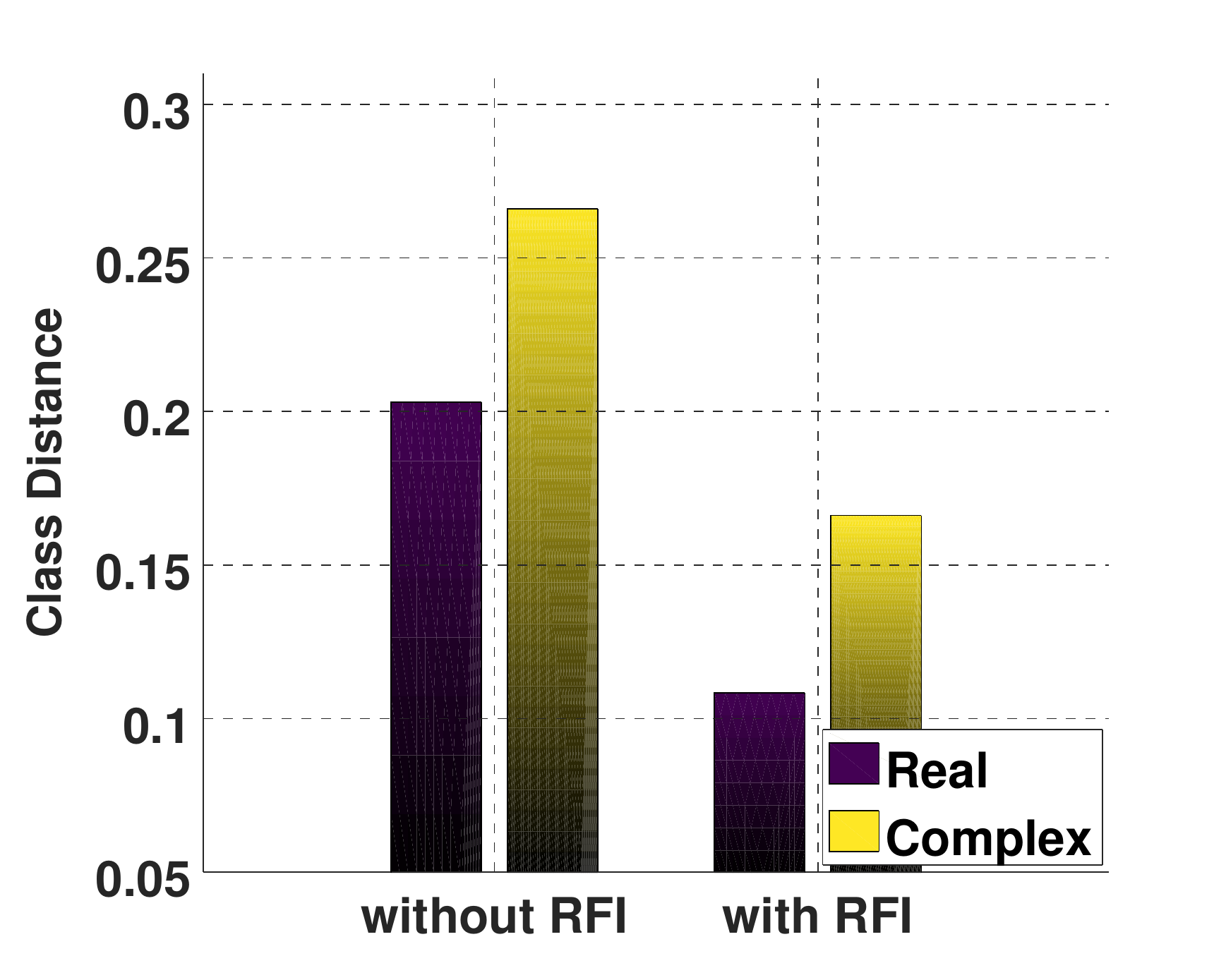}
	\caption{Class difference in ``Channel 157 without RTI'' and ``Channel 157 with RTI'' }\label{fig:disdiff}
\end{figure}

Fig. ~\ref{fig:disdiff} shows the  \emph{class distances} in the datasets ``\emph{Channel 157 without RTI}'' and ``\emph{Channel 157 with RTI}'' by using real-valued CSI and complex-valued CSI. 
Please note the real-valued SRC is used for real-valued CSI based classification. 
% The class distances increase significantly when using complex-valued CSI. 
In the dataset ``\emph{Channel 157 without RTI}'', the class distance is 0.203 using real-valued CSI. The class distance rises 31.0\% to 0.266 when using complex-valued CSI. The class distance decreases to 0.108 when using real-valued CSI in the dataset ``\emph{Channel 157 with RTI}'' due to the existence of RFI. When using the complex-valued, the class distance significantly rises 52.0\% to 0.166 when using complex-valued. There are two observations: (1) RFI can reduce the class distance, which results in the reduction of the recognition performance. (2) Complex-valued CSI can increase the class distance significantly, and improve the classification results in return. 

We will further confirm this using the true detection rates. 
Table~\ref{tab:cxrealcomclean} shows the true detection rate in the dataset ``\emph{Channel 157 without RTI}''.
%In other words, when design a metric of distance, we need to consider distance 
%We use four data sets as well, i.e. ``\emph{whole bandwidth without RFI}'', ``\emph{whole bandwidth with RFI}'', ``\emph{Channel 157 without RFI}'' and ``\emph{Channel 157 with RFI}''. 
%For all datasets, we use window size $ws = 5$ and bandwidth window $B = 20$~MHz.
%Table~\ref{tab:cxrealcomcleanwhole} and 
It shows the accuracy stays similar without RFI when using complex-valued and real-valued based CSI. 
%The difference is not obvious neither in the dataset ``\emph{whole bandwidth with RFI}'' as shown in Table \ref{tab:cxrealcomnoisewhole}, because only 20~{MHz} out of 125~{MHz} bandwidth is affected by RFI. 
%We show the performances between complex CSI based and real CSI based classification methods in Table \ref{tab:cxrealcomclean} and Table \ref{tab:cxrealcomnoise}.  
%Table \ref{tab:cxrealcomclean} indicates real CSI based methods achieve the accuracy 93.5\%, 93.0\%, and 90.9\% using $\ell_1$-$weighting$, $\ell_1$-$sumup$ and $\ell_1$-$voting$ respectively. When these data fusion methods apply on real CSI, they can achieve very similar performances. 
%
When looking at the dataset ``\emph{Channel 157 with RTI}'' as shown in Table \ref{tab:cxrealcomnoise}, the accuracy improves significantly using complex-valued CSI. When using complex-valued CSI, the true detection rate of $\ell_1$-$win1$ increases from 55.8\% to 61.2\%. Applying data fusion methods, real-valued CSI based $\ell_1$-$weighting$, $\ell_1$-$sumup$ and $\ell_1$-$voting$ can achieve 80.1\%, 80.7\% and 75.2\% respectively. When using complex-valued CSI, their accuracy increases to 85.1\%, 85.4\% and 77.0\%. 
% 
% TODO: add and discuss confusion metrics 
This shows the complex-valued CSI can help improve recognition performance with RFI. 
Fig.~\ref{fig:cm_win10_clean}, Fig.~\ref{fig:cm_win10_noise}, Fig.~\ref{fig:cm_win10_clean_real} and Fig.~\ref{fig:cm_win10_noise_real} demonstrate the confusion matrices of these settings, which also confirms the fact that complex-valued CSI achieves better recognition performance. 
complex-valued CSI can supply phase information additional to amplitude information. When the environment is without RFI, the amplitude information is capable to show clear distinguishable patterns, so the performance stays similar in the dataset ``\emph{Channel 157 without RTI}''. However, the additional phase information is important for the scenario with RFI and help improve the recognition accuracy, where CSI amplitude are not sufficiently informative.

\subsubsection{ Effect of Sanitised CSI }\label{subsec:processed}

\begin{figure}[htb]
	\centering
	\includegraphics[scale = 0.4]{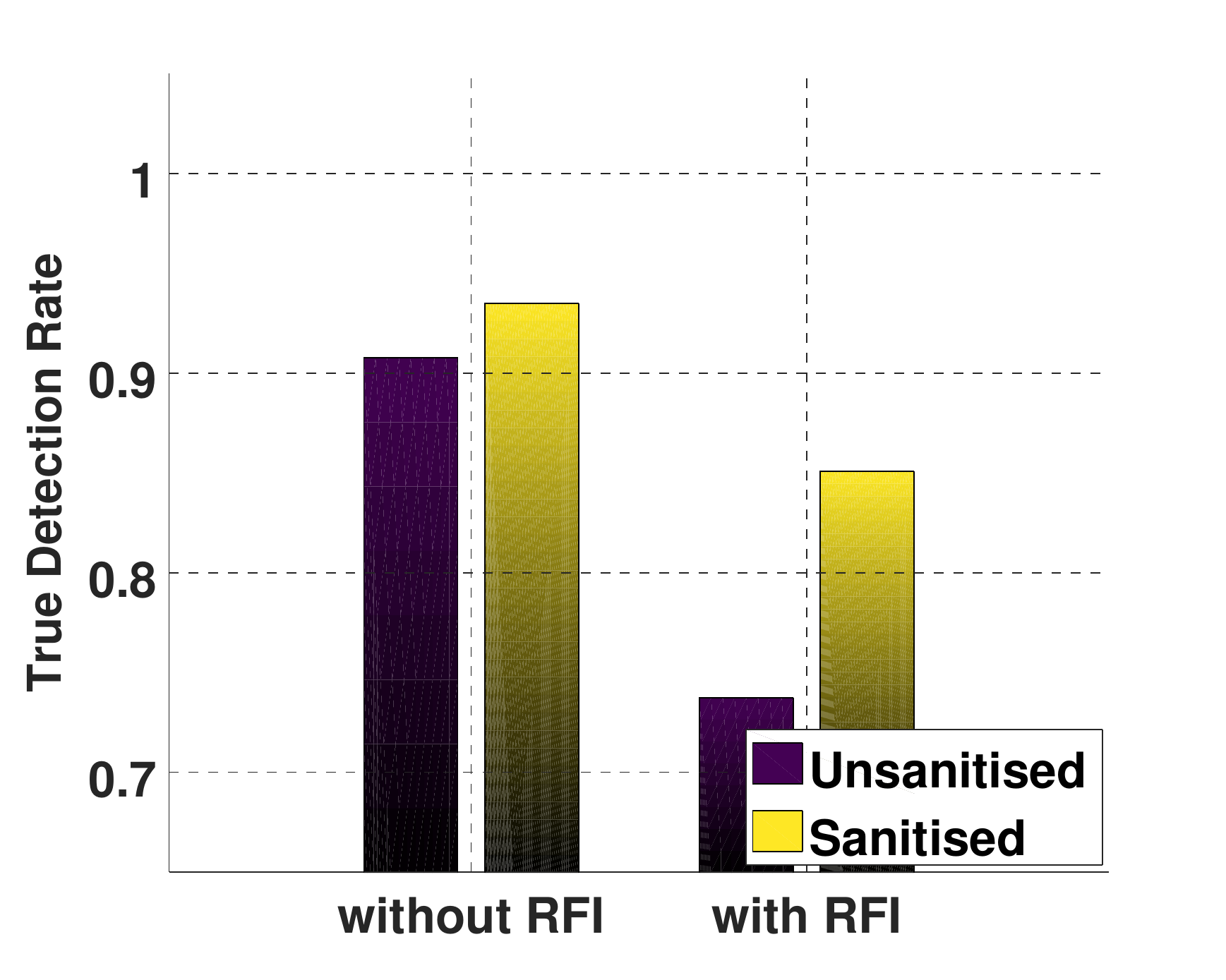}
	\caption{The comparison between sanitised and unsanitised complex-valued CSI in ``Channel 157 without RTI'' and ``Channel 157 with RTI'' }\label{fig:processed}
\end{figure}

To show the effect of Sanitised CSI, we also use ``Channel 157 without RFI'' and ``Channel 157 with RFI'' datasets and the default window size $ws$=0.5 second and bandwidth window $B$=5 MHz. For unsanitised CSI, we directly uses the raw CSI vectors without any processing. 
Fig. \ref{fig:processed} shows the true detection rate using sanitised and unsanitised CSI. In the dataset ``\emph{Channel 157 without RFI}'', the accuracy increases from 90.76\% to 93.48\%. When introducing RFI in the dataset ``\emph{Channel 157 with RFI}'', the accuracy drops to 73.73\% with unsanitised CSI. After sanitising, the accuracy increases to 85.08\%. This shows it can significantly improve performance using sanitised CSI.

\subsubsection{Effect of Complex-valued SRC }\label{subsec:complexsrc}
\begin{figure}[htb]
	\centering
	\includegraphics[scale = 0.4]{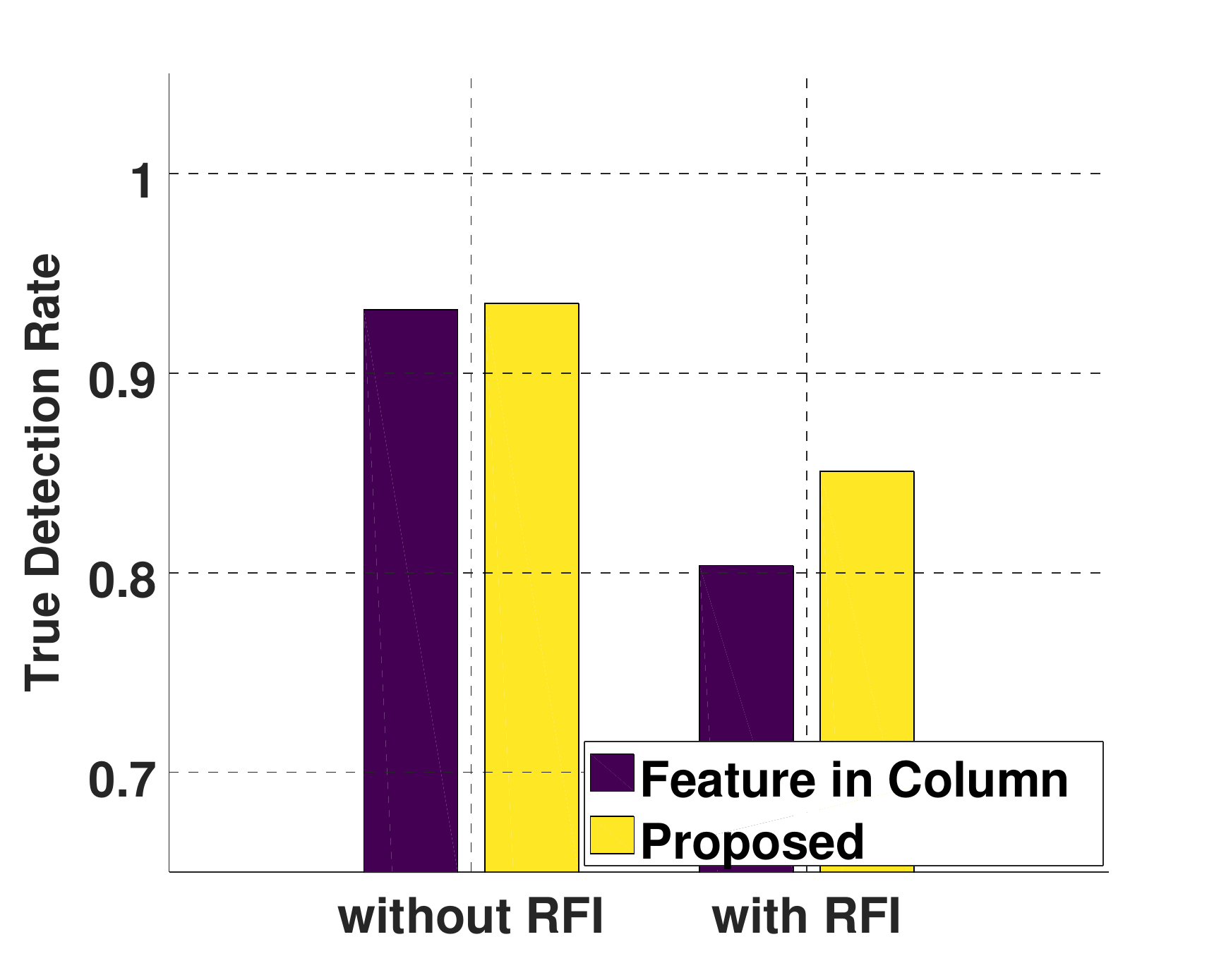}
	\caption{The comparison between sanitised complex-valued CSI and CSI features in one column  in ``Channel 157 without RTI'' and ``Channel 157 with RTI'' }\label{fig:featuresinrow}
\end{figure}
In this section, we discuss the effect of SRC using complex-valued CSI. We are the first to apply complex-valued $\ell_1$ minimisation for classification tasks. To compare with complex-valued SRC, we combine phase vectors and amplitude vectors into one column as inputs (\textit{``Features in Column''}) for a real-valued SRC. Figure \ref{fig:featuresinrow} shows the true detection using in the dataset ``\emph{Channel 157 with RFI}'' and ``\emph{Channel 157 without RFI}''. In the dataset ``\emph{Channel 157 without RFI}'', the accuracy slightly increases from 93.17\% to 93.48\%. With RFI in the dataset ``\emph{Channel 157 with RFI}'', the accuracy decreases to 85.08\% using our proposed method, while the true detection rate is only 80.36\% using \textit{``Features in Column''} method.

}

%% file: Tex/related.tex
\vspace{-2mm}
\section{Related work}
\label{sec:related}
We have already discussed the most related work, i.e. work on using CSI for activity classification, in Section \ref{sec:intro}. \vtbo{In this section, we discuss work in three areas: activity recognition, pattern recognition from radio data and the application of compressed sensing in wireless networks. }
\vspace{-2mm}
\subsection{Activity recognition}
% acceleration sensor
Activity recognition forms the basis of many context aware pervasive computing applications. We can broadly classify activity recognition according to whether they are sensor-based or camera-based. Many sensor-based activity recognition systems have been proposed. Acceleration sensor is one of the most frequently used sensors \cite{kwapisz2011activity,bao2004activity,ravi2005activity}. This is because miniature acceleration sensors are cheap and readily available, and they can be found on all smartphones. For example, Keally et al.~\cite{keally2011pbn} used sensors smartphone as well as sensors wore on wrists, ankles and head to distinguish between walking, cycling, sitting and other activities. Recently, a number of wristbands, that are equipped with acceleration sensor, are available in the market. Products such as Jawbone Up \cite{jawboneup} and Xiaomi Mi Band \cite{xiaomiwristband} can achieve good activity recognition accuracy but they require the users to wear the devices. 

Microphone is another sensor that has been used in activity recognition. Hao et al.~\cite{hao2013isleep} present a method to monitor sleep quality using microphone; however, it is not sure whether the same method will function in daily activity recognition in a noisy environment. Yatani and Truong~\cite{yatani2012bodyscope} designed a  wearable acoustic sensor which can be used to record the sound near the throat of the user, and use the measurements for activity recognition. Again, the issue is that the subject has to wear a sensor.

% camera based activity recognition

Cameras are also widely used for activity recognition, localisation and tracking \cite{oliver2002layered,harville2004fast,zhao2005real,cohen2003inference,kinect,Leap,Shen2012Efficient}. 
An advantage of camera sensors is that they free the subjects the need to wear or to remember to wear a device. Another advantage is that they provide very rich data which can be used to distinguish between many different activities. However, the Achilles' heel of using camera for activity recognition is privacy concern. Also, cameras can only cover a limited area. For monitoring in an apartment, a camera is needed in each room. On the contrary, activity recognition using radio signals can cover a much wider area and can ``see" through walls, while cameras cannot. We have therefore chosen to use device-free radio-based activity recognition which does not need subjects to carry a device and has no privacy concerns.

\vspace{-2mm}
\subsection{Radio based pattern recognition}
% RSSI localization

% device free localization

The propagation of radio waves in an environment is affected by the objects and people in the environment, through reflection, diffraction, constructive and destructive interference and so on. There is much interest in using the received signal characteristics to infer about the attributes of people and objects in an environment. The received signal characteristics used can be coarse or fine grained. 

An example of coarse grained radio signal feature is RSSI which measures the received signal power. RSSI has been successfully used in device-free localisation 
\cite{WilsonRTI:2010,ZhaoNoise2011,KaltiokallioBP12,ZhaoKRTI:2013,WeidRTI:2015,Xu:2012:sensys}. This is because a person standing in an area between the transmitter and receiver can attenuate, reflect, scatter the radio waves. These effects create a characteristic pattern in RSSI which can be used to infer the location of people in the environment. 

% fine-grained pattern recignition

Unfortunately, only limited information on the environment can be inferred from RSSI. There is a growing interest to use fine grained features of radio signals for inference. This is also fuelled by the availability of API to query CSI from WiFi chipsets such as Intel 5300~\cite{Halperin_csitool} and Atheros 9390~\cite{sen2013avoiding}. CSI has been used for many pattern recognition problems, including localisation~\cite{sen2012you}, human detection~\cite{zhou2013omnidirectional}, activity recognition~\cite{wang2014eyes,sigg2014rf}, fine-grained gesture recognition~\cite{melgarejo2014leveraging} , ``lip-read" \cite{wang2014we}, emotion recognition \cite{zhao2016emotion}, identity recognition\cite{zeng2016wiwho,zhang2016wifi,wang2016gait}. As mentioned in Section \ref{sec:intro}, our work differs from earlier work on using CSI for activity recognition in that we take RFI into consideration while earlier work did not. 

%boosts the application of CSI for pattern recognition, such as  and mouth motion recognition~\cite{wang2014we}. Melgarejo etc. use the WARP board to extract the similar CSI from Intel 5300 and Atheros 9390, and takes the amplitude and phase information from CSI for 

Radio signals have also been used to perform gesture recognition. WiSee designed by Adib et al.~\cite{Adib:2013:STW:2486001.2486039} and WiVi designed by Pu et al.~\cite{pu2013whole} used software-defined radio to extract the Doppler effect caused by the gesture. In order to reduce energy consumption, Kellogg et al.~\cite{Kellogg:2014} built AllSee which uses RFID tags and power-harvesting sensors for gesture recognition. However, these work can only recognise dynamic gestures and are not able to detect static activities because they rely on Doppler effect. 
To further improve the resolution of the radio signal based localisation and gesture recognition, Adib et al.~\cite{Adib:2014} designed WiTrack and obtain time-of-flight from the Frequency Modulated Carrier Wave~(FMCW) technology for localisation in 3 dimensions. Witrack has high resolution for localisation~(approximately 10~cm), but needs to use a bandwidth of 1.69~GHz. However, we show in this paper as few as 20~MHz of bandwidth can be used to distinguish static activities with good accuracy. Moreover, none of these works designed their systems with RFI, while our work takes RFI into consideration. 

\subsection{Application of Compressed Sensing on Wireless Sensor Networks}
Recently, compressed sensing has been applied to wireless sensor networks. SRC proposed by Wright et al. \cite{Wright:09} is one of the applications of compressed sensing which helps increase the recognition performance. Wei et al. \cite{wei2013real} developed an acoustic classification method on wireless sensor networks by applying SRC to increase the recognition performance and decrease the computation time to meet the requirement of real-time classification. \cxbo{Shen et al. \cite{Shen2014Face,shen2017learn}} optimised the SRC to boost the face recognition performance in smartphones. 

Besides recognition, compressed sensing is also applied to background \cxbo{subtraction \cite{Shen2012Efficient,shen2016real,xu2016sensor}}, data compression for in-situ soil moisture sensing \cite{Wu2012In}, and cross-correlation for acoustic ranging \cite{Misra2012Efficient} and GPS ranging \cite{misra2014energy}.

Compared with these works, this paper is the first to investigate the feasibility of SRC for radio-based activity recognition. Furthermore, our work also takes advantage of SRC for activity recognition with RFI in present.